\documentclass[sigconf]{acmart}

\AtBeginDocument{%
  }

\copyrightyear{2025} 
\acmYear{2025} 
\setcopyright{cc}
\setcctype{by-nc-sa}
\acmConference[CHI '25]{CHI Conference on Human Factors in Computing Systems}{April 26-May 1, 2025}{Yokohama, Japan}
\acmBooktitle{CHI Conference on Human Factors in Computing Systems (CHI '25), April 26-May 1, 2025, Yokohama, Japan}\acmDOI{10.1145/3706598.3713137}
\acmISBN{979-8-4007-1394-1/25/04}

\begin{document}

\title[Generative AI as a Playful yet Offensive Tourist]{Generative AI as a Playful yet Offensive Tourist: Exploring Tensions Between Playful Features and Citizen Concerns in Designing Urban Play}

\author{Peng-Kai Hung}
\affiliation{%
  \institution{Department of Industrial Design, Eindhoven University of Technology}
  \city{Eindhoven}
  \country{The Netherlands}
  \institution{National Taiwan University of Science and Technology}
  \city{Taipei}
  \country{Taiwan}
}
\email{p.hung@tue.nl}
\orcid{0000-0002-0300-9604}

\author{Janet Yi-Ching Huang}
\affiliation{%
  \institution{Department of Industrial Design, Eindhoven University of Technology}
  \city{Eindhoven}
  \country{The Netherlands}}
\email{y.c.huang@tue.nl}
\orcid{0000-0002-8204-4327}

\author{Rung-Huei Liang}
\affiliation{%
 \institution{Department of Design, National Taiwan University of Science and Technology}
 \city{Taipei}
 \country{Taiwan}}
\email{liang@mail.ntust.edu.tw}
\orcid{0000-0002-7294-8154}

\author{Stephan Wensveen}
\affiliation{%
  \institution{Department of Industrial Design, Eindhoven University of Technology}
  \city{Eindhoven}
  \country{The Netherlands}
}
\email{s.a.g.wensveen@tue.nl}
\orcid{0000-0001-8804-5366}

\renewcommand{\shortauthors}{Hung et al.}

\begin{abstract}
Play is pivotal in fostering the emotional, social, and cultural dimensions of urban spaces. While generative AI (GAI) potentially supports playful urban interaction, a balanced and critical approach to the design opportunities and challenges is needed. This work develops iWonder, an image-to-image GAI tool engaging fourteen designers in urban explorations to identify GAI's playful features and create design ideas. Fourteen citizens then evaluated these ideas, providing expectations and critical concerns from a bottom-up perspective. Our findings reveal the dynamic interplay between users, GAI, and urban contexts, highlighting GAI's potential to facilitate playful urban experiences through \textit{generative agency}, \textit{meaningful unpredictability}, \textit{social performativity}, and the associated offensive qualities. We propose design considerations to address citizen concerns and the `tourist metaphor' to deepen our understanding of GAI's impacts, offering insights to enhance cities' socio-cultural fabric. Overall, this research contributes to the effort to harness GAI's capabilities for urban enrichment.
\end{abstract}




\maketitle

\begin{figure*}
  \includegraphics[width=\textwidth]{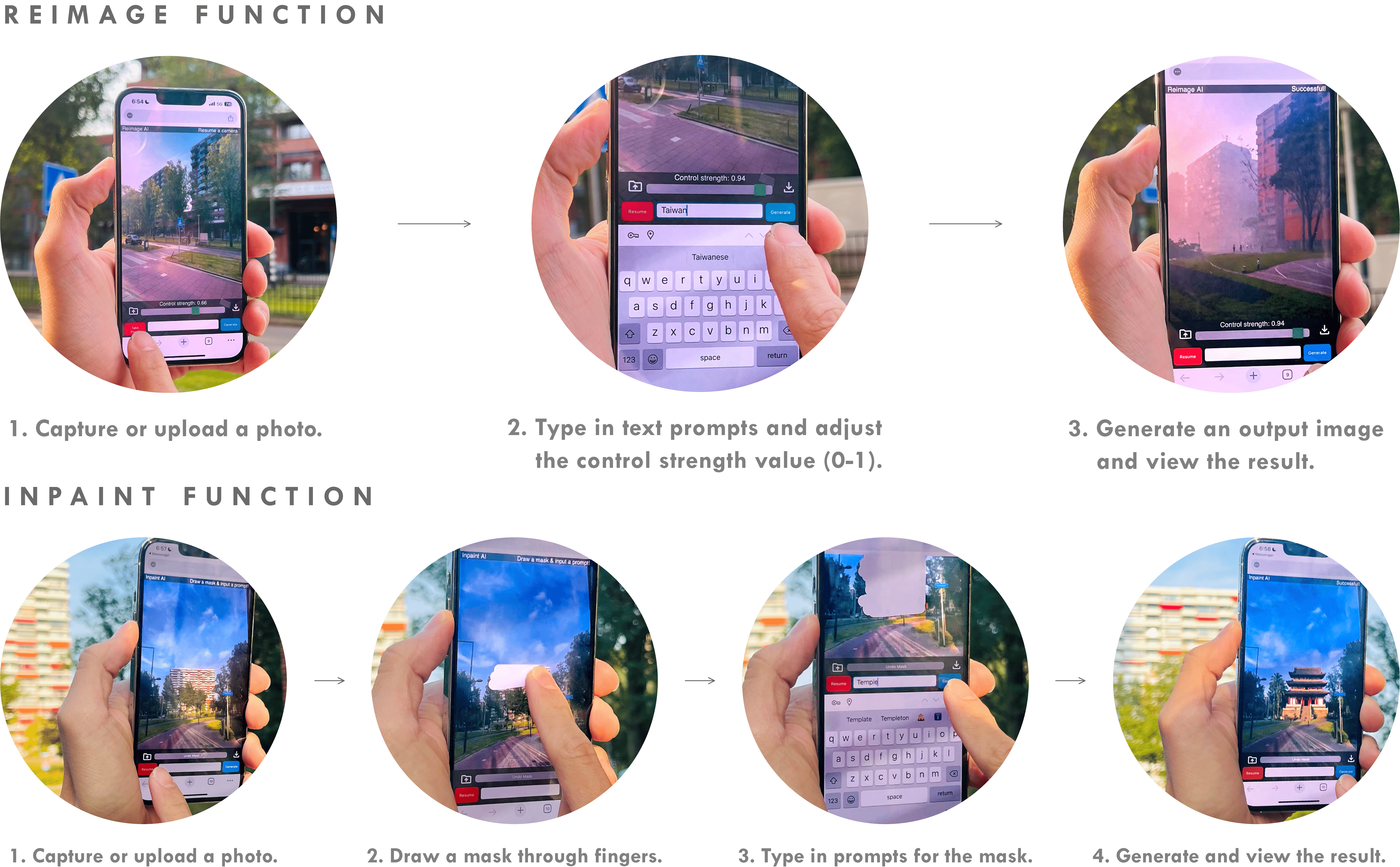}
  \caption{The interfaces of our image-to-image GAI tool: iWonder.}
  \Description{The interfaces of our image-to-image GAI tool: iWonder.}
  \label{fig:figure1_gai_tool}
\end{figure*}

\section{Introduction}
Play has been considered an essential quality that contributes to the emotional, social, and cultural flourishing of urban spaces \cite{Chew:OzCHI2021,Innocent:2018, Pang:2020}. The research domain of urban play focuses on leveraging technology to facilitate deliberate or spontaneous playful urban behaviors \cite{Nijholt:Springer2017, Bertran:2022}. This endeavor caters to citizens' needs and expectations to experience positive emotions, reimagine and learn from their urban surroundings, enhance social cohesiveness, and/or develop a sense of agency \cite{Bertran:2022}. However, technology-mediated urban play may also cause problems, such as conflicts between play and city operations, dictatorship of playfulness \cite{Bertran:2022}, insecurity, and disturbance to local communities \cite{Wetzel:2016,Wagner-Greene:2017aa}. Game and play researchers have underscored the need to scrutinize both opportunities and citizens' attitudes when incorporating technology into playful urban interactions \cite{Bertran:2022, Nijholt2020}. In response, this research seeks to understand the tensions of playful urban interaction enabled by the recent emerging Generative Artificial Intelligence.

Generative Artificial Intelligence (GAI) technologies have become increasingly prevalent in city life~\cite{Xu:2024,Lin:CC2023,Xu:2023, zhu:2023,Luusua:2020} and offer novel design possibilities for urban play. GAI refers to AI systems capable of producing new and plausible content across diverse modalities (e.g., text and image data)~\cite{Muller:GenAICHI2022}. Some of these systems can process data from urban environments and generate inspiring results on a wide range of topics~\cite{Sun:CHI2024,Lau:DIS2024,Walsh:CSCWWiP2019,Chiou:2023,Fu:DIS2024}, which may foster storytelling of places and joyful dialogues between individuals and city spaces~\cite{Lin:CC2023,Fu:DIS2024}. Furthermore, several GAI tools enable users to easily modify street photos (e.g.,~\cite{Lau:DIS2024,DutchCyclingLifestyle2024,shi2024}), making it possible to engage the public in generating interesting content that facilitates connectivity and discussions around city development~\cite{Lau:DIS2024}. 

Despite the potential of GAI for bringing playfulness, there is a growing attention to AI systems' risks and harms in public applications, warranting further investigation and consideration~\cite{Guridi:2024,Mlynar:CHI2022,Alkfairy:2024,Lin:CC2023,Fu:DIS2024,Moreno-Ibarra2024,zhu:2023,Luusua:2023aa}. GAI may produce socially biased outcomes, be misused, or generate misleading information \cite{Bird:AIES2023,Hung:DISWiP2024,Laura:FACCT2022,Weisz:CHI2024}. These issues, along with new tensions and societal implications, might escalate as more GAI tools incorporate data from urban environments and are optimistically promoted for playful purposes~\cite{Googlepixel,Adobe,Brazil}. For instance, the image-to-image GAI tool embedded in Google Pixel 9 encouraged users to playfully ``reimagine'' surroundings. Yet, they could creatively produce convincing pictures of fabricated accidents or crime scenes, which amuses certain users but disturbs other citizens~\cite{Googlepixel}. Although prior work has begun examining the ethical challenges accompanied by creative manipulations of GAI in public venues~\cite{Fu:DIS2024,Lin:CC2023,Guridi:2024}, few GAI tools and studies conducted inquiries within situated city contexts~\cite{Petridis:2024}, limiting the understanding of playful experiences enabled by GAI and their pitfalls in urban spaces. This gap causes ambiguity among designers and city stakeholders about how the introduced AI systems enhance and undermine the creation of engaging and inclusive public spaces~\cite{Long:2019,zhu:2023,Guridi:2024,Luusua:2020,Luusua:2023aa}. Therefore, this work aims to investigate GAI through a balanced and critical approach which uncovers its situated potential for playful urban interaction while addressing public apprehensions.  Particularly, we inquire into tensions between GAI playfulness and citizen considerations in urban play, including three research questions: What playful features emerge when interacting with GAI in public spaces? How could GAI be integrated into urban play? What are citizens’ concerns and risks associated with the GAI-enabled applications?

\begin{figure*}
  \centering
  \includegraphics[width=\linewidth]{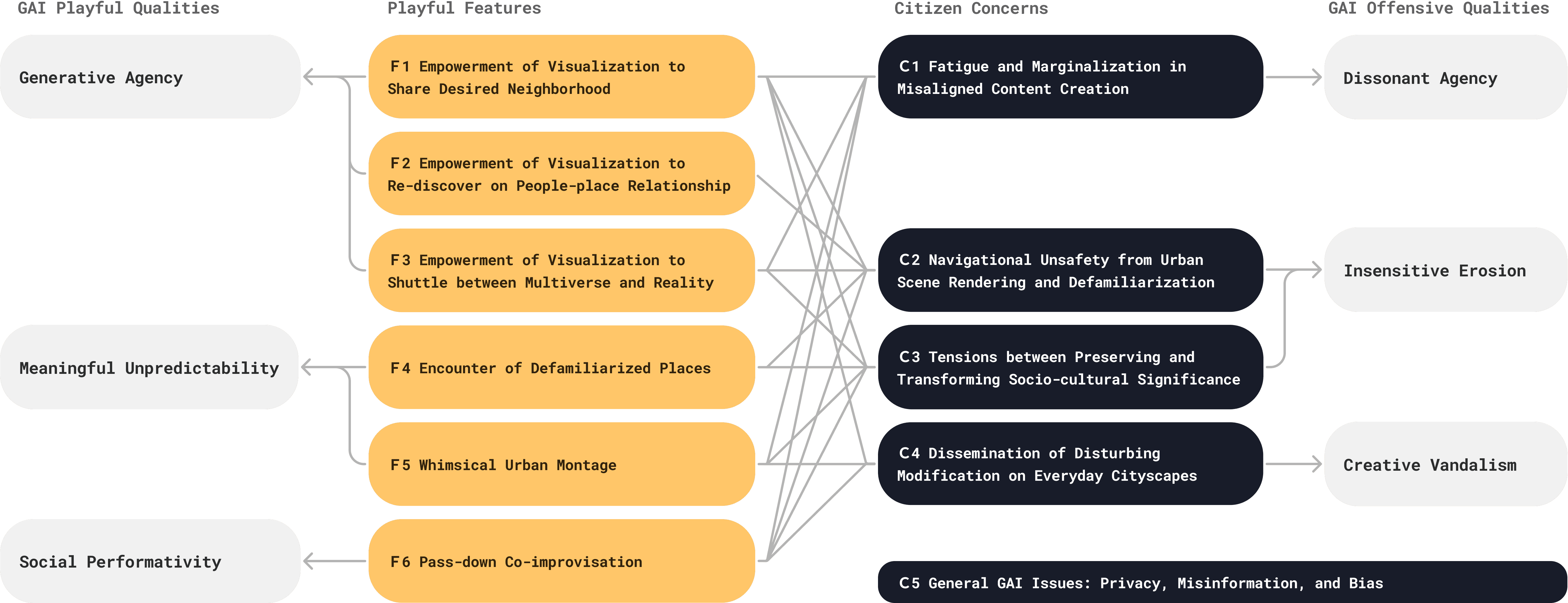}
  \caption{A Summary Illustrating the Tensions of GAI-enabled Urban Play.}
  \Description{A Summary Illustrating the Tensions of GAI-enabled Urban Play.}
  \label{fig:figure20}
\end{figure*}

We designed and developed a situated GAI tool, named iWonder (Figure~\ref{fig:figure1_gai_tool}). It empowers people to explore, create, and evaluate their ideas in the wild using image-to-image models with their inputs. iWonder allows participants to capture an image of their surroundings, input text prompts, and use two functions: 1) Reimage: transforming photo styles and elements of the image while retaining its overall composition, and 2) Inpaint: painting over designated areas of the image and replacing those with GAI-generated content (Figure~\ref{fig:figure1_gai_tool}). Then, we invited 14 designers to engage in the design exploration workshop in which they wandered through urban environments, took photographs, and interacted with our tool. Next, they participated in group interviews about their experiences and the playful features of GAI in cities, followed by brainstorming sessions on GAI-enabled urban play ideas. These ideas were discussed with 14 citizens with various socio-cultural backgrounds in the citizen evaluation workshop, where they reviewed the ideas, prototyped them with our tool, and provided their opinions and suggestions. The analysis of their feedback indicated how the playful features embedded in these ideas facilitated urban play qualities while possibly leading to concerns. Finally, extracted from playful features and their concerns, we discussed three categories of GAI playful and offensive qualities, illustrated their relations to each other (Figure~\ref{fig:figure20}), and further proposed the metaphor ``GAI as a playful yet offensive tourist.'' 

This work contributes to uncovering tensions between design opportunities and challenges of GAI-enabled urban play, advancing a creative yet critical approach to city enrichment. Specifically, we offer the following contributions: 1) We developed a situated GAI tool that empowers designers in exploring new playfulness in situ and creating design ideas, while also empowering citizens to experience them and express related concerns, surfacing contextual tensions. 2) Our empirical findings from two workshops identify six Playful Features and five Citizen Concerns, deepening the HCI community's understanding of the dynamic relationship and friction between urban contexts, GAI, and users. 3) We contribute the GAI playful qualities (i.e., \textit{generative agency}, \textit{meaningful unpredictability}, \textit{social performativity}), offensive qualities (summarized in Figure~\ref{fig:figure20}), alongside the `tourist metaphor.' These insights serve as key takeaways and implications, guiding designers and researchers in crafting playful experiences and placemaking with GAI while remaining mindful of their potential issues in public deployment. This work extends broader urban research efforts to bridge AI characteristics with empirical realities and citizen perspectives in public spaces, fostering city stakeholders' evaluation and reflection of AI technology for city flourishing. 

\section{Related Work}
\subsection{Technology-mediated Urban play and Playfulness}
  \label{subsection2.1}
Although the concepts of urban play and playfulness `are difficult to define' and `substantial overlap'~\cite{Nijholt:Springer2017}, an increasing amount of work in this area shares the same initiatives for `making urban space and infrastructure more playable and citizens more playful'~\cite{Chisik:2022}. According to urban play literature, `playfulness' links to `playful play.' This concept `is accompanied by a particular positive mood state' in which people tend to behave and think `in a spontaneous and flexible way'~\cite{Nijholt:Springer2017}. Chisik et al. framed `Urban play' as `spontaneous and non-instrumental play' and `digital play' mediated by technology in city environments~\cite{Chisik:2022}. Urban play holds a similar perspective to Playable Cities, which considers the enjoyment of play a necessity rather than a selective quality for urban prosperity~\cite{Bertran:2022}. Citizens are not work-producing machines but are driven by social and emotional relationships, pleasure, and agency~\cite{Bertran:2022,Brown:2015}. Therefore, these values should be carefully considered when incorporating technologies into urban objects and spaces~\cite{Bertran:2022}. 

Instead of regarding technology as an independent layer of the metropolis that produces efficiency and functionality invisibly~\cite{Aoun:2013,Mora:2017,Dameri:2014,Paulos:2005}, urban play expects it to meet the playful applications imagined by citizens~\cite{Innocent:2018}. Meanwhile, citizens could reorganize their stories through them and participate in the city network~\cite{Nijholt:Springer2017}. This bottom-up vision is also highlighted and manifested in urban play studies. They inquire into citizens' experiences of urban technology, invite them to comment on design proposals, or empower them to influence design decisions. These civic engagements in the design process avoid an authoritarian design model that solely caters to the expectations of city administrators, while crafting experiences genuinely benefiting the public~\cite{Bertran:2022}. 

In HCI communities, there is a growing interest to investigate the capability of new technologies to create playfulness and urban play forms, including mixed reality games~\cite{Wetzel:2016,Wetzel:2008,Flintham:CHI2003,Karpashevich:MUM2016}, virtual reality applications \cite{Harley:DIS2019,Hung:2024}, pervasive and location-based games~\cite{Stenros:2007,Benford:2005,Kasapakis:PCI2013,Karpashevich:MUM2016,Capra:Multimedia2005,Pang:2020}, urban planning games~\cite{Ampatzidou:2018,Poplin:ICCSA2011}, the application of game techniques for solving urban problems~\cite{Thibault:2019}, and other playful installations in city spaces~\cite{Tobias:2010,urbanimals:2024,Long:2019,Stokes:CHIPLAY2020,Tobias:2010,helloLampPost}. Based on the playful characteristics of technologies and these play forms, their qualities have been proposed and analyzed~\cite{Bertran:2022,OHara:CHI2008,Innocent:2019,Innocent:2021,Chew:OzCHI2021}. Here, we focus on four dimensions of urban play qualities: emotional, intellectual, social, and agential. The emotional aspect centers on generating positive emotions and making the tedious situations more enjoyable and entertaining. By turning ordinary spaces into temporary playgrounds, play provides moments of fun, laughter, and healthy experiences that benefit our life quality~\cite{Bertran:2022}. The intellectual facet describes the cognitive stimulation and of play which foster a sense of curiosity, creativity, and discovery, thereby obtaining a deeper understanding, reflection, reimagination, or new meanings of the urban environment~\cite{Chew:OzCHI2021,Wang:RTD2019,DeLange:2015,Hung:DISWiP2024,Hung:2024}. Regarding the social aspect, play participates in building meaningful connections and communities both physically and virtually. It creates opportunities for shared experiences among citizens and towards places, contributing to social cohesiveness and collective city-making~\cite{Bertran:2022,Pang:2020,Procyk:2014,Lancel:2019}. Finally, the agential dimension highlights how play can empower individuals by providing a sense of agency and meaningful action. Possessing a disruptive and appropriative nature~\cite{Bertran:2022,Innocent:2021,Korte:2018}, play allows individuals to creatively engage with and challenge the status quo of their surroundings, introducing freedom and possibility without causing complete upheaval. 

Following urban play's ongoing efforts to uncover the play potentials of novel technologies, this work explores the empirical features and interaction forms of GAI tech that may support the above-mentioned qualities. 

\subsection{GAI as a Design Material in Public Venues}
GAI has become increasingly integrated into urban lives, attributed to the growing release of highly accessible tools, such as ChatGPT and Stable Diffusion. In the HCI community, GAI tools are considered design materials for exploration and experimentation~\cite{Benjamin:2023,Chiou:2023,Brand:CHI2023,Blythe:2023}. Investigating their various applications, user experiences, and value to society becomes a crucial consideration~\cite{Muller:GenAICHI2022}.  

A trajectory of work starts to explore GAI’s capabilities to support diverse experiences and designs for the public, implying its potential to enrich the socio-cultural fabrics of cities. For example, Suh et al. examined AI’s role in collaborative music composition, showcasing the influence of AI on social dynamics in creative processes~\cite{Suh:CHI2021}. Several studies crafted meaningful interactions with GAI in museum contexts. Sivertsen and L{\o}vlie employed a deep generative model to design an interactive drawing table in their research-through-design project, seeking to assist museum visitors in exploring the aesthetics of Edvard Munch's art~\cite{Sivertsen:DIS2024}. Fu et al. curated an AI-generated content exhibition and conducted an empirical study of how text-to-image AI inspires personal narratives with emotions and reflection on cultural heritages~\cite{Fu:DIS2024}. Although concentrating less on playfulness in urban environments, they highlight the design opportunity of GAI to enhance citizens’ creativity and cultural understanding. Some projects provide GAI tools for people to visualize alternative cities, which stimulates conversations and co-speculation about the future of cityscapes. For instance, Wander 2.0~\cite{Sun:CHIEA2023} invites visitors to generate future narratives or visualizations of realities about specific locations. A few studies also developed platforms for citizens to modify urban scenes~\cite{Epstein:2022,Lau:DIS2024,Rafner:CC2021,Guridi:2024}. They demonstrate how GAI may allow citizens to express opinions and reimagine public spaces with pleasure. However, situated explorations of their use and empirical data gathered from designers and citizens are limited, restricting our understanding of the complex relationships intertwined between GAI, users, and the socio-cultural dynamics of urban places. 

Finally, an increasing body of work on GAI’s joyful applications in outdoor urban settings, motivating us to further understand its capability for constructing playful urban interaction. For instance, recent projects proposed playful interactions with GAI that reimagine urban scenes, including a website from the Netherlands Board of Tourism that adds cycling pathways to street views~\cite{DutchCyclingLifestyle2024}, a camera that generates alternative appearances of places based on location data~\cite{Karmann:2023}, and a speculation design where residents input local stories as prompts to create paintings on murals~\cite{Lin:CC2023}. Hung et al. investigated how users engage in city exploration with image-to-image AI and highlighted GAI's potential role in encouraging discoveries and speculations of places~\cite{Hung:DISWiP2024}. Furthermore, Long et al. reviewed previous studies and their work to derive the principles of technology, interaction, and research design for crafting co-creative Al in city areas~\cite{Long:2019}. Similarly, Li et al. developed an Al-embedded prototype which synthesized citizens’ sketches to encourage their participation in creating street art~\cite{Li:CHIEA2020}. Their efforts focused on cultivating creativity among citizens and inspiring designers to enhance the creative city agenda with AI.

Drawing on the endeavors of these studies, this work seeks to uncover how GAI could playfully reconfigure cities and understand how citizens respond to the innovative play forms enabled by GAI. 

\subsection{Challenges of GAI Applications}
Although GAI holds great potential as a design material to playfully reconfigure cities, its limitations may cause negative experiences and even societal harm. Studies have shown that GAI's inaccuracy and inconsistency in generated content led to dissatisfaction, frustration, confusion, and disengagement~\cite{Fu:DIS2024,Weisz:CHI2024,Weisz:CoRR2023}. These shortcomings diminish the intended experience and alienate users who expect more precise and meaningful interactions. Furthermore, the ethical issues surrounding GAI's use in public spaces are of growing concern. The United Nations has emphasized the need to safeguard human dignity and well-being when deploying AI technology, especially in public and civic contexts~\cite{UNESCO:ethics}. A series of studies have investigated the ethical implications of GAI and identified critical risks, such as privacy violations, discrimination, toxic content, and potential misuse~\cite{Hagendorff:CoRR2024,Kenthapadi:KDD2023,Chu-Ke:2024,Alkfairy:2024}. Particularly, literature on GAI applications in public spaces found that due to the historical and cultural biases in training data for generative models, GAI might generate biased and false information, perpetuating existing inequities and stereotypes~\cite{Fu:DIS2024,Epstein:2022}.

Notably, while the above-mentioned research examines GAI issues in various contexts, it focuses less on the challenges of incorporating GAI into urban interaction design. Although existing studies address broader concerns of urban technologies in shaping Playable cities (e.g.,~\cite{Bertran:2022,Araake:2021,Carter:2020}), the specific ways in which GAI interaction might conflict with citizens' needs, deter public participation, or introduce risks remain underexplored. Therefore, it is worth investigating whether the issues previously identified also manifest in the urban play context, aiding designers in making comprehensive decisions and preventing potential harm to the public. 
 
\section{Methods: Overview and iWonder Development}
This work aims to explore how GAI could be used for transforming urban cities into playful environments and further understand how citizens respond to these new forms of play. We present our design research process sequentially as it occurred. The whole study was approved by the Ethical Review Board. All the collected data was anonymized to protect participants' privacy. Initially, we developed a GAI tool tailored for the subsequent phases of our study. We carried out a design exploration where designers used the tool to explore urban spaces, shared experiences, and created design concepts (Section 4). We detailed the outcomes as six playful GAI features and GAI-enabled urban play design ideas (Section 5). Next, we conducted a citizen evaluation that engages citizens to discuss, prototype, and refine the proposed design ideas using our GAI tool (Section 6).

\subsection{iWonder: Situated Playful Reimagination with GAI}
To facilitate the explorative and evaluative activities of designers and citizens, respectively, we developed a GAI tool capable of recreating images based on provided inputs, including an image, a text prompt, and/or a control value. The tool features: 1) Reimage function: generating images by maintaining the structure of the input image but altering the content based on provided inputs. The tool is designed as a mobile interface that allows users to generate images in the wild, and 2) Inpaint function: modifying images by filling in or replacing a specific area with new content.  

\subsubsection{Reimage Function}
The Reimage interface contains an image capture button, an input field, an image generation button, a control strength, and a display area (Figure~\ref{fig:figure1_gai_tool}, top).  Users can 1) capture an image using the back camera or upload a photo from the phone's photo library, 2) describe the desired changes using a text prompt, 3) adjust the control strength value from (0 to 1) to determine how much (structure) control on the image generation, 4) generate an output image, and 5) view the result in the display area.

\subsubsection{Inpaint Function}
The Inpaint interface consists of an image capture button, an input field, and a display area (Figure~\ref{fig:figure1_gai_tool}, bottom). The interface has a drawing function that allows users to draw a mask to indicate a specific area to be filled in or replaced on the image display area. When using the Inpaint function, users can 1) capture an image using the back camera or upload a photo from the photo library on the phone, 2) draw a mask with their fingers to specify an area, 3) describe what they want to fill in or replace with a text prompt, 4) generate an output image, and 5) view the result in the display area.

\subsubsection{Implementation} 
We implemented the GAI tool using p5.js\footnote{https://p5js.org} and StabilityAI REST API (v2beta)\footnote{https://platform.stability.ai/docs/api-reference} on the platform developed by our department. This platform supports the necessary web infrastructure, including data storage, website hosting, and remote API access. The ``Structure'' API was used for the reimage function, and the ``Inpaint'' API for the inpaint function under the Stable Image editing service, with each API call costing USD 0.03. Prior to the two studies, the researchers had informed the participants that their prompts and photos would be transmitted to the external server when using iWonder. All participants consented to this arrangement before participating in the studies. 

\section{Design Exploration}
\subsection{Research objective} 
The aim of our design exploration is to explore how GAI can support urban playfulness. Design activities include in-situ explorations, group interviews, and ideation sessions, designed to engage designers to play with AI, design with AI, and prototype with AI `in the wild'~\cite{Malsattar:2019}. This approach prioritizes the situated and embodied nature of user experience, encouraging designers to engage with AI in real-world contexts. This interaction allows them to leverage GAI as a tool to make diverse interpretations of the context, understand GAI's capabilities, and initiate idea generation that can be promptly iterated. We believe this approach is instrumental in uncovering the dynamic between GAI, users, and city environments, revealing new design opportunities for playful urban experiences. 

\subsection{Designer Participants and Recruitment} 
\begin{figure*}
  \centering
  \includegraphics[width=\linewidth]{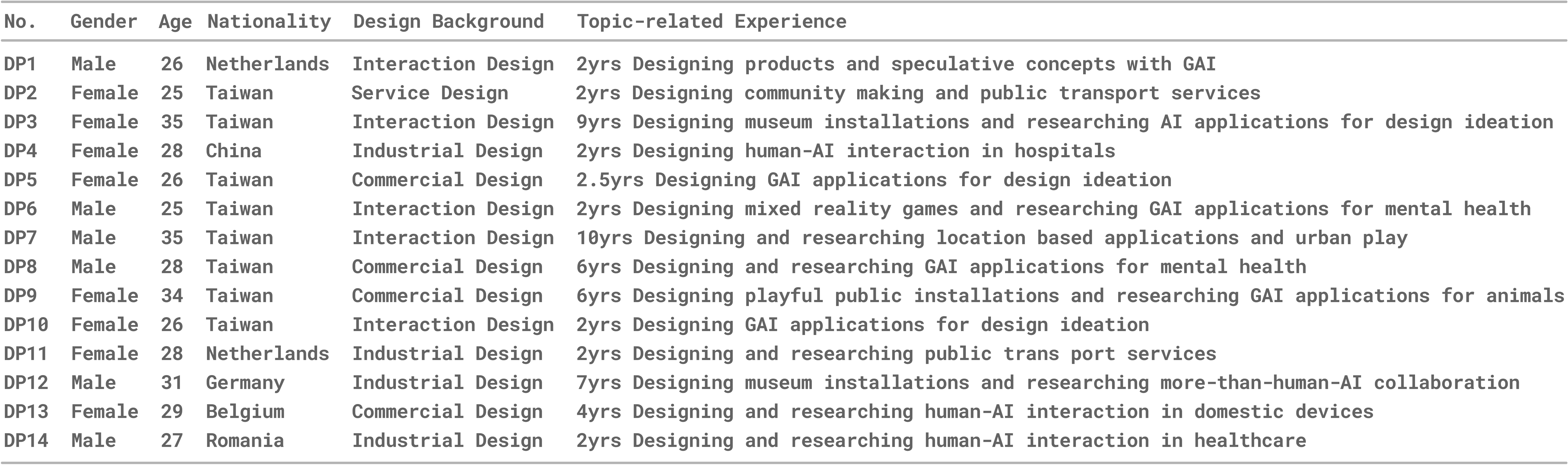}
  \caption{The Backgrounds of Designer Participants.}
  \Description{The Backgrounds of Designer Participants.}
  \label{fig:figure22}
\end{figure*}

We recruited 14 designers through social media in Taiwan and the Netherlands, with their demographic and backgrounds shown in Figure~\ref{fig:figure22}. Each of them received a compensation of €30. We use the abbreviation DP to refer to designer participants in the paper (e.g., DP1 means designer participant 1). They have previous design experience related to urban interaction or human–AI interaction projects. We divided them into four groups for the design exploration. The first two groups, comprising DP1-4 and DP11-14, conducted their exploration in the Netherlands, while the other two groups, DP5-7 and DP8-10, carried out theirs in Taiwan. The group size of three or four enabled the researchers to effectively manage and coordinate the exploration activities while making our participants more comfortable sharing thoughts and collaborating.

\subsection{Procedure}

\subsubsection{Exploratory activity in urban spaces}
Prior to the activity, the participants completed informed consent and were requested to read Altarriba et al.'s play potentials of urban spaces \cite{Bertran:2022}. This framework outlines 15 play forms that citizens already enjoy in ordinary city lives. We believe it can assist our participants in exploring GAI’s design possibilities that respond to citizens’ socio-emotional needs. The exploratory activity began with the tutorial and trial of our GAI tool in a meeting room. Then, the designers were asked to go outside and wander in the city, freely observe city elements, take photographs, and play with our tool in the city center for 2 hours. They were invited to apply their expertise---such as sensitivity, creativity, and adaptability---to unfold GAI’s playfulness in city spaces. Yet, they were also advised to prioritize safety by staying aware of their surroundings, limiting tool usage to secure locations, and avoiding areas with hazardous conditions. 

\subsubsection{Group Interview}

After the exploratory activity, participants joined a 1.5-hour interview in a meeting room, which was audio-recorded. The interview focused on their exploratory experiences and the discussion of playful features. Drawing on playfulness descriptions from literature, we identified the playful feature as the empirical features that evoke positive emotions and encourage spontaneous, flexible behavior and thinking. We asked our participants to use this definition to describe the playful features of GAI they recognized during the exploration. They also provided detailed accounts of their situated experiences related to these features, including their feelings and thoughts on photographing, prompting, and viewing AI-generated images in city environments. 

\subsubsection{Ideation Session}
\begin{figure}
  \centering
  \includegraphics[width=\linewidth]{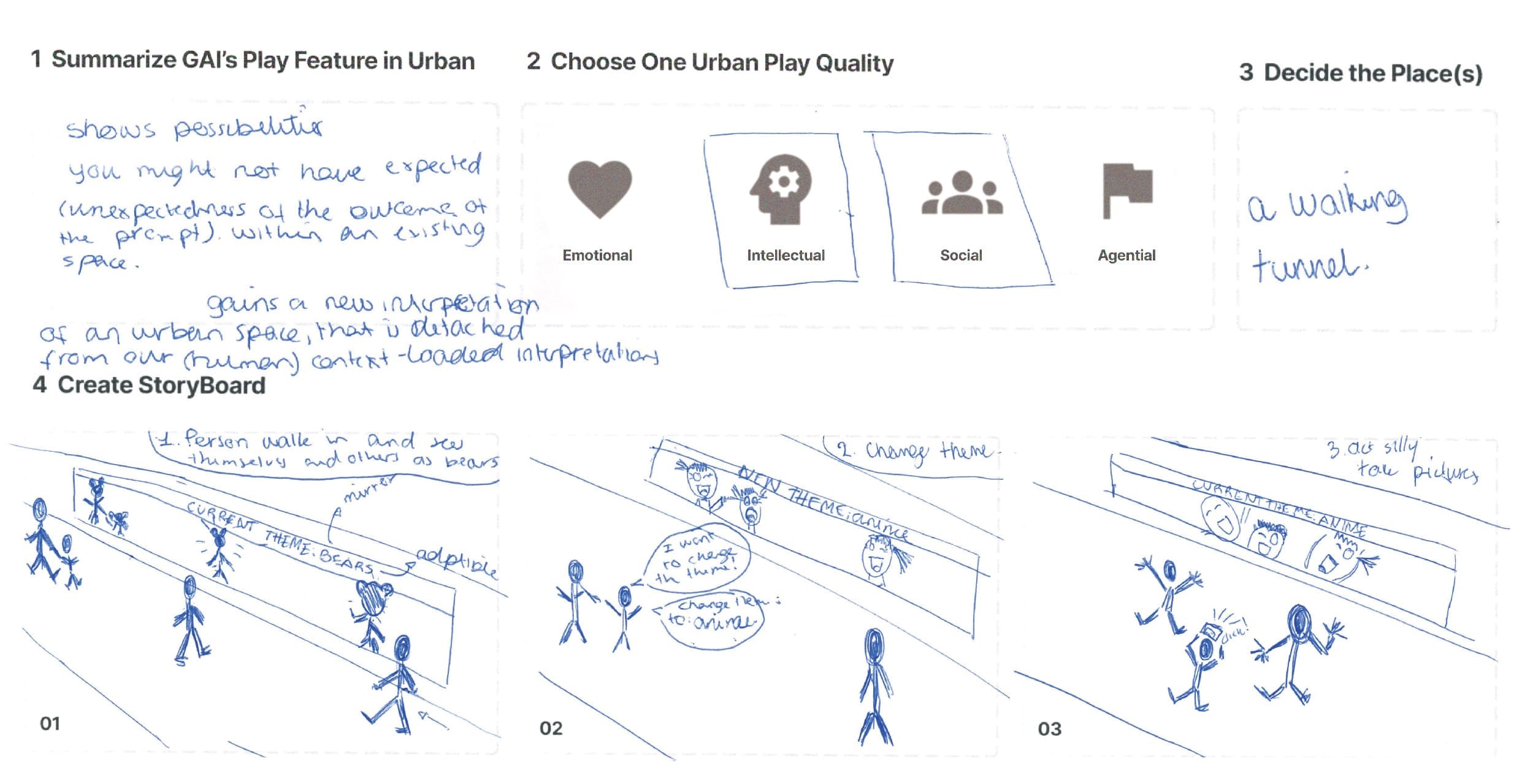}
  \caption{An example of the canvas in the ideation session.}
  \Description{An example of the canvas in the ideation session.}
  \label{fig:figure2}
\end{figure}

The session, lasting for 2.5 hours, took place after the interviews and a one-hour break. To foster creative exchange and enhance the quality of ideas, we divided participants into a group of two (DP1-2, 3-4, 11-12, 13-14) or three (DP5-7, 8-10). Each group received a design canvas with instructions for the ideation process. Groups began by summarizing one playful feature they discovered during the exploratory activity (see Section 1 in Figure~\ref{fig:figure2}). The authors then introduced four dimensions of urban play qualities (described in Section~\ref{subsection2.1}) and asked each group to select at least one (see Section 2 in Figure~\ref{fig:figure2}). The selected dimensions, combined with the summarized playful feature, served as the design goal. Each group was asked to brainstorm an idea of GAI-enabled urban play which allowed citizens to experience the summarized playful feature while potentially promoting the chosen quality. This ideation method aims to support designers to create concepts that not only leverage GAI’s strengths but also consider the emotional, social, and cultural aspects of cities.

After confirming the design goals, the groups decided the place(s) for their designs and generated initial concepts with storyboards (see sections 3 and 4 in Figure ~\ref{fig:figure2}). They then shared these concepts and collectively discussed how their design might foster playful urban experiences. Following this, participants gave suggestions to each other and refined their concepts. Finally, they generated images via our tool to prototype the AI-generated content that citizens may experience in their design scenarios and provided titles and descriptions for their ideas. In total, our participants created six design ideas in this session. The information of these ideas was then compiled and formatted into posters as materials for citizen evaluation.

\subsection{Analysis}
\subsubsection{Analysis Procedure}
After completing the design exploration, we conducted a comprehensive analysis of the collected data, including 273 photos, 367 AI-generated images and their corresponding prompts from the exploratory activity, 3 interview transcripts, and 6 scanned design canvases with descriptions from the ideation session. We used reflexive thematic analysis~\cite{Byrne:2022aa,Braun:2019} to analyze the collected data. The goal of this analysis was to identify GAI's playful features that direct new playfulness situated in urban contexts and inspire design opportunities while potentially connecting to issues. This approach helped hint at the tensions inherent in GAI-enabled urban play and enhance the understanding of GAI's dual nature.

The analysis began with data familiarization and open coding by the first, second, and fourth authors, allowing the codes to represent original meaning as communicated by participants and remain open to capturing a broader spectrum of potential insights. Next, the first author reviewed codes, organized them into an affinity diagram, and generated initial themes, followed by three iterative rounds conducted by all authors to review potential themes and examine them with the codes. After the citizen evaluation workshop, all authors engaged in two additional rounds of iterations to re-examine codes, address conflicts over codes and thematic boundaries until reaching the final themes. During the iterations, themes were refined based on the following reflective moments: 1) We reflected on whether and how potential themes aligned with our analytical goal, such as corresponding with our definition of playful features (described in Section 4.3.2), possibly supporting at least one urban play quality, and demonstrating innovative applications of GAI, which offered inspirational value from a design perspective. 2) We also compared tentative themes and excluded those with broader applicability to private or indoor settings rather than reflecting urban-specific dynamics. 3) After the citizen workshop, we found that tentative themes lacked connections to citizen concerns, which inadequately addressed our analytical goal. This reflection prompted us to emphasize more on keeping the tensions alive within the themes, retaining underlying associations of playful features with risks and avoiding them being one-sidedly positive or idealized. 4) In the last two iterations, we endeavored to ensure the themes adhered to reflexive thematic analysis criteria~\cite{Byrne:2022aa}, such as offering a cohesive and internally consistent narrative distinct from those captured by other themes. The analysis outcome was a list of six Playful Features (shown in Figure~\ref{fig:figure20}). 

Followed by the analyses of design exploration and citizen evaluation, the authors revisited the original design ideas and labeled which of our finalized Playful Features they may support. This labeling was informed by both designer and citizen participants' discussions regarding the playful experiences these ideas could create. Once this process was complete, two of the authors added the labels to the design idea posters to facilitate readability. 

\subsubsection{Positionality Statements}
The reflexive thematic analysis accounts for the researchers’ perspectives and experiences, which we outline in the following statement: The first and fourth authors specialize in industrial design, with the second and third authors having computer science backgrounds. Sharing the qualitative research domains in urban contexts, the first author is devoted to designing and researching technology-mediated playful urban interactions and their user experiences, whereas the third author contributes expertise in place theories, urban computing, and critical design research. With a shared focus on AI studies, the second author specializes in the technical mechanisms of AI and the development of human-AI interaction systems, and the fourth author brings expertise in the aesthetics and ethics of AI technologies. These expertise areas inform our data analysis and theme development, enabling a nuanced focus on qualitative experiences of interacting with GAI tools situated in city spaces, alongside tensions between their design possibilities and challenges.

\section{Results of Design Exploration}

\subsection{Six Playful Features of GAI in Urban Spaces}
We present six Playful Features, along with the experiences of our designer participants, their photographs, and AI-generated images. These findings demonstrate how GAI characteristics contribute to playfulness in public spaces. All quotes included in this paper were translated into English by the first author and verified by the co-authors. To aid readers in identifying changes made by GAI, we provide the original photos and their AI-generated images using the Reimage function. For the results from the Inpaint function, we include the original photos, the images highlighting the areas painted by the participants, and the final AI-generated outcomes. 

\subsubsection{Empowerment of Visualization to Share Desired Neighborhood (F1)}

\begin{figure}
  \centering
  \includegraphics[width=\linewidth]{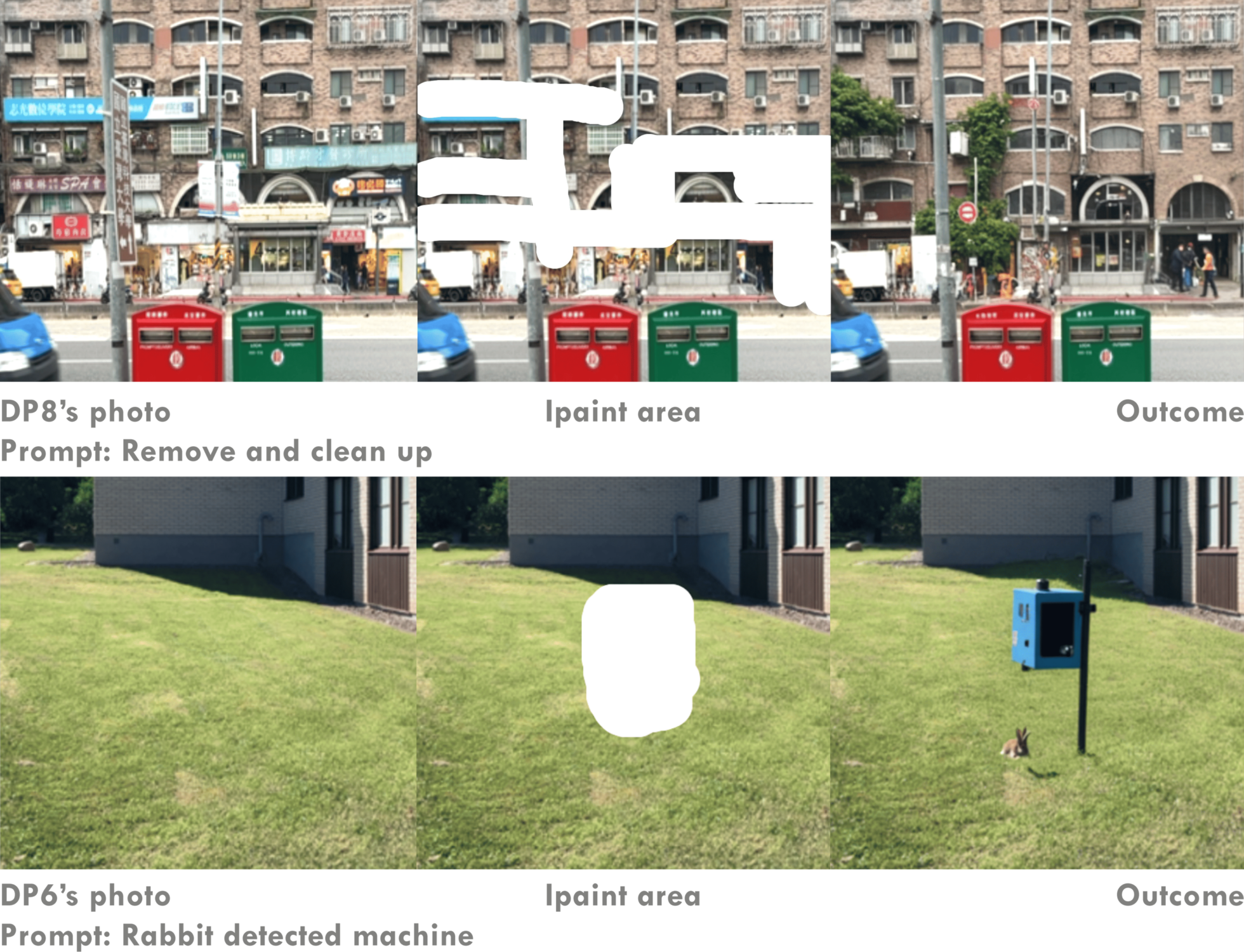}
  \caption{Top: DP8 painted over the unsightly signage on the building and generated a street view without them. Bottom: DP6 noticed the open lawn and generated a device detecting rabbit activities. This creation integrated his place memories.}
  \Description{Top: DP8 painted over the unsightly signage on the building and generated a street view without them. Bottom: DP6 noticed the open lawn and generated a device detecting rabbit activities. This creation integrated his place memories.}
  \label{fig:figure5_DP6_quote}
\end{figure}

Participants became aware of how the GAI tool enhanced their visualization ability to modify urban environments, fostering a sense of agency and enjoyment in reimagining cityscapes. This empowerment encouraged them to share the desired appearances of their surroundings. The accessible use of the tool enables individuals to playfully express their imaginations of urban settings with ease. Tasks once considered ``a big hurdle to many people'' (DP1), such as creating personalized cityscapes, are now both ``convenient'' (DP4-7, 11-14) and ``joyful'' (DP3, 8-11, 13). Participants emphasized that GAI allowed them to ``visualize'' (DP1) and ``experience various imagined cities that only existed in mind before'' (DP6), bridging the gap between abstract ideas and tangible representations. Beyond the intrinsic satisfaction, this interaction evoked a heightened sense of agency among participants, motivating them to actively engage in city-making activities through the sharing of ideal community environments. As DP1 observed, ``This AI encourages people to actively make better versions of neighborhoods and share them with others'' (DP1). Participants suggested urban improvements by envisioning alternatives to existing challenges (DP1-2, DP6, DP8), such as removing unsightly signage from street views (Figure~\ref{fig:figure5_DP6_quote}, top). By democratizing the process of visualizing and proposing reimaginations of city conditions, GAI could transform passive observers into active contributors, facilitating collective engagement in shaping improved urban futures.

\begin{figure*}
  \centering
  \includegraphics[width=\linewidth]{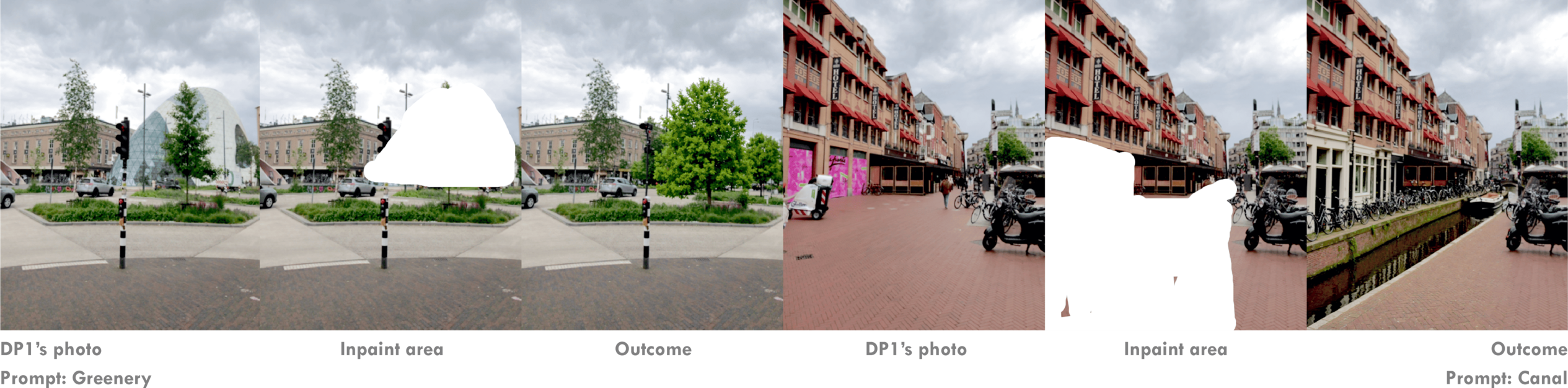}
  \caption{DP1 rediscovered the city's characteristics through modifying photos---removing landmarks to create ordinary scenes (left) and adding Dutch-style canals to streets (right).}
  \Description{DP1 rediscovered the city's characteristics through modifying photos---removing landmarks to create ordinary scenes (left) and adding Dutch-style canals to streets (right).}
  \label{fig:figure6_DP1_quote}
\end{figure*}

\subsubsection{Empowerment of Visualization to Re-discover People-place Relationship (F2)} 

As participants spent more time interacting with GAI in situ, they gradually began to see the city through the lens of ``editing urban scenes'' (DP5), facilitating re-discoveries on relationships between them and the place. The empowerment of visualizing urban scenes changed their perspective on cities from a fixed entirety to what could be added, removed, transformed, or reinterpreted. For example, DP6 noted, ``With this tool, you can better observe and fill in the city's gaps... I started looking for blank walls or open streetscapes.'' He then realized that `` the entire city has empty spaces that could be filled.'' Similarly, DP5 stated, ``I began to pay attention to the sky or the ground---spaces where things could be inserted, or chaotic areas that could be tidied up.'' This shift in perspective allowed them to deepen the observation and unravel the significance of the visited places. Figure~\ref{fig:figure5_DP6_quote} illustrates an example from DP6, where he noticed an open patch of grass near his home, remembered seeing rabbits there before, and used the tool to generate an installation signaling their presence. 

In addition to creatively viewing the city, comparing the real scene with the AI-generated version brought participants both enjoyment and reflection. By ``removing city elements and observing the differences'' (DP1), they were able to identify iconic aspects of their cities, discover which elements dominated their perspective and made them overlook others, and reflect on what they wished to preserve (DP1, 5-7). This is exemplified by DP1's exploration in Figure~\ref{fig:figure6_DP1_quote} and his remark:  

``I was comparing this city with other cities, trying to see what makes this city unique, and then remove it with AI to see if it's still unique... Currently, there’s no canal here and what if this city has canals like some famous Dutch cities... Or what if the famous building of this city is not there and becomes just a normal square? Maybe there are other things that define the city's character besides the buildings I removed, but I never noticed them because they are always there in reality.''  

\begin{figure}
  \centering
  \includegraphics[width=\linewidth]{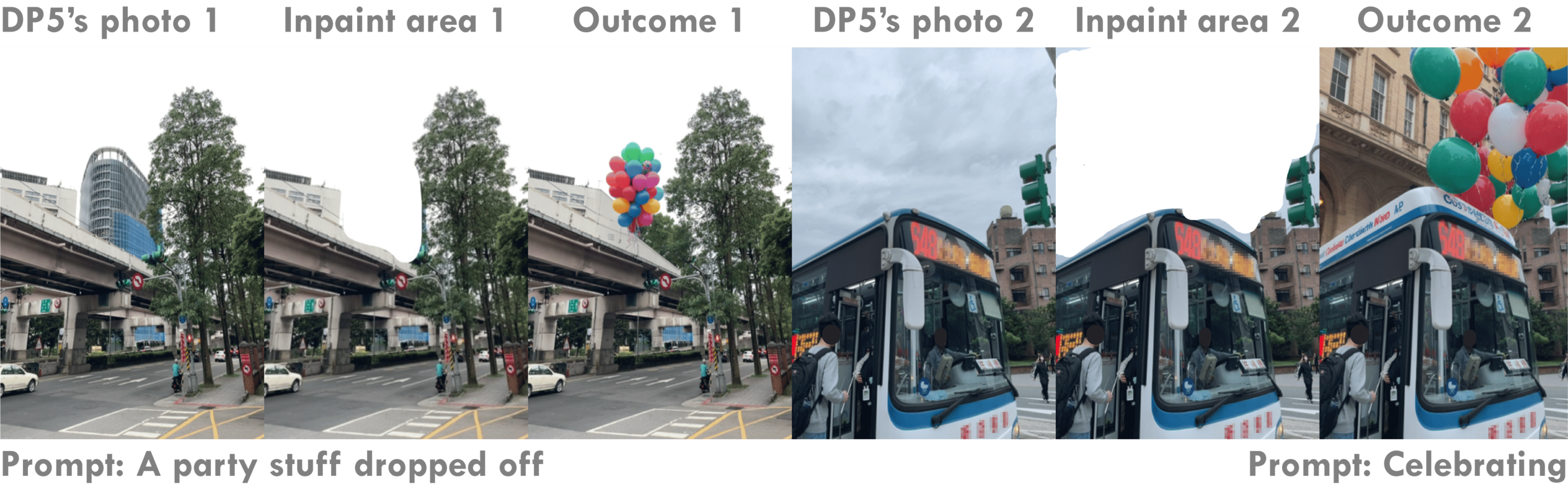}
  \caption{DP5 reflected on her impression of the city after Inpainting the building and sky with the prompt `party.'}
  \Description{DP5 reflected on her impression of the city after Inpainting the building and sky with the prompt `party.'}
  \label{fig:figure7_DP5_quote}
\end{figure}

Moreover, by comparing real cityscapes, prompts, and generated results, participants reflected on their relationships with the city facilitated by GAI's interpretation of the urban environment (DP2, 5-7, DP10-12). For instance, DP5 used the prompt ``party'' to generate several party-themed streetscapes (Figure~\ref{fig:figure7_DP5_quote}) and shared the following reflection:  

``I noticed that when I input `party,' the outcomes often included balloons; this happened three times... It seems that AI associates a happy city with balloons, but when I feel happy in this city, it doesn't necessarily involve that element. Instead, it's the sounds of passersby, the music in the square, the smell of food, and so on.''

\begin{figure}
  \centering
  \includegraphics[width=\linewidth]{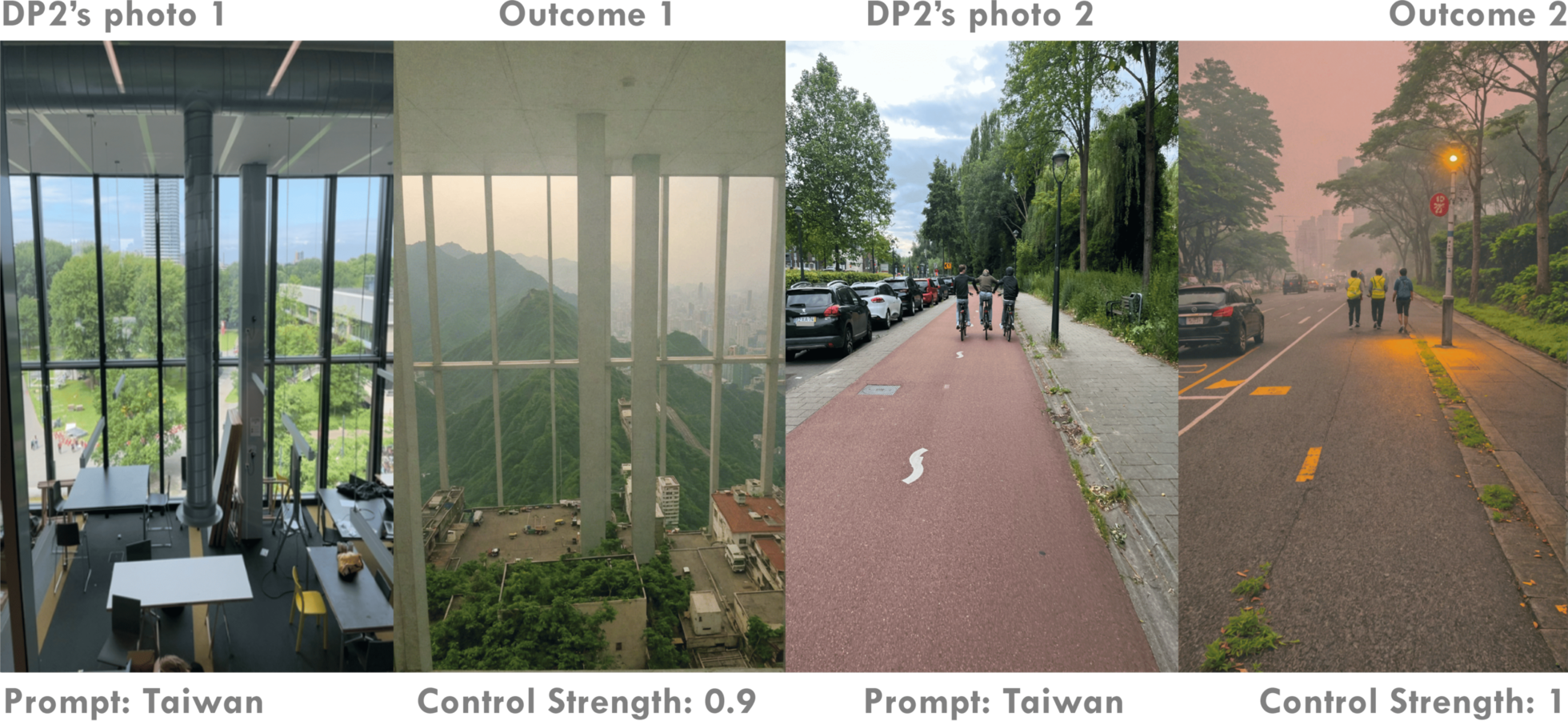}
  \caption{DP2 altered Dutch scenes, with the generated mountains evoking memories of her hometown in Taipei.}
  \Description{DP2 altered Dutch scenes, with the generated mountains evoking memories of her hometown in Taipei.}
  \label{fig:figure3_DP2_quote}
\end{figure}

\begin{figure*}
  \centering
  \includegraphics[width=\linewidth]{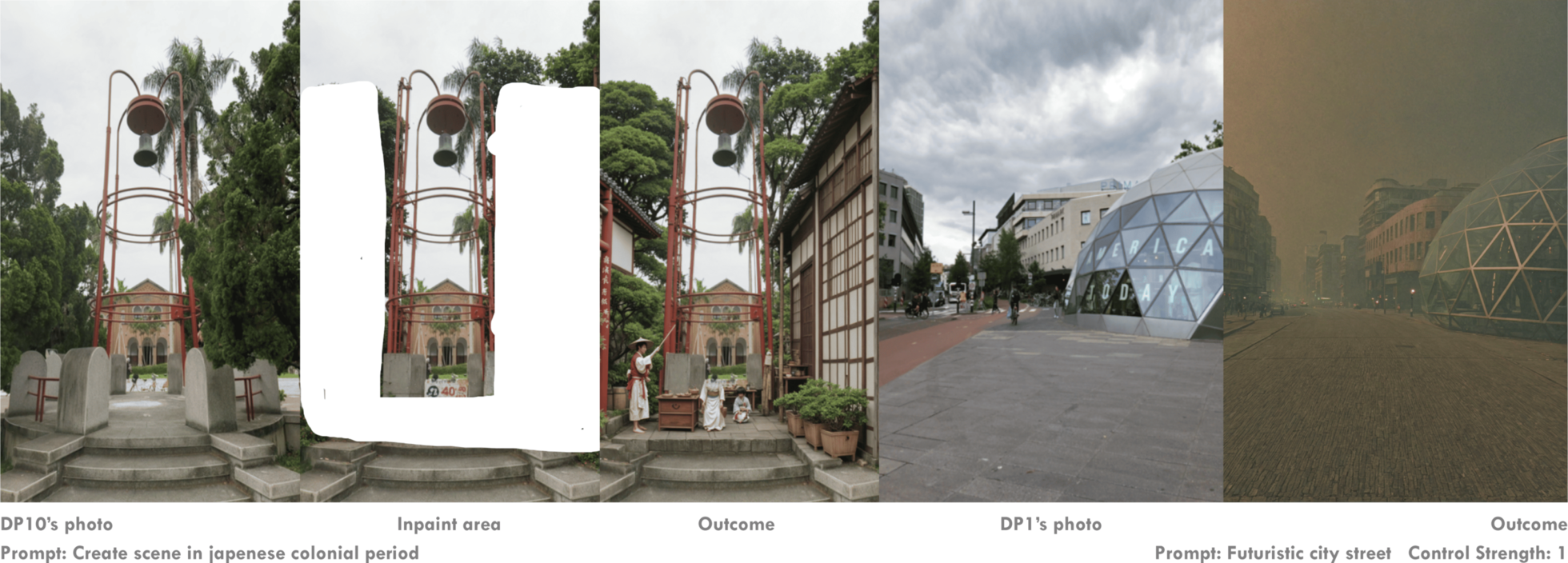}
  \caption{DP10 and DP1 defamiliarized the places by transforming the trees into Japanese buildings and characters (left) and reimagining the workplace as a sci-fi setting (right).}
  \Description{DP10 and DP1 defamiliarized the places by transforming trees into Japanese buildings and characters (left) and reimagining the workplace as a sci-fi setting (right).}
  \label{fig:figure8_DP10_quote}
\end{figure*}

\subsubsection{Empowerment of Visualization to Shuttle between Multiverse and Reality (F3)}

The visualization ability empowered by the GAI tool allowed participants to bridge their locations to diverse and imagined worlds in their mind, contributing to the experiences of momentary travel through multiverses. Some participants described our tool as a ``portal'' they could carry with them when exploring the city, attributed to its portability and the wide range of generation themes it offered (DP7-9). They could use the tool anywhere in the city and temporarily immerse themselves in the bend of time and space. DP9 and DP8 captured this experience in their remarks: ``After AI replaced certain parts of the photo, it's like a portal transporting us to another time of this place'' (DP9). ``The Reimage function changed cities' appearance while preserving the structure similar to the original scene, with similar building shapes, object outlines, and even people's movements... feeling like bumping into another universe'' (DP8). 

GAI became an intersection between different worlds, offering emotional resonance as participants traversed these altered spaces. Some enjoyed transforming the temporal and spatial dimensions of their surroundings, merging their current locations with the past or future of another real or fantasy world, which sparked greater imagination about the places (DP1, 7, 9, 10) or provided a temporary escape from the present moment (DP7, 12). Interestingly, DP2 connected their location to their hometown in Taiwan, experiencing a novel way of reminiscing (Figure~\ref{fig:figure3_DP2_quote}). She noted, ``AI generated many mountains; it made me nostalgic... really felt like returning to my hometown... as if visiting a hillside cabin near Taipei 101'' (DP2). When she photographed and generated a Taiwanese version of her surroundings, she reflected, ``Though AI generates something inaccurate, it catches the impression of places... Looking at the pictures helps recall more memories in Taiwanese cities'' (DP2). 

Furthermore, this multiverse travel encouraged people to explore different cultures. For instance, when DP2 and DP11 compared their locations with the AI-generated streetscapes that blended elements from another country, they became curious about the cultural aspects of both places. DP2 observed that GAI ``can make the Netherlands more like Taiwan or make Taiwan more Dutch'' (DP2), allowing her to sense the historical differences between the two places. DP11 commented that GAI ``links the cultures of familiar and unfamiliar places'' (DP11), prompting her to seek a deeper understanding of local knowledge. 

\subsubsection{Encounter of Defamiliarized Places (F4)}
Several participants defamiliarized the spaces they visited with iWonder, playfully exploring the aesthetic and experiential boundaries between familiarity and unfamiliar, as well as between real and fictional spaces. Defamiliarization is a strategy which can be employed in designing urban play~\cite{Hung:DISWiP2024,Wang:RTD2019}, where familiar places are rendered unfamiliar to create abnormal but plausible urban landscapes that arouse pleasure and curiosity. Participants experimented with iWonder to modify the familiarity of places, discovering new ways to appreciate city landscapes (DP1, 3, 7, 10). For example, in Figure~\ref{fig:figure8_DP10_quote}, DP10 ``kept the symbolic clock'' of a Taiwanese location, which ``still preserved the local essence'' (DP10) and maintained a relation to the original site. Notably, she let iWonder integrate unfamiliar Japanese elements into the surrounding scenery to experience a semi-familiar city (DP10). Interestingly, the streetscapes generated with this technique might evoke an inexplicable sense of familiarity, akin to d\'ej\`a vu. These altered scenes, which blur the lines of familiarity and reality, gave some participants the feeling they had already experienced the space before (DP7, 8) or allowed them to suspend disbelief and imagine themselves in parallel worlds (DP2, 7). Figure~\ref{fig:figure8_DP10_quote} presents an example from DP1’s working place. Although the generated images alienated the location into a sci-fi setting, he still recognized it as his working place, immersing himself in the parallel version. As he described: 

``I really like this image because it looks like it was still the city, but very apocalyptic... You can see the shapes of the city still there... It's almost like our city after an atomic bomb… having a shade around it, dust particles, and smog in the air, which is kind of mysterious but familiar.''

\begin{figure}
  \centering
  \includegraphics[width=\linewidth]{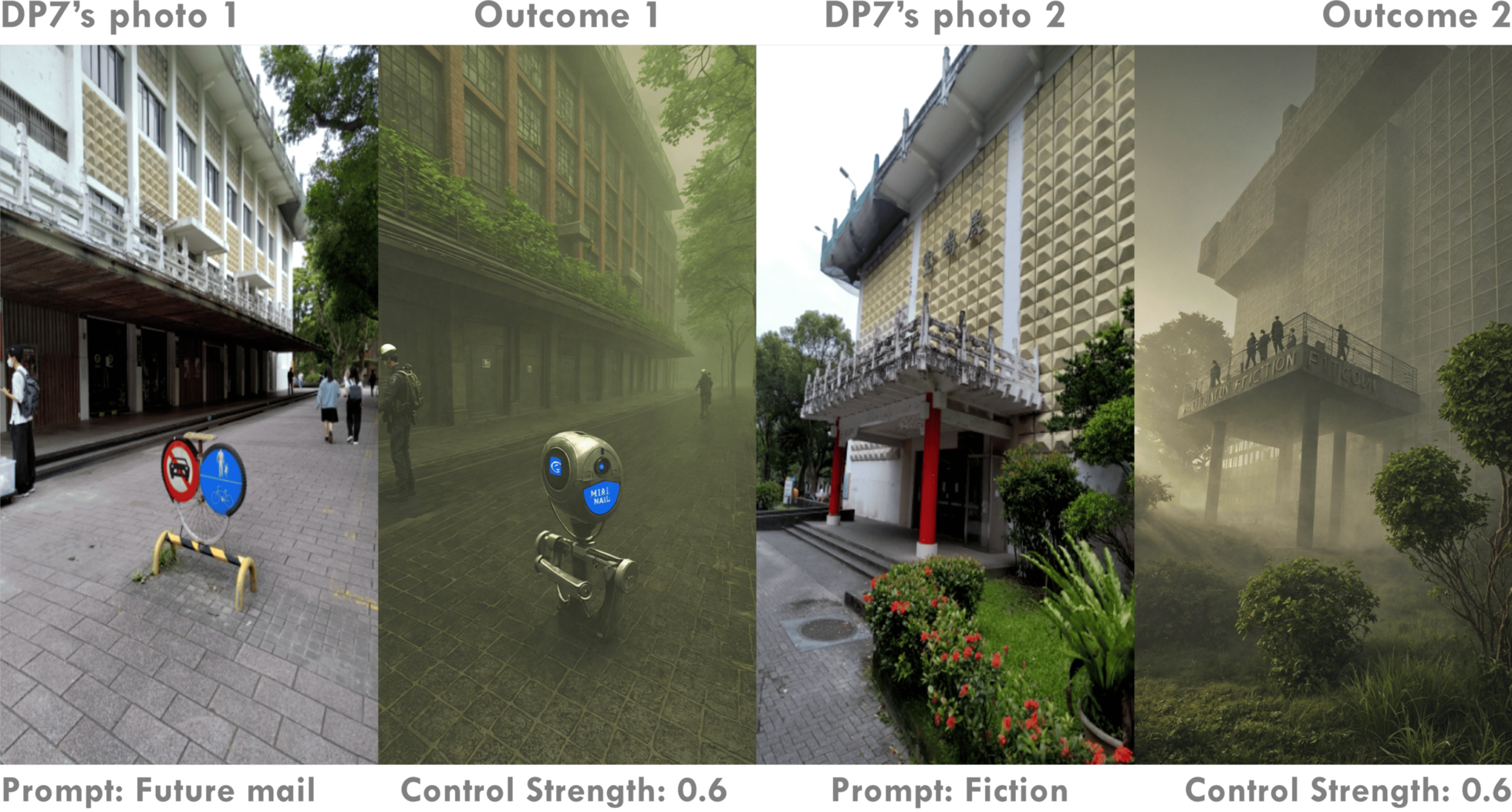}
  \caption{The sci-fi streetscapes generated by DP7---depicting a roadblock transformed into a futuristic mailbox (left) and a roof into a terrace (right)---arouse a feeling of d\'ej\`a vu.}
  \Description{The sci-fi streetscapes generated by DP7---depicting a roadblock transformed into a futuristic mailbox (left) and a roof into a terrace (right)---arouse a feeling of d\acute{e}j\grave{a} vu.}
  \label{fig:figure4_DP7_quote}
\end{figure}

Similarly, iWonder incorporated elements of a sci-fi world into the streets near DP7's workplace (Figure~\ref{fig:figure4_DP7_quote}). The outcome momentarily made him believe his familiar street had once housed a future mailbox, giving him ``a sense of d\'ej\`a vu'' (DP7). However, he noticed that achieving this feeling was fortuitous, as it depended on the precise and coincidental alignment of several factors, such as personal familiarity with the location, the specific world referenced by the prompts, the structure of prompts, and the inherent uncertainty of AI-generated content (DP7). This case highlights the distinction between the playful feature in this section and shuffling between multiverse and reality (F3): Rather than using GAI to predeterminedly navigate either familiar or novel multiverses, participants appeared to grant GAI greater freedom to generate perceptible yet unpredictable worlds for playful defamiliarization. This granting nudges them to a serendipitous encounter of alienating yet connected cityscapes.

\subsubsection{Whimsical Urban Montage (F5)}
\begin{figure}
  \centering
  \includegraphics[width=\linewidth]{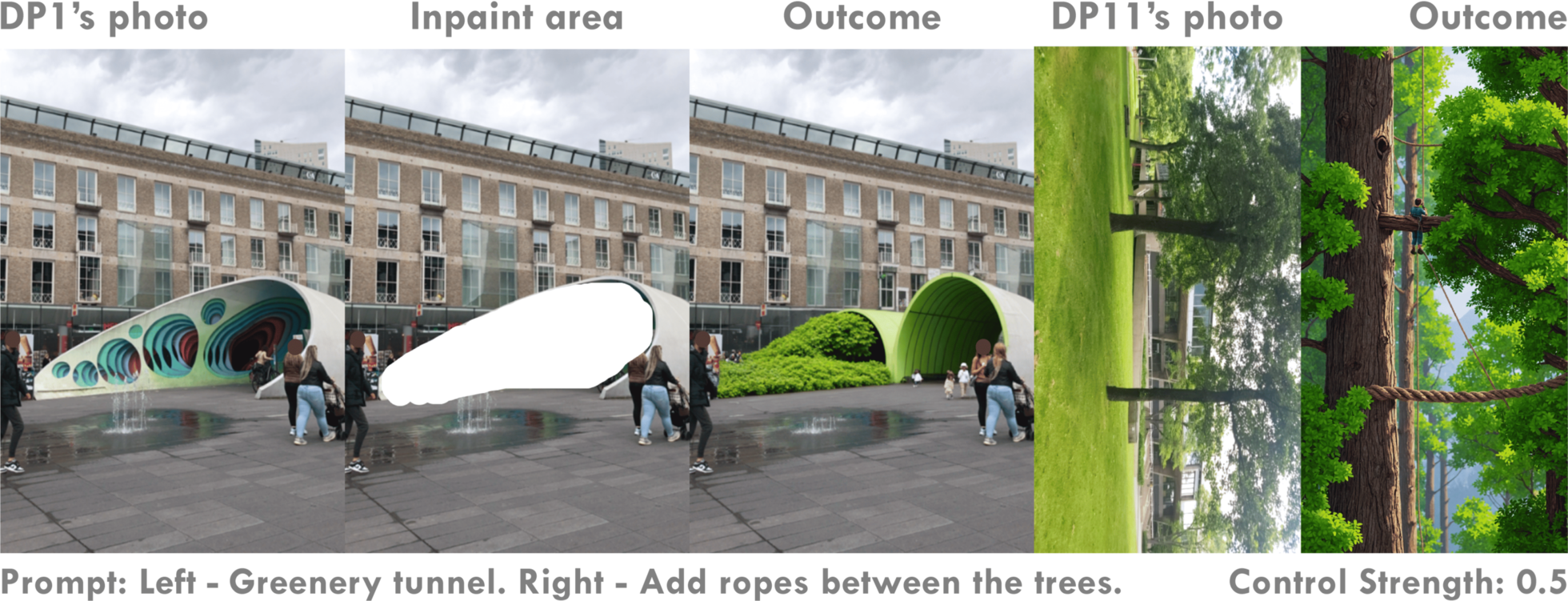}
  \caption{Left: iWonder whimsically filled the tunnel with greenery and added another tunnel. Right: iWonder reinterpreted DP11’s horizontal photo vertically, transforming the ground and trees into a thick trunk with branches.}
  \Description{Left: iWonder whimsically filled the tunnel with greenery and added another tunnel. Right: iWonder reinterpreted DP11’s horizontal photo vertically, transforming the ground and trees into a thick trunk with branches.}
  \label{fig:figure9_DP1_DP11_quote}
\end{figure}

Participants found that some AI-generated outcomes are ``out of the blue'' (DP5). Different from encountering defamiliarized places (F4), where participants reached a consensus on the intent of defamiliarization, these outcomes completely defied their usual expectations, deviated from what they associated with prompts, and brought a sense of surprise. These whimsical results often reflected a non-human perspective, intriguingly diverging from conventional spatial logic and ``human’s context-loaded perception of urban settings'' (DP13). This is exemplified in DP1's generation (Figure~\ref{fig:figure9_DP1_DP11_quote}, left) and his experience: 

``Initially, I imagined making the tunnel more greenery... I, as a human, expected adding plants to its wall and still allowing entry. But look, what AI thought was filling the entire tunnel with plants and added another one… Though AI didn't show what I expected, it’s quite interesting and introduces ideas I never thought of.''

DP2 and DP11 also encountered images where the sense of space was unexpectedly altered. When DP2 transformed a Dutch scene into a Taiwanese version, she noticed that the scale of the space had changed (Figure~\ref{fig:figure3_DP2_quote}, left). She observed, ``The cars in the image are smaller than the chair in the original one, and the houses are the size of the table here... The space suddenly becomes enormous.'' Similarly, DP11's horizontally captured photo was reinterpreted by the AI as a vertical image (Figure~\ref{fig:figure9_DP1_DP11_quote}, right), momentarily evoking ``a sense of dreamlike space'' (DP11).  

\begin{figure}
  \centering
  \includegraphics[width=\linewidth]{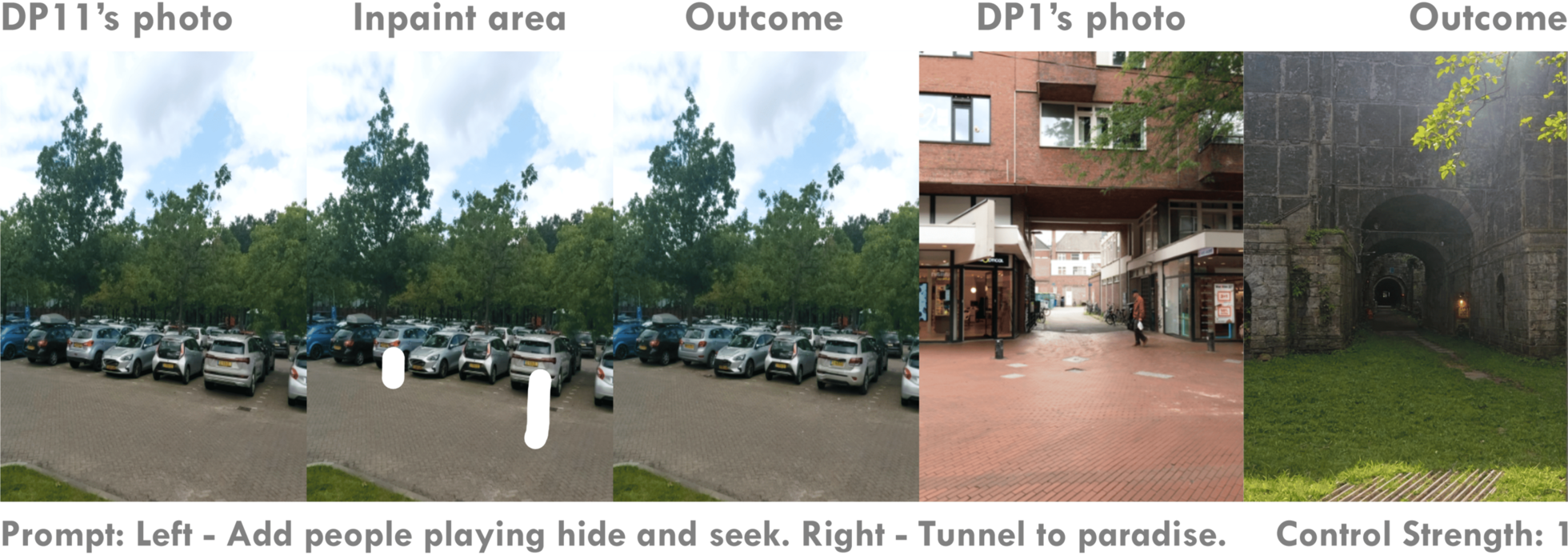}
  \caption{Left: DP11 intended to depict people playing hide-and-seek, but the image remained unchanged. Right: A bright alleyway unexpectedly became a shadowy tunnel.}
  \Description{Left: DP11 intended to depict people playing hide-and-seek, but the image remained unchanged. Right: A bright alleyway unexpectedly became a shadowy tunnel.}
  \label{fig:figure10_DP11_DP1_quote}
\end{figure}

Although these AI-generated whims were initially considered ``failed generations'' (DP13), participants believed that they could stimulate ``positive emotions'' (DP11-13) and ``creative imagination'' (DP14). Some noted that this unpredictability sparked surprise and anticipation. They ``look forward to seeing the unforeseen results'' (DP12, 13) and were willing to give away the control to let AI lead them into imaginative, surprising journeys. These whimsical outcomes could also be ``integrated with people's imaginations of original places to think of another possibilities'' (DP4). For instance, DP11 attempted to generate an image of children playing hide-and-seek, but iWonder made no changes to the scene (Figure~\ref{fig:figure10_DP11_DP1_quote}, left). She then figured out that: ``if people are playing hide and seek, they are hidden. That's why you don't see anyone in this image... People are hiding well'' (P11). DP1 also noticed an alley that appeared unusually bright and planned to transform it into a paradise (Figure~\ref{fig:figure10_DP11_DP1_quote}, right). However, our tool darkened the entire alley, prompting him to speculate that it was discouraging the idea and suggesting an untold story (DP1). These creative interpretations merged participants' personal imaginations with whims generated by GAI, sparking new ideas (P3, P14) and offering novel perspectives of city spaces.

\begin{figure}
  \centering
  \includegraphics[width=\linewidth]{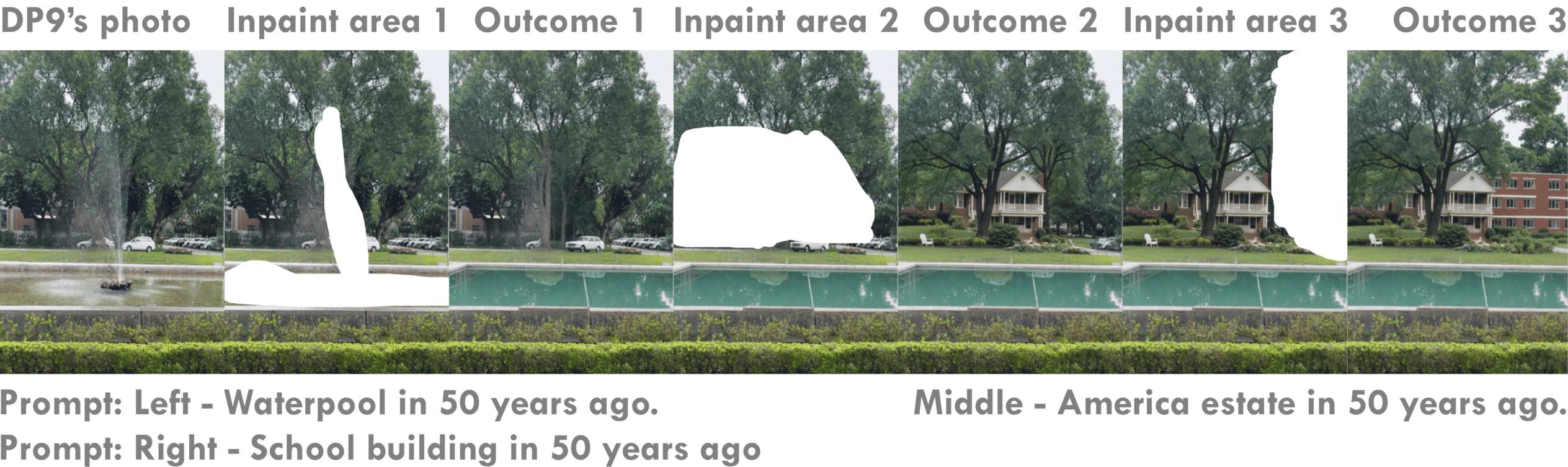}
  \caption{DP9 and DP10 impromptu cooperated with iWonder to create a vintage fountain with American buildings.}
  \Description{DP9 and DP10 impromptu cooperated with iWonder to create a vintage fountain with American buildings.}
  \label{fig:figure11_DP9_DP10_quote}
\end{figure}

\subsubsection{Pass-down Co-improvisation (F6)}
The interaction of iteratively generating images with GAI could act as a ``social catalyst,'' facilitating playful conversations and co-improvisation within communities (DP5). The participants found that ``previous generated outcomes can be used by others as inputs for the next generation'' (DP3). This iterative and accumulative process allows diverse people to integrate their ideas of the place into a coherent work, potentially fostering collaboration and dialogue within the local community. As DP1 noted, ``Residents can see what we generated, share thoughts about them or this place that we missed, and generate new iterations with AI. Their work then inspires us to rethink and do another iteration... Eventually, it'll be a community thing'' (DP1). Through this process, co-creation becomes a living and improvisational ``evolution tree upon the neighborhood,'' which joyfully conveys ``what different people find nice with'' and ``establish connections with others'' (DP1). Notably, GAI's capability to impromptu generate diverse elements from prompts offered inspirations and made the participants view it as an active collaborator in the co-creation process (DP1, 8-10, 12-14). DP9 and DP10, for example, collaborated with iWonder without deliberate preparation to transform a Taiwanese streetscape into an American-style estate (Figure~\ref{fig:figure11_DP9_DP10_quote}). They described the experience: 

``Initially, we merely thought about how the location might have looked fifty years ago, but AI's first transformation inspired us to envision having the houses in American dramas. After AI's second generation, which really changed the place into the drama-like style, we imagined adding an American-style dormitory.'' 

The participants conveyed their urban imaginations through these images, and by collectively discussing the differences between the generated images and their surroundings, they were able to develop new ideas for subsequent iterations. This process ``continually sparks new imagination and enjoyment'' (P9), making the outcomes ``a co-creation among people, GAI, and urban'' (DP1). 

\subsection{Six Design Ideas of GAI-enabled Urban Play}

\begin{figure}
  \centering
  \includegraphics[width=\linewidth]{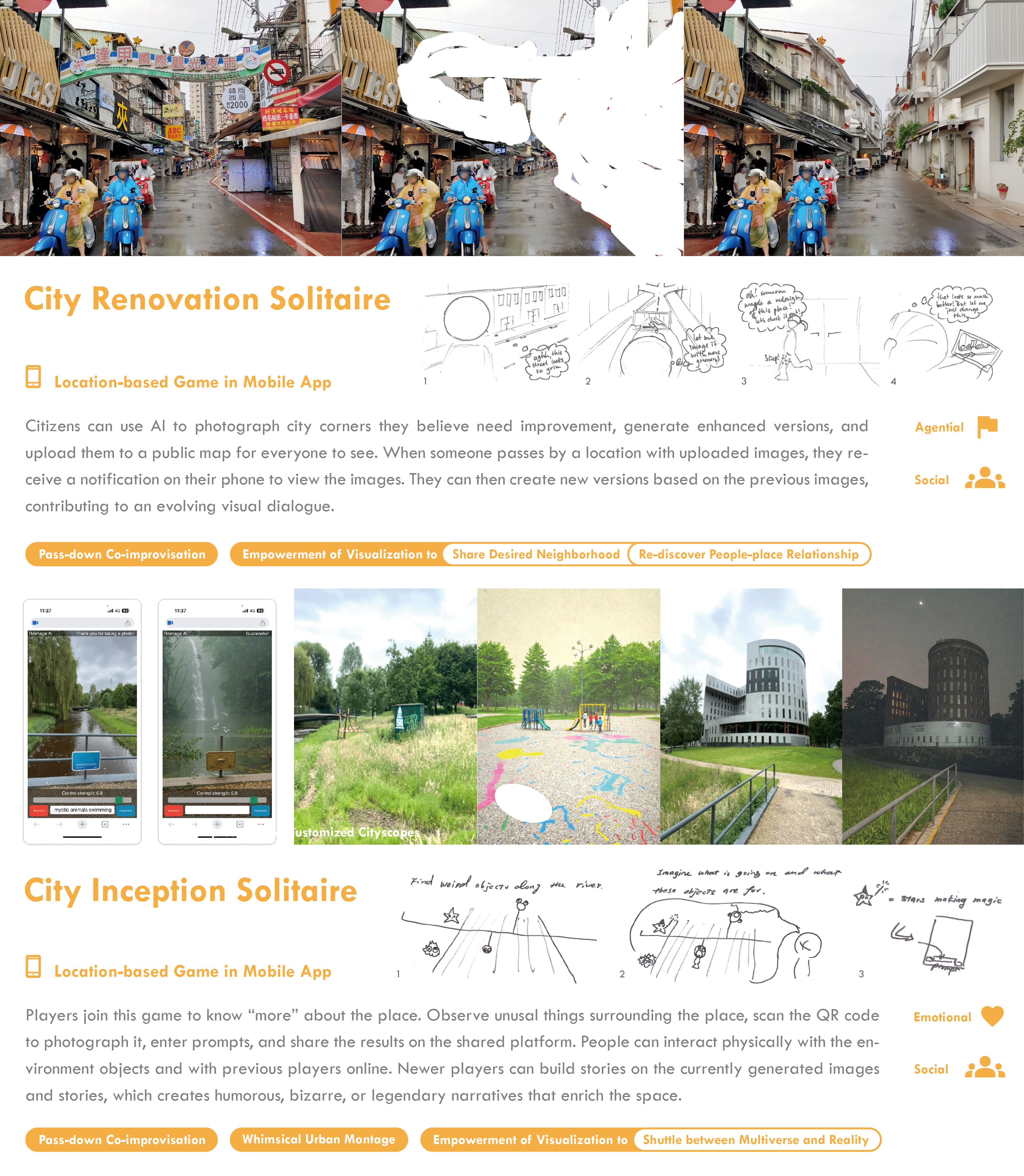}
  \caption{The posters of the design ideas ``City Renovation Solitaire'' and ``City Inception Solitaire.''}
  \Description{The posters of the design ideas ``City Renovation Solitaire'' and ``City Inception Solitaire.''}
  \label{fig:figure12}
\end{figure}

\begin{figure}
  \centering
  \includegraphics[width=\linewidth]{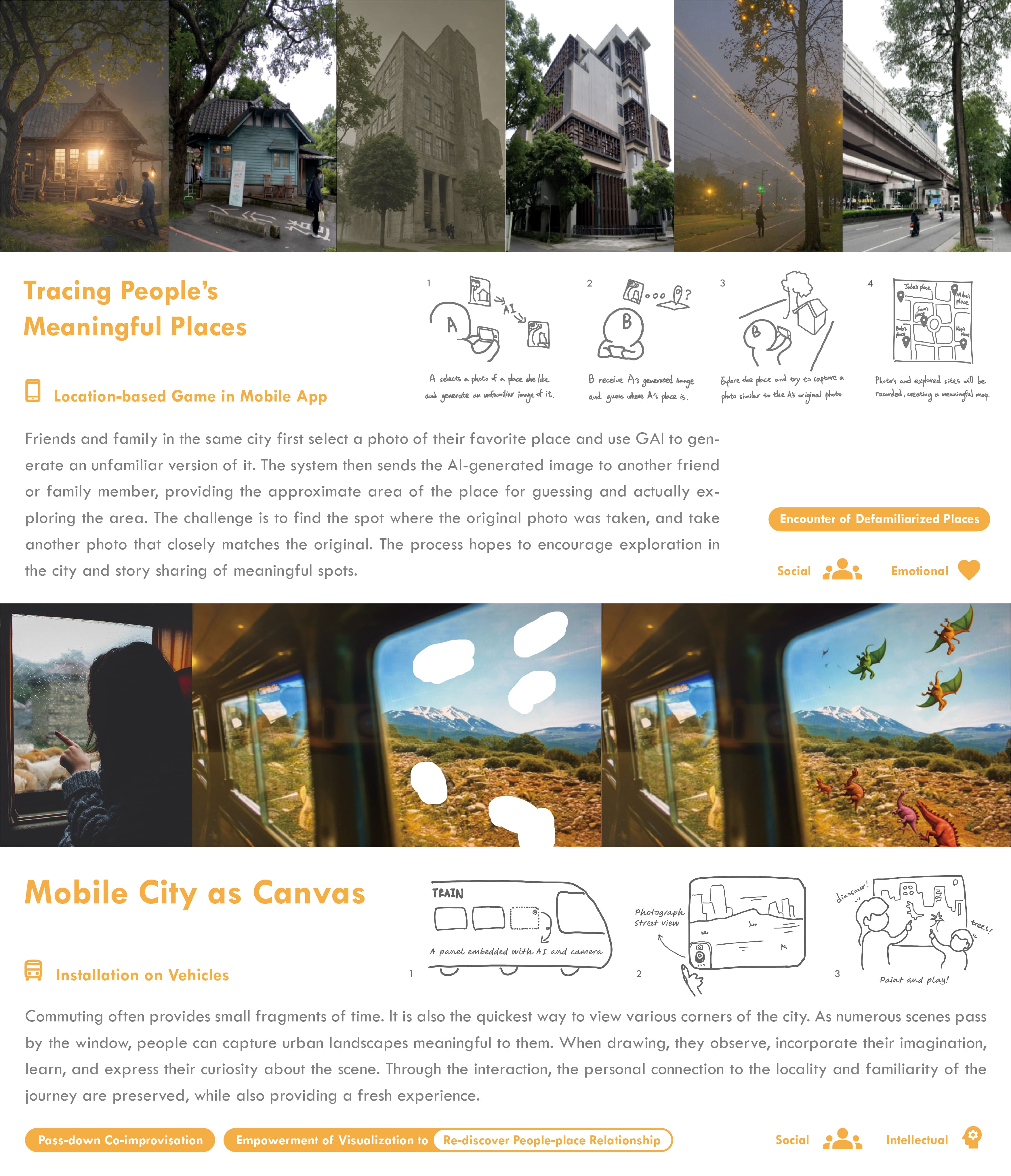}
  \caption{The posters of the design ideas ``Tracing People's Meaningful Places'' and ``Mobile City as Canvas.''}
  \Description{The posters of the design ideas ``Tracing People's Meaningful Places'' and ``Mobile City as Canvas.''}
  \label{fig:figure13}
\end{figure}

\begin{figure}
  \centering
  \includegraphics[width=\linewidth]{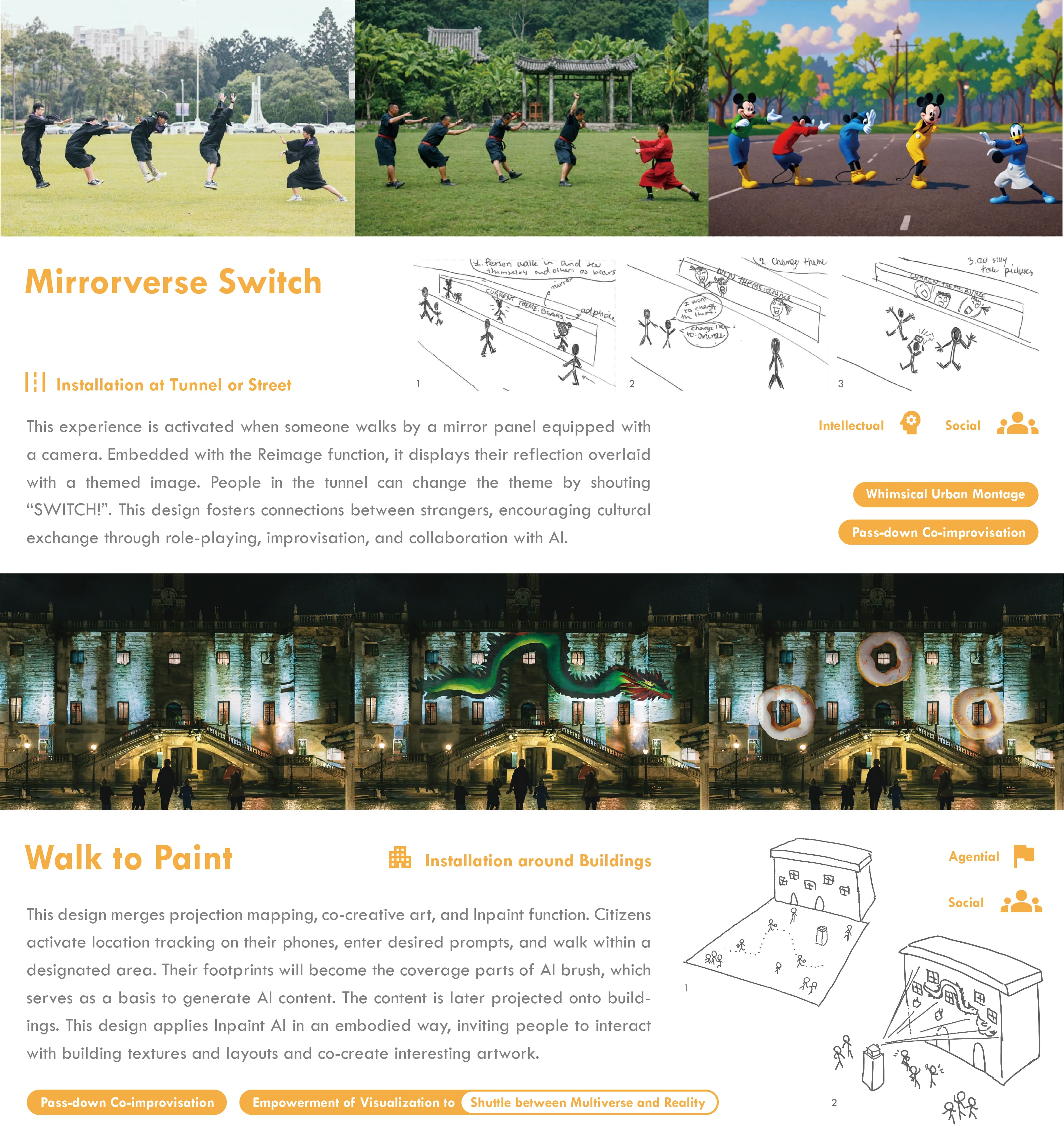}
  \caption{The posters of the design ideas ``Mirrorverse Switch'' and ``Walk to Paint.''}
  \Description{The posters of the design ideas ``Mirrorverse Switch'' and ``Walk to Paint.''}
  \label{fig:figure14}
\end{figure}
We showcase six design ideas developed by our participants in the ideation session. These ideas explore various ways to integrate GAI with urban environments to create playful interactions. The posters of these concepts are depicted in Figure ~\ref{fig:figure12}, ~\ref{fig:figure13}, and ~\ref{fig:figure14}. Each poster contains an idea introduction, a storyboard drawn by the participants, original images and the modified versions generated by our tool, the targeted urban play qualities, and the Playful Features that citizens could experience through the design.

``City Renovation Solitaire'' and ``City Inception Solitaire'' both involve citizens using AI to digitally alter or enhance urban spaces; however, ``City Renovation Solitaire'' focuses on improving city corners and sharing those improvements on a public map, while ``City Inception Solitaire'' encourages storytelling and legend-building around specific locations by generating and sharing AI-altered images. ``Tracing People's Meaningful Places'' and ``Mobile City as Canvas'' share a common theme of blending familiar urban landscapes with imaginative elements. ``Tracing People's Meaningful Places'' creates a playful guessing game that combines AI-generated imagery with real-world exploration, while ``Mobile City as Canvas'' encourages commuters to capture and creatively reinterpret urban scenes during their journeys, preserving personal connections to the city while offering new perspectives. ``Mirrorverse Switch'' and ``Walk to Paint'' emphasize embodied interaction with the urban surroundings. ``Mirrorverse Switch'' engages people in spontaneous role-playing and cultural exchange through AI-enhanced mirror images in public spaces, while ``Walk to Paint'' combines physical movement with AI-generated art, allowing citizens to co-create digital artwork projected onto buildings by walking within a designated area and transforming the footprints into the brush for Inpainting. While all six designs leverage AI to enrich urban experiences, they differ in their focus---ranging from collaborative storytelling and cultural exchange to personal exploration and public art creation. 

\section{Citizen Evaluation}
\subsection{Research objective}

After the design exploration, we conducted workshops to evaluate the six design ideas with citizens from various social and cultural backgrounds. This bottom-up approach is instrumental in surfacing barriers to playful urban experiences, thereby informing design refinements that respond to citizens’ needs and expectations~\cite{Bertran:2022}. Including citizens’ diverse perspectives also sparked a broad spectrum of discussions and helped uncover potential issues related to the designs~\cite{Bertran:2022}, potentially mitigating negative impacts when deploying GAI in public spaces. 

\subsection{Citizen Participants and Recruitment}

\begin{figure}
  \centering
  \includegraphics[width=\linewidth]{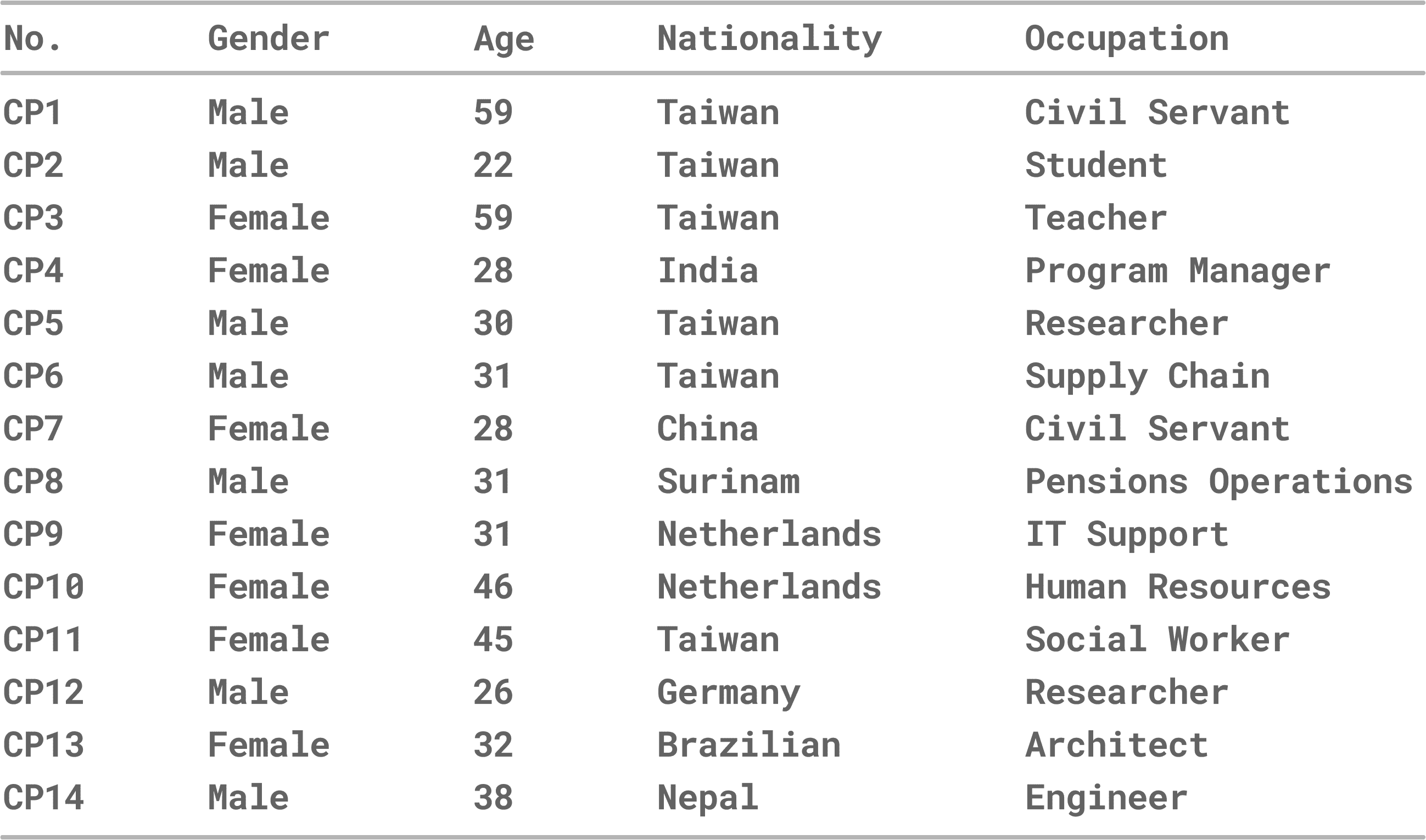}
  \caption{The Backgrounds of Citizen Participants.}
  \Description{The Backgrounds of Citizen Participants.}
  \label{fig:figure23}
\end{figure}

We recruited 14 participants through advertisements and word of mouth from cities in Taiwan and the Netherlands. Each participant received a compensation of €15. In this paper, we refer to the participants as CP (e.g., CP1 means citizen participant 1). The average age was 35, with participants ranging from their early 20s to early 60s. Their nationality and occupation varied (presented in Figure ~\ref{fig:figure23}).

\subsection{Procedure}

Before the workshop, participants signed informed consent forms and were instructed to take two photos of the cities they lived in, which were emailed to the first author and presented during the workshop. These photos were used in a prototyping activity to generate images, aiding them in identifying potential issues when integrating the designs into their urban lives. Each workshop took 2 hours and was moderated by the first author. The session began with a demonstration and trial of our GAI tool to familiarize participants with its capabilities. They then reviewed the six design idea posters and discussed concerns and frictions that might arise upon deployment and implementation. During the discussion, participants were invited to prototype the design ideas by uploading their photos or pictures from open source platforms (e.g., Unsplash), entering prompts that they may use in the context of design ideas, and reviewing the generated images. This activity served as a basis for them to evaluate the ideas and express their preferences for AI-generated content. Finally, the first author summarized the issues raised by the participants and collectively discussed potential solutions to these issues. The whole workshop was audio-recorded and transcribed for analysis. 

\subsection{Analysis}
We analyzed the workshop data, including five interview transcripts, 28 photos taken by participants prior to the workshops, and 107 AI-generated images along with their prompts. Our analytical objective was to identify the potential of design ideas that aligned with citizens' needs for playful engagement in public spaces and those they viewed with reservations. Particularly, we sought to distill the concerns associated with the Playful Features in design ideas, illuminating the tensions in integrating GAI into urban play.

Employing reflexive thematic analysis~\cite{Byrne:2022aa,Braun:2019}, the first, second, and fourth authors familiarized themselves with the data and conducted the open coding process. The first author then reviewed all codes, clustered them into positive, negative, and remaining citizen comments, and created an affinity diagram with initial themes. Subsequently, all authors participated in three iterations to examine codes, potential themes, and their naming, resulting in one theme describing the aspects of design ideas appreciated by citizens and four themes related to their concerns. After finalizing the Playful Features, all authors collaborated through two additional iterations, involving reflective engagements to ensure the themes effectively addressed our analytical objective. For instance, we realized that the candidate themes lacked links to the Playful Features, which insufficiently unveiled tensions of GAI's playfulness in urban situations. This realization led us to revisit the codes representing positive and negative citizen comments and examine how they were (not) connected to specific Playful Features, followed by theme reorganization and refinement. Finally, we settled on one theme outlining design ideas' positive potential to support citizens' needs and its associated Playful Features (Section 6.5.1) and five themes presenting Citizen Concerns (summarized in Figure{~\ref{fig:figure20}}). Among the concerns, four elaborate on distinct challenges and their relations to the Playful Features (Section 6.5.2 to 6.5.5), and the final theme reports general GAI issues across all design ideas (Section 6.5.6).

\subsection{Results of Citizen Evaluation}
\subsubsection{Enliven Cites' Socio-cultural Fabrics with GAI}
\begin{figure}
  \centering
  \includegraphics[width=\linewidth]{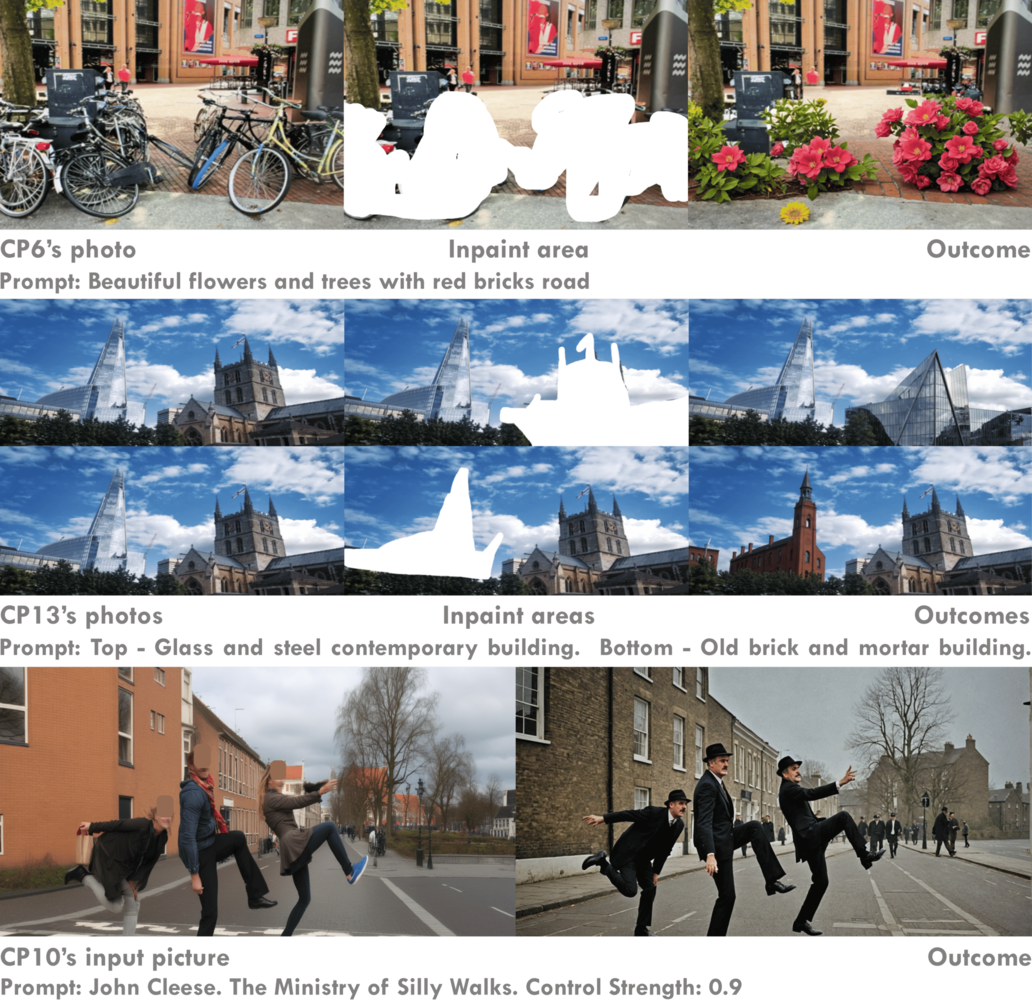}
  \caption{Top: CP6 changed a clutter of bikes into a flowerbed. Middle: CP13 altered the old architecture into a modern structure and vice versa. Bottom: CP10's images demonstrated how ``Mirrorverse Switch'' engaged people with Monty Python's ``silly walks,'' turning their poses into the distinctive comedic style.}
  \Description{Top: CP6 changed a clutter of bikes into a flowerbed. Middle: CP13 altered the old architecture into a modern structure and vice versa. Bottom: CP10's images demonstrated how ``Mirrorverse Switch'' engaged people with Monty Python's ``silly walks,'' turning their poses into the comedy's distinctive style.}
  \label{fig:figure15}
\end{figure}

The participants generally resonated with the design ideas and were attracted by the embedded playful features, recognizing their potential to infuse cities with dynamic and socio-cultural values that responded to citizens' needs. In ``City Renovation Solitaire,'' participants recognized the pleasure and meaning associated with the empowerment of visualization to share desired neighborhood (F1) and re-discover the people-place relationship (F2). They appreciated how the design enabled them to ``observe the city in a critical way'' (CP10) and creatively express their visions for improvement, thereby encouraging them to participate in placemaking (CP1-3, 5-8, 10-14). For example, after transforming a cluttered corner into a flowerbed (Figure~\ref{fig:figure15}, top), CP6 remarked, ``In the past, we often criticized the city through complaints or harsh words, but now we’re empowered to modify the city with elegance and joy... It becomes an incentive to improve urban environments together.'' Moreover, the generated images were considered ``a valuable compilation of visions from bottom-up'' which could ``facilitate conversations between citizens and municipalities'' (CP13) and ``enhance citizens' influence on shaping the city'' (CP1).   

Our participants also saw emotional and intellectual value in ``City Inception Solitaire,'' connecting to shuttling between multiverse and reality (F3). It allowed them to ``travel between different worlds with fun'' (CP9) and be motivated to ``understand more about the city'' (CP2, CP8). For instance, CP13 enjoyed transforming the time of two architectures built in different eras and expressed curiosity about their styles (Figure~\ref{fig:figure15}, middle). By merging neighborhoods with historical cityscapes, such as Taiwan's military villages, this design idea offered new opportunities ``for the elderly to reminisce and dream again'' (CP11). If mainly featuring whimsical urban montage (F5), ``City Inception Solitaire'' could also create a bizarre journey of non-human urban legends. CP8 noted that this journey made him ``more likely to get out of the house'' to ``see what random AI transformations will occur'' (CP8). Additionally, linked to the encounter of defamiliarized places (F4), ``Tracing People's Meaningful Places'' created a mysterious city treasure hunt (CP5, CP6, CP12). Generating or experiencing unfamiliar streetscapes with GAI made participants ``more interested in observing and exploring every corner of the familiar city'' (CP5), thus ``noticing things they had not noticed before'' (CP12).

Finally, the participants gravitated towards ideas featuring pass-down co-improvisation (F6), which delightfully fostered connections and cultural exchange through collaborative inputs. They were attracted to the way ``Mobile City as Canvas'' created shared memories for friends' and families' journeys by iteratively generating images together (CP1-4, 14), while in ``Mirrorverse Switch'' and ``Walk to Paint,'' impromptu co-creating content with prompts related to cultural festivals served as an ``icebreaker'' for diverse conversations (CP5-7) and made ``expats feel welcome in the city'' (CP10). The examples from CP10 demonstrated how ``Mirrorverse Switch'' could involve passersby in collectively interact with classic artworks (Figure~\ref{fig:figure15}, bottom). Overall, most participants valued the playfulness these designs introduced to public spaces and supported their further development to enrich urban lives. 

\begin{figure*}
  \centering
  \includegraphics[width=\linewidth]{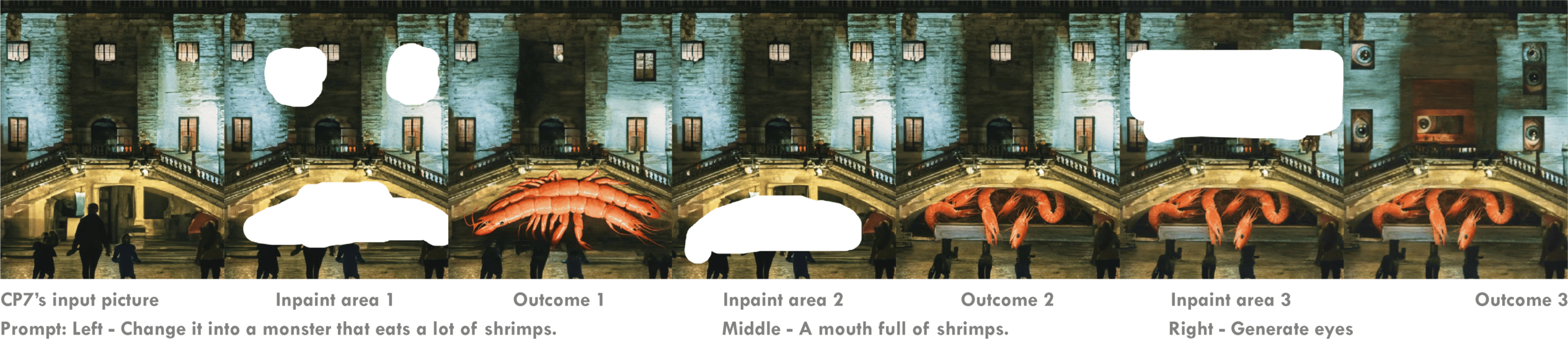}
  \caption{CP7 and CP9 attempted thrice to generate a face eating shrimp, but the results only included shrimp and eyes.}
  \Description{CP7 and CP9 attempted thrice to generate a face eating shrimp, but the results only included shrimp and eyes.}
  \label{fig:figure17}
\end{figure*}

\subsubsection{Fatigue and Marginalization in Misaligned Content Creation (C1)} Although our participants anticipated engaging in the designs in urban settings, they indicated the issues of fatigue and marginalization stemming from the mismatch between users' skills and GAI's functionalities, which should be addressed before achieving playful experiences. The misalignment includes the limitations of GAI’s training data and users’ prompting ability. Due to the lack of datasets associated with specific prompts, GAI may misinterpret the prompts and fill the outcomes with whims. For instance, CP7 and CP9 attempted to transform a building into a monster eating shrimps, but iWonder struggled to accurately generate the elements mentioned in the prompts, requiring multiple attempts to achieve a partially satisfactory result (Figure~\ref{fig:figure17}). Moreover, users’ inability to articulate appropriate prompts, such as illiteracy, language barriers, and insufficient strategies for prompt construction, can cause dissatisfying outcomes (CP1, 10, 11, 13).

The workshop discussion revealed that the misalignment in urban play situations could lead to citizen fatigue. For example, CP7 and CP9 expressed frustration when encountering mismatches. They considered that such experiences, if happening during the ``Walk to Pain'' activity, would demand greater mental and physical effort to generate prompts, thereby causing exhaustion. As they commented, ``Suppose we play it in an actual square. We already walked and had imagined how it should look, but when it doesn't match, we'll be disappointed and tired.'' This problem also highlighted the tension created by whimsical urban montages (F5), demonstrating how overwhelming whims---in this case, the misaligned AI-generated content---can disengage participants and undermine the playfulness in empowerment of visualization (F1-3). 

Additionally, misalignment can occur within specific groups during pass-down co-improvisation (F6), potentially amplifying marginalization. Participants observed that cultural minorities often faced more misaligned experiences, implicitly excluding them from collective play. For example, CP11 noted the relative ease of transforming a neighborhood into an Iron Man-themed setting compared to the difficulty of recreating a Taiwanese Indigenous village. This discrepancy reflects the under-representation of certain cultural contexts in training datasets, which tend to favor culturally dominant groups. Furthermore, AI models predominantly trained in English may exacerbate mismatches for non-English speakers, further diminishing inclusivity in such activities. This issue became evident during our prototyping exercise of ``City Renovation Solitaire'' and ``City Inception Solitaire,'' where CP1, CP10, CP11, and CP13 noted frequent misalignment between outcomes and the prompts described in Chinese, Dutch, or Portuguese, hindering their experience of modifying cityscapes in relays. When discussing ``Mirrorverse Switch,'' C10 also pointed out disparities in prompting abilities across communities with varying literacy levels, stating, ``There's a big divide between highly educated people, kids, and those who're illiterate in cities. They may struggle with adding prompts.'' Such challenges increase the likelihood of mismatches, creating barriers to meaningful engagement in collaborative play with GAI. Participants suggested solutions for mitigation, such as implementing prompt language switching and incorporating voice input. As CP10 elaborated, ``Children's words aren't precise... It's good if they use simple terms, and ChatGPT refines them into complete prompts more compatible with the image generator.''

\subsubsection{Navigational Unsafety from Urban Scene Rendering and Defamiliarization (C2)}

Participants emphasized the importance of preventing navigational danger associated with GAI in urban contexts, including directing players into hazardous situations or disrupting non-players' commutes, thereby protecting physical safety in all GAI-enabled urban play. For example, while CP1, CP3, and CP5 appreciated the playfulness of empowered visualization, they worried that it might worsen the problem of ``phone zombies'' due to the immersion of this power. This concern also conveyed how they prioritized safety above GAI's empowerment (F1-3). Additionally, in the ``Tracing People's Meaningful Places,'' AI's defamiliarization of real-world locations (F4) could inadvertently ``lead people into unsafe areas'' (CP12) or ``private neighborhoods'' (CP2), which threatened both personal and residential safety. The placement of the designs, whether involving single or multiple players (F6), should also be carefully planned to avoid putting citizens at risk or affecting urban functions. CP3 and CP10 specifically advised against using such activities in high-traffic areas, where they might distract drivers. These concerns underscore the need for robust safeguards when designing GAI-enabled urban play to prevent harm in cities.

\subsubsection{Tensions between Preserving and Transforming Socio-cultural Significance (C3)}  

When participants evaluated ``City Inception Solitaire,'' they expressed two opposing expectations about how citizens might engage with GAI, potentially leading to tensions: the inclination to embrace playful features that could distort, mock, or even replace socio-cultural origins, against the desire to curb those features to learn about and preserve the origins.  

\begin{figure*}
  \centering
  \includegraphics[width=\linewidth]{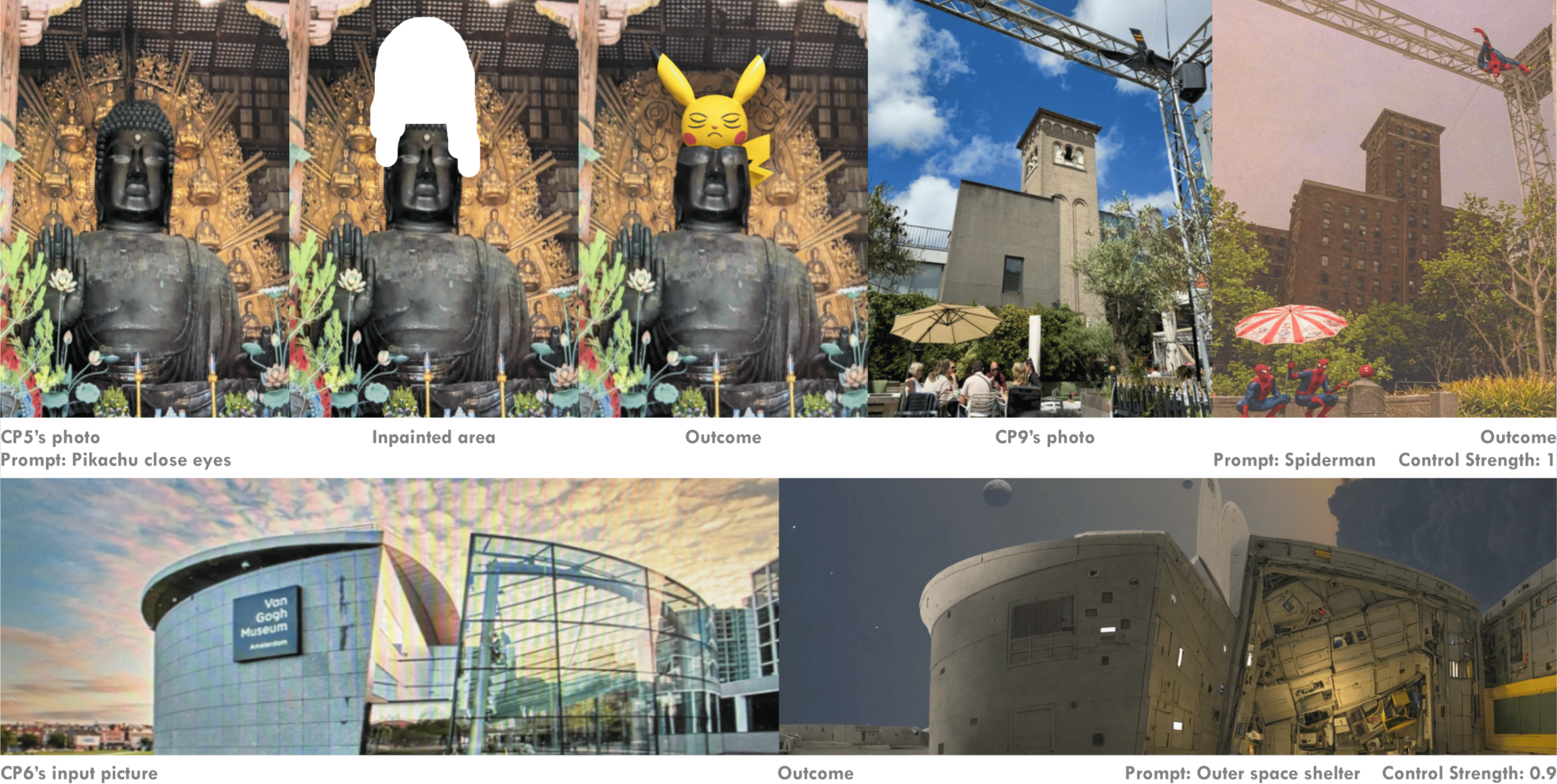}
  \caption{Generated images reconfigured socio-cultural elements. Top: CP5 added Pikachu to a Buddha statue and the crowds in CP9's photo became Spider-Man characters. Bottom: CP6 turned the Van Gogh Museum into a space station.}
  \Description{Generated images reconfigured socio-cultural elements. Top: CP5 added Pikachu to a Buddha statue and the crowds in CP9's photo became Spider-Man characters. Bottom: CP6 turned the Van Gogh Museum into a space station.}
  \label{fig:figure19}
\end{figure*}

Some participants saw urban spaces as dynamic, open to playful and creative reinterpretation (CP2-6, 8, 10). They considered the solitaire a liberating force that ``anyone can reshape their environment with fun'' (CP5). This force connected to the empowerment of visualization (F1-3) and the possibility of achieving pass-down co-improvisation (F6), which transformed cities into canvases for ``joyful experimentation'' (CP6) and playfully produced new cultural references. They also appreciated how whimsical and defamiliarized elements added by GAI could disrupt rigid narratives, cross cultural boundaries, and invite diverse urban experiences (F4, 5). These participants enjoyed parodying, deconstructing, or merging socio-cultural elements with GAI to introduce new layers of meaning and challenge traditional interpretations (Figure~\ref{fig:figure19}). They embraced the potential of GAI to ``reimagine the city and create new stories'' (CP4), rendering the urban surroundings more vibrant and transformative. 

Conversely, for other participants, city spaces are significant vessels of history, memory, and identity, and they viewed GAI's transformations with a mix of fascination and caution (CP1, 7, 9-11). They emphasized the need in ``City Inception Solitaire'' to ``deepen connections to the past rather than replace them'' (CP7) and help citizens understand and appreciate ``urban’s socio-cultural roots'' (CP9). This need implied their expectation on amplifying the empowerment of visualization to re-discover people-place relationship (F2) while restricting the playfulness from other visualization purposes (F1, 3), GAI’s whims and defamiliarization (F4, 5), and co-improvisation (F6), which oriented toward the preservation of existing socio-cultural structures. There was a concern that the boundless playfulness might trivialize or distort the precious meanings embedded in the city’s architecture, landmarks, and streets. As CP7 noted, ``In the past, it's hard to change cityscape images… now we can just type in prompts and change it, but the historical meaning behind it is lost… I'm afraid people only focus on the fantasy version.'' They feared that, if incautious and unchecked, GAI could disrupt cultural authenticity, replace meaningful elements with superficial interpretations, and potentially disconnect people from the historical and social narratives that define their community. As CP9 echoed this sentiment, ``I wouldn't change my hometown in the solitaire because I want to see how it actually is... It's about preserving the emotions and memories of places. I don't want to mix up the stories that happened here'' (CP9). 

\subsubsection{Dissemination of Disturbing Modification on Everyday Cityscapes (C4)}

The prototyping with our tool elicited participants' concerns about disseminating altered daily urban scenes that irritated audiences, implying how GAI's playful features are maliciously leveraged in urban play scenarios. City environments could become materials for misusing the empowerment of sharing desired neighborhood (F1) to spread sensitive content. For instance,  CP14 recognized the high accessibility and mobility of iWonder and highlighted the risk of sharing uncomfortable news about places by photographing and Inpainting. He stated, ``This AI makes photoshopping streetscapes easy. People can capture them from everywhere nearby and share modified versions... very valuable but also capable of creating scams or panic if not managed well.'' CP12 also found that our tool could unpredictably produce street views with disasters, highlighting that GAI’s playful whims (F5) might inspire or promote misuse practice. For example, when he prototyped ``Mobile City as Canvas,'' a prompt intended to create a cartoonish explosion resulted in a realistic depiction resembling a terrorist attack (Figure~\ref{fig:figure18}, left). He noted, ``If players on the train find it funny and start generating this kind of scene all the time, the passengers might feel scared and think there's something wrong outside.''  

Apart from the misapplications by an individual, they could be collectively performed to ``test the limits of your design'' (CP8). CP3 and CP10 warned that young people might engage in relays to generate discomfort content, such as ``bloody monsters'' or ``naked ladies crossing the street,'' which disturbed passersby. These examples revealed that the unlimited pass-down co-improvisation (F6) can harm city society. To mitigate these risks, participants recommended implementing content moderation measures, though they acknowledged the challenge of filtering content in real time. For instance, during the Inpainting in ``Walk to Paint'' or ``Mobile city as Canvas,'' CP9 worried that people could try out various pass-down possibilities to promote inappropriate ideologies before they are effectively blocked, as ``sometimes you don't know what they're drawing until they finish it'' (CP9).  

\subsubsection{General GAI Issues: Bias, Misinformation, and Privacy (C5)}
In addition to the specific GAI concerns in urban play situations (presented in Sections 6.5.2-6.5.5), this study observed that general ethical issues of GAI possibly occur in the play, encompassing introducing bias, creating misinformation, and invading privacy. 

\begin{figure}
  \centering
  \includegraphics[width=\linewidth]{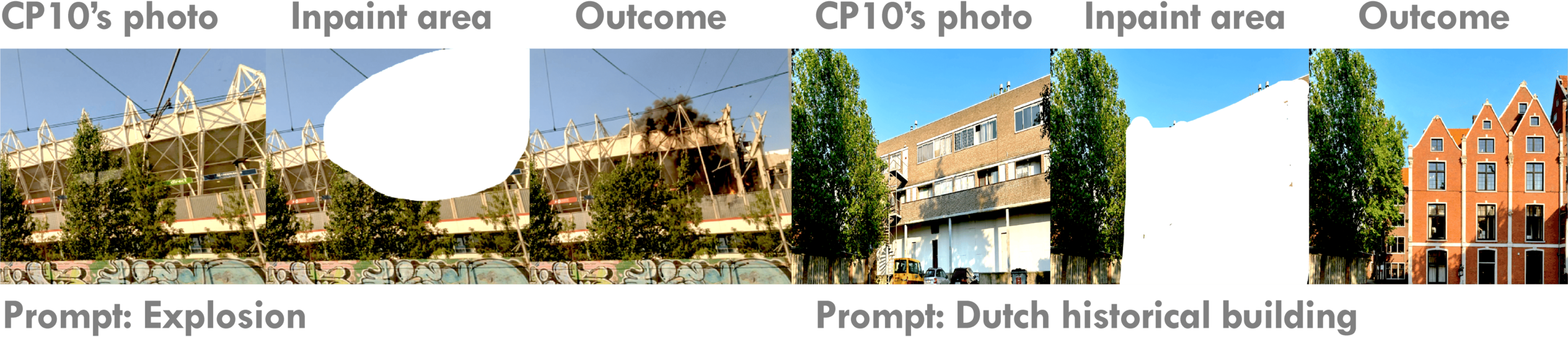}
  \caption{Left: iWonder depicted a realistic explosion resembling a terrorist attack. Right: CP10 observed that iWonder misinterpreted Dutch historical buildings in a modern style.}
  \Description{Left: iWonder depicted a realistic explosion resembling a terrorist attack. Right: CP10 observed that iWonder misinterpreted Dutch historical buildings in a modern style.}
  \label{fig:figure18}
\end{figure}

Regarding bias and misinformation problems, participants noted that GAI might perpetuate cultural misunderstandings or reinforce stereotypes. For example, when CP10 attempted to generate a ``historical'' Dutch scene, the result was a modern image, highlighting a bias in the GAI's understanding of Dutch history (Figure~\ref{fig:figure18}, right). Similarly, CP11 pointed out iWonder's lack of knowledge regarding native Taiwanese plants, leading to inaccurate representations that could mislead users unfamiliar with the local context and potentially reinforce false perceptions. Moreover, connected to the dissemination issue in Section 6.5.5, the disturbing cityscape modification may also become ``realistic but fake disasters'' (CP14), ``fabricated news,'' and ``scams'' (CP12) which deluded viewers. The participants suggested clear indicators in the play to inform users that the pictures are AI-generated and unreal (CP12-14).  

Privacy was another concern raised by participants, particularly regarding the use of photographs. CP10 voiced concern about unintentional privacy and violations, noting that ``other people appear in the background without their consent.'' CP13 also questioned data handling, asking, ``Where is that data going? Who is managing this data?'' To mitigate these risks, participants suggested that the AI should ``delete photos immediately after generation'' (CP5) or automatically blur people’s faces (CP7-10, 13, 14). Clear communication and guidance were also called for, with CP13 recommending, ``You need to let people know they can enter information confidently and first guide them to play.'' Importantly, participants felt that citizens should be able to easily opt out of AI-enabled activities. CP4, CP9, and CP10 stressed the need for mechanisms allowing individuals to avoid being photographed if they choose not to participate. These concerns underline the necessity for comprehensive examinations and clear instructions to ensure that AI-enabled urban play remains responsible and respectful of citizens’ privacy. 

\section{Discussion}
\subsection{Fostering Urban Play Qualities with GAI}
In this work, we engaged designers in situated exploration with a GAI tool (iWonder) and design ideation to uncover six Playful Features and ideas of GAI-enabled urban play. Our mobile-compatible tool allows users to wander into urban areas, capture images, and edit (parts of) those images with GAI through textual prompts. While similar tools exist (e.g.,~\cite{Epstein:2022,Lau:DIS2024,Rafner:CC2021}), our tool's mobility, simplicity in functionality, and ease of use, combined with the power of GAI, sparked the designers' imagination and ideation to explore playful features of using GAI in city spaces. We also conducted the citizen workshop involving discussion on the design ideas and their use of our tool to gather bottom-up feedback. The empirical results demonstrated how GAI can facilitate playful urban engagement and introduce new playfulness. Here, we propose three playful qualities of GAI distilled from the Playful Features (mapped in Figure~\ref{fig:figure20}) and discuss the urban play qualities they potentially support. 

Among the urban play qualities, GAI appears to particularly enhance the agential aspect~\cite{Bertran:2022}. Linked to the empowerment of visualization (F1-3), GAI provides citizens with \textit{generative agency}: the power to pleasurably create reimaginations of cities' every corner, fostering a sense of control and motivation to express, reflect, and recreate their impressions of places. In most prior forms of urban play, such as mixed reality games and interactive installations, content is primarily crafted by designers. Although some designs invite citizens to contribute, these contributions are restricted to text~\cite{helloLampPost,Ruiz:2011}, photographs~\cite{Stokes:CHIPLAY2020}, or graffiti projections with specified themes~\cite{Tobias:2010}. In comparison, GAI empowers citizens to co-create their own play by offering the ability to render everyday cityscapes into alternatives with various themes. They can apply this ability to share desired neighborhood (F1), which not only obtains intriguing experiences but also constructively expresses opinions about cities---shifting from negative or abstract language to positive and concrete imagery that depicts their vision of ideal living cities. Citizens could also use GAI to re-discover the people-place relationship (F2) and shuttle between multiverse and reality (F3). Through modifications of urban scenes, they can playfully reflect on the essence of cities and their attachment to places, explore diverse urban cultures, and momentarily immerse in time traveling and reminiscing, additionally contributing to the emotional and intellectual aspects of urban play qualities. 

The \textit{meaningful unpredictability} of AI-generated content is another playfulness that enriches the emotional and intellectual quality. While similar to location-based and augmented reality technologies which connect various urban locations and virtual worlds, GAI distinctively infuses uncertainty into the connection and opens spaces for meaning-making. Citizens are invited to experiment wtih how it might mediate the familiarity of places and guide them to the encounter of defamiliarized places (F4), sparking curiosity and serendipitously evoking rich feelings such as d\'ej\`a vu. They can also appreciate the whimsical urban montage (F5), where GAI brings surprising and creative imaginations that joyfully challenge conventional spatial logic and urban understanding.

Lastly, GAI supports playful urban behaviors that foster social quality through \textit{social performativity}. Evidenced in the design ideas of ``Mirrorverse Switch'' and ``Walk to Paint,'' GAI possesses great potential to facilitate performative activities such as urban art creations and interactive engagement with strangers~\cite{Bertran:2022}, where the city itself becomes a stage for spontaneous expression and interaction. Uniquely, GAI enhances these activities through pass-down co-improvisation (F6): an iterative and accumulative interaction that introduces and invites impromptu generated content during the social play in public, adding dynamism and amplifying the enjoyment of the performative experiences. 
  
Collectively, this research serves as a starting point to illuminate GAI's potential as a design material to construct playable cities, contributing to the emerging design space of technology-mediated urban play. Moreover, our findings advance the HCI community's understanding of the relationship between GAI's characteristics, urban environments, and users. 

\subsection{Considerations Before Design Implementation in Urban Settings}

Through the citizen workshop, this work reveals concerns and risks of GAI's playful features in city situations, surfacing the tensions of GAI-enabled urban play. We invited citizen participants to comment on design idea posters and engage in our GAI tool. These materials make the abstract GAI technology and design concepts relatable and experiential, enabling constructive opinions on how GAI should make cities more playful. The findings reveal potential issues of GAI in urban contexts from bottom-up perspectives, providing designers and researchers with critical insights into the obstacles to playful urban experiences and guiding informed decision-making before deploying GAI. Based on the Citizen Concerns, we conclude three offensive qualities, \textit{dissonant agencies}, \textit{insensitive erosion}, \textit{creative vandalism}, to encapsulate how GAI-enabled playfulness might be limited, mishandled, or misemployed. Below we elaborate on the offensiveness, compare the involving concerns to those of prior literature, and provide pre-deployment considerations.   

First, a major challenge of applying GAI to urban play is the misalignment between human skills and GAI capabilities, termed \textit{dissonant agency}. This dissonance, encompassing limitations of user abilities of prompting and GAI's training database, can lead to inaccurate and inconsistent content generation. Such discrepancies result in disappointment, puzzlement, and discouragement---issues echoed in recent GAI literature~\cite{Fu:DIS2024,Weisz:CHI2024,Weisz:CoRR2023}. Notably, the mismatch in GAI-enabled urban play requiring social engagement or physical effort tends to amplify fatigue with higher user frustration or disengagement (C1). Moreover, since urban play involves users of diverse literacy levels, ages, and languages, reducing barriers in prompt creation is crucial to prevent marginalization (C1). Several studies have suggested potential solutions, including the adjustment of model themes and parameters~\cite{Weisz:CoRR2023}, provision of multiple output options to increase satisfaction likelihood~\cite{Fu:DIS2024}, and automatic suggestion of text prompts based on parsed user inputs~\cite{Dang:ArXiv2022}. In addition, we recommend the exploration of intuitive instructions and interactions for prompting, such as built-in scenarios of transferrable styles, voice input with speech-to-text AI systems, and the integration of large language models, which makes the play more inclusive and supportive.  

Next, \textit{insensitive erosion} involves GAI's unawareness of nuances of urban contexts and the inadvertent use of GAI playfulness, along with their consequences of degrading physical safety and socio-cultural authenticity. Apart from limited considerations on placement of play that causes unsafe situations {\cite{Bertran:2022}}, the immersion and defamiliarization enhanced by GAI could displace users’ attention from surroundings and lead to ``phone zombie'' behaviors or navigational dangers, eroding urban functionality and public safety (C2). When playing with GAI's transformative power, the insensitivity of socio-cultural origins risks distorting their meanings and replacing authentic narratives with superficial reinterpretations, potentially eroding the emotional and historical ties between communities and their places (C3). To mitigate the offensiveness of \textit{insensitive erosion}, it is essential to examine the navigational safety of both players and non-players and explore design approaches that balance creative GAI transformation with socio-cultural preservation. 

Moreover, GAI-enabled playfulness could disrupt urban environments through \textit{creative vandalism}, where everyday cityscapes are playfully modified and disseminated to provoke discomfort or spread harmful ideologies (C4). GAI seems to lower the threshold for creating landscapes with inappropriate elements, enable coordinated sharing of them, and unpredictably offer misuse inspirations, which escalates societal disturbance (C4). The suggestions for mitigating this issue include clearly labeling AI-generated content as unreal, regulating usage {\cite{Bertran:2022}}, ensuring accountability, stress-testing, and continued oversight of algorithms {~\cite{Fu:DIS2024,Laura:FACCT2022}}, and incorporating ethical intervention in the prompts{\cite{Hritik:2022}}. Lastly, our findings align with prior research on GAI's ethical concerns, such as privacy, bias, and misinformation (C5). The privacy concerns observed in this study, such as the inclusion of bystanders in the database and unclear data handling, prompted calls for automatic face-blurring, immediate photo deletion, opt-out options, and the establishment of robust data policies {\cite{Bertran:2022}}. Regarding bias and misinformation issues, the realistic yet false depictions of neighborhoods by GAI, along with its inaccurate illustrations of historical and regional elements, may misguide people or perpetuate cultural misunderstandings. To address these issues, it is beneficial to apply the solutions for~\textit{creative vandalism} or the principles from literature, including selecting more comprehensive datasets {~\cite{Fu:DIS2024,Laura:FACCT2022}}, articulating GAI's capability, and elaborating rationales for outputs {\cite{Weisz:CHI2024}}.   

In addition to the above recommendations, it is suggested to conduct field-testing and multi-stakeholder workshops before implementing and deploying GAI-enabled urban play. The field-testing approach is crucial for understanding the technological performance in the wild, specific social dynamics, and nuances of place-specific contexts difficult to capture in indoor settings \cite{Long:2019}. Furthermore, designers and researchers can hold workshops to discuss urban play proposals with stakeholders such as city planners, technologists, educators, and artists~\cite{Innocent:2019,Innocent:2021}, facilitating design evaluation with a balanced lens on city layers \cite{Bertran:2022}. The workshop can also involve prototyping of the proposals through situated GAI tools, allowing them to constructively evaluate the designs based on contextual GAI experiences. Together, the field-testing and workshops inform design refinements that align with different concerns, needs, and values, thereby preventing harm and enabling contextually sensitive urban play experiences.  

\subsection{GAI as a Playful yet Offensive Tourist}
In Figure~\ref{fig:figure20}, we mapped the relationships between GAI-enabled Playful Features and Citizen Concerns situated in city contexts and connected them with corresponding playful and offensive qualities, illustrating the tensions of incorporating GAI into playful urban interaction. The reflection on these findings led us to propose GAI as a ``playful yet offensive tourist.'' This metaphor encapsulates the overarching contradictions and complexities inherent in GAI-enabled urban play and offers a valuable perspective for the HCI community regarding GAI applications across domains. 

Firstly, the tourist metaphor captures the tensions between the playfulness and offensiveness GAI could introduce in city spaces. GAI's playful qualities, such as \textit{generative agency} and \textit{meaningful unpredictability}, become a force that infuses cities with new meanings and redefines socio-cultural boundaries, akin to an experienced yet curious tourist who---``wanders'' through a city, creatively assembles new elements with past experiences, and seeks serendipity. This process generates intriguing reimaginations and interpretations that are unfamiliar yet inspirational, surprising, or thought-provoking to locals. Through \textit{social performativity}, GAI further fosters the transformation of urban spaces into improvised stages, analogous to a tourist participating in local festivities, enhancing the vibrancy of collective performances and impromptu co-creating meanings. On the other hand, GAI's playful interventions could be seen as intrusive. \textit{Dissonant agency} in GAI-enabled play parallels language barriers between tourists and locals, hindering rich dialogues and creating rifts. Analogous to a tourist unaware of the subtle socio-cultural and geographical nuances, GAI might contribute to \textit{insensitive erosion}, unwittingly disrespecting local customs, flattening the deeper and lived meanings of a place into surface-level entertainment, or leading to safety concerns. GAI can also become a radical tourist who misapplies the exploratory freedom to present inappropriate content, giving rise to \textit{creative vandalism} that propagates harmful ideologies and disrupts local order. Through the metaphor, we aim to sensitize designers, researchers, and city stakeholders to recognize GAI's dual nature in enriching urban experiences and encourage future work on leveraging GAI’s playful potential while mitigating its offensive tendencies. 

Secondly, the tourist metaphor provides a new lens for incorporating GAI into various HCI domains. GAI is capable of grasping the general essence of a domain and delivering a fresh and engaging experience, similar to a tourist who quickly and joyfully learns about the major landmarks of a city. Like a tourist interpreting new sites through diverse prior travel experiences, GAI introduces an outsider's perspective that injects novel ideas and unexpected connections into the domain. These attributes can stimulate creative innovation, challenge traditional boundaries, and open new possibilities for exploration. However, akin to a tourist who might overlook or misinterpret the intricacies of a culture, GAI can fail to capture the nuanced meanings embedded in a domain, resulting in outcomes that, while innovative, may also feel reductive or even disrespectful to those deeply familiar with the domain’s subtleties. Thus, this metaphor underscores the need for design considerations and discourses that explores GAI's playful and transformative potential while carefully considering the domain’s nuances and critically examining GAI's impacts on them.  

\section{Conclusion}
This study follows the continuing endeavors of playfully re-signifying urban environments through technologies while critically examining the accompanying risks. We unfold how designers interact with an image-to-image GAI tool in urban settings and conclude them as six Playful Features, followed by interviews and design ideation producing ideas of GAI-enabled urban play. Next, we collect citizens’ responses through a citizen evaluation of the ideas, capturing their desires and concerns related to Playful Features. Finally, we distill playful and offensive qualities from our findings, offer design considerations for addressing the concerns, and propose the tourist metaphor. This work acknowledges that more comprehensive studies and expertise are necessary to mitigate GAI's risks and implement responsible design. We anticipate future research to investigate design possibilities and challenges of playful urban interaction facilitated by GAI, contributing to invigorating cities and harmonizing the tensions between users, AI, and urban contexts.




\begin{acks}
We sincerely appreciate the efforts from our reviewers and participants. We are thankful to our funders: Taiwan National Science and Technology Council under the grant number MOST 113-2917-I-011-004, alongside the Department of Industrial Design and the Eindhoven Artificial Intelligence Systems Institute at Eindhoven University of Technology.
\end{acks}

\bibliographystyle{ACM-Reference-Format}
\bibliography{gai_urban_play}


\begin{thebibliography}{88}


\ifx \showCODEN    \undefined \def \showCODEN     #1{\unskip}     \fi
\ifx \showDOI      \undefined \def \showDOI       #1{#1}\fi
\ifx \showISBNx    \undefined \def \showISBNx     #1{\unskip}     \fi
\ifx \showISBNxiii \undefined \def \showISBNxiii  #1{\unskip}     \fi
\ifx \showISSN     \undefined \def \showISSN      #1{\unskip}     \fi
\ifx \showLCCN     \undefined \def \showLCCN      #1{\unskip}     \fi
\ifx \shownote     \undefined \def \shownote      #1{#1}          \fi
\ifx \showarticletitle \undefined \def \showarticletitle #1{#1}   \fi
\ifx \showURL      \undefined \def \showURL       {\relax}        \fi
\providecommand\bibfield[2]{#2}
\providecommand\bibinfo[2]{#2}
\providecommand\natexlab[1]{#1}
\providecommand\showeprint[2][]{arXiv:#2}

\bibitem[Al-kfairy et~al\mbox{.}(2024)]%
        {Alkfairy:2024}
\bibfield{author}{\bibinfo{person}{Mousa Al-kfairy}, \bibinfo{person}{Dheya Mustafa}, \bibinfo{person}{Nir Kshetri}, \bibinfo{person}{Mazen Insiew}, {and} \bibinfo{person}{Omar Alfandi}.} \bibinfo{year}{2024}\natexlab{}.
\newblock \showarticletitle{Ethical Challenges and Solutions of Generative AI: An Interdisciplinary Perspective}.
\newblock \bibinfo{journal}{\emph{Informatics}} \bibinfo{volume}{11}, \bibinfo{number}{3} (\bibinfo{year}{2024}).
\newblock
\showISSN{2227-9709}
\urldef\tempurl%
\url{https://doi.org/10.3390/informatics11030058}
\showDOI{\tempurl}


\bibitem[Altarriba~Bertran et~al\mbox{.}(2022)]%
        {Bertran:2022}
\bibfield{author}{\bibinfo{person}{Ferran Altarriba~Bertran}, \bibinfo{person}{Laura Bisbe~Armengol}, \bibinfo{person}{Cameron Cooke}, \bibinfo{person}{Ivy Chen}, \bibinfo{person}{Victor Dong}, \bibinfo{person}{Binaisha Dastoor}, \bibinfo{person}{Kelsea Tadano}, \bibinfo{person}{Fyez Dean}, \bibinfo{person}{Jessalyn Wang}, \bibinfo{person}{Adri\`{a} Altarriba~Bertran}, \bibinfo{person}{Jared Duval}, {and} \bibinfo{person}{Katherine Isbister}.} \bibinfo{year}{2022}\natexlab{}.
\newblock \showarticletitle{Co-Imagining the Future of Playable Cities: A Bottom-Up, Multi-Stakeholder Speculative Inquiry into the Playful Potential of Urban Technology}. In \bibinfo{booktitle}{\emph{Proceedings of the 2022 CHI Conference on Human Factors in Computing Systems}} (New Orleans, LA, USA) \emph{(\bibinfo{series}{CHI '22})}. \bibinfo{publisher}{Association for Computing Machinery}, \bibinfo{address}{New York, NY, USA}, Article \bibinfo{articleno}{534}, \bibinfo{numpages}{19}~pages.
\newblock
\showISBNx{9781450391573}
\urldef\tempurl%
\url{https://doi.org/10.1145/3491102.3501860}
\showDOI{\tempurl}


\bibitem[Ampatzidou et~al\mbox{.}(2018)]%
        {Ampatzidou:2018}
\bibfield{author}{\bibinfo{person}{Cristina Ampatzidou}, \bibinfo{person}{Katharina Gugerell}, \bibinfo{person}{Teodora Constantinescu}, \bibinfo{person}{Oswald Devisch}, \bibinfo{person}{Martina Jauschneg}, {and} \bibinfo{person}{Martin Berger}.} \bibinfo{year}{2018}\natexlab{}.
\newblock \showarticletitle{All Work and No Play? Facilitating Serious Games and Gamified Applications in Participatory Urban Planning and Governance}.
\newblock \bibinfo{journal}{\emph{Urban Planning}} \bibinfo{volume}{3}, \bibinfo{number}{1} (\bibinfo{year}{2018}).
\newblock


\bibitem[Aoun(2013)]%
        {Aoun:2013}
\bibfield{author}{\bibinfo{person}{Charbel Aoun}.} \bibinfo{year}{2013}\natexlab{}.
\newblock \bibinfo{booktitle}{\emph{The Smart City Cornerstone: Urban Efficiency}}.
\newblock \bibinfo{type}{{T}echnical {R}eport}. \bibinfo{institution}{Schneider Electric}.
\newblock


\bibitem[Araake et~al\mbox{.}(2021)]%
        {Araake:2021}
\bibfield{author}{\bibinfo{person}{Koichi Araake}, \bibinfo{person}{Michinari Kono}, \bibinfo{person}{Eiji Iwata}, {and} \bibinfo{person}{Norio Sasaki}.} \bibinfo{year}{2021}\natexlab{}.
\newblock \showarticletitle{Playful Engagement for Public Spaces: A Case Study on a Mall Escalator}.
\newblock \bibinfo{journal}{\emph{Proc. ACM Hum.-Comput. Interact.}} \bibinfo{volume}{5}, \bibinfo{number}{ISS}, Article \bibinfo{articleno}{498} (\bibinfo{date}{nov} \bibinfo{year}{2021}), \bibinfo{numpages}{19}~pages.
\newblock
\urldef\tempurl%
\url{https://doi.org/10.1145/3488543}
\showDOI{\tempurl}


\bibitem[Bansal et~al\mbox{.}(2022)]%
        {Hritik:2022}
\bibfield{author}{\bibinfo{person}{Hritik Bansal}, \bibinfo{person}{Da Yin}, \bibinfo{person}{Masoud Monajatipoor}, {and} \bibinfo{person}{Kai-Wei Chang}.} \bibinfo{year}{2022}\natexlab{}.
\newblock \showarticletitle{How well can Text-to-Image Generative Models understand Ethical Natural Language Interventions?}. In \bibinfo{booktitle}{\emph{Proceedings of the 2022 Conference on Empirical Methods in Natural Language Processing}}, \bibfield{editor}{\bibinfo{person}{Yoav Goldberg}, \bibinfo{person}{Zornitsa Kozareva}, {and} \bibinfo{person}{Yue Zhang}} (Eds.). \bibinfo{publisher}{Association for Computational Linguistics}, \bibinfo{address}{Abu Dhabi, United Arab Emirates}, \bibinfo{pages}{1358--1370}.
\newblock
\urldef\tempurl%
\url{https://doi.org/10.18653/v1/2022.emnlp-main.88}
\showDOI{\tempurl}


\bibitem[Benford et~al\mbox{.}(2005)]%
        {Benford:2005}
\bibfield{author}{\bibinfo{person}{Steve Benford}, \bibinfo{person}{Carsten Magerkurth}, {and} \bibinfo{person}{Peter Ljungstrand}.} \bibinfo{year}{2005}\natexlab{}.
\newblock \showarticletitle{Bridging the physical and digital in pervasive gaming}.
\newblock \bibinfo{journal}{\emph{Commun. ACM}} \bibinfo{volume}{48}, \bibinfo{number}{3} (\bibinfo{date}{mar} \bibinfo{year}{2005}), \bibinfo{pages}{54--57}.
\newblock
\showISSN{0001-0782}
\urldef\tempurl%
\url{https://doi.org/10.1145/1047671.1047704}
\showDOI{\tempurl}


\bibitem[Benjamin et~al\mbox{.}(2023)]%
        {Benjamin:2023}
\bibfield{author}{\bibinfo{person}{Jesse~Josua Benjamin}, \bibinfo{person}{Heidi Biggs}, \bibinfo{person}{Arne Berger}, \bibinfo{person}{Julija Rukanskaitundefined}, \bibinfo{person}{Michael~B. Heidt}, \bibinfo{person}{Nick Merrill}, \bibinfo{person}{James Pierce}, {and} \bibinfo{person}{Joseph Lindley}.} \bibinfo{year}{2023}\natexlab{}.
\newblock \showarticletitle{The Entoptic Field Camera as Metaphor-Driven Research-through-Design with AI Technologies}. In \bibinfo{booktitle}{\emph{Proceedings of the 2023 CHI Conference on Human Factors in Computing Systems}} (Hamburg, Germany) \emph{(\bibinfo{series}{CHI '23})}. \bibinfo{publisher}{Association for Computing Machinery}, \bibinfo{address}{New York, NY, USA}, Article \bibinfo{articleno}{178}, \bibinfo{numpages}{19}~pages.
\newblock
\showISBNx{9781450394215}
\urldef\tempurl%
\url{https://doi.org/10.1145/3544548.3581175}
\showDOI{\tempurl}


\bibitem[Bird et~al\mbox{.}(2023)]%
        {Bird:AIES2023}
\bibfield{author}{\bibinfo{person}{Charlotte Bird}, \bibinfo{person}{Eddie Ungless}, {and} \bibinfo{person}{Atoosa Kasirzadeh}.} \bibinfo{year}{2023}\natexlab{}.
\newblock \showarticletitle{Typology of Risks of Generative Text-to-Image Models}. In \bibinfo{booktitle}{\emph{Proceedings of the 2023 AAAI/ACM Conference on AI, Ethics, and Society}} (Montr\'{e}al, QC, Canada) \emph{(\bibinfo{series}{AIES '23})}. \bibinfo{publisher}{Association for Computing Machinery}, \bibinfo{address}{New York, NY, USA}, \bibinfo{pages}{396--410}.
\newblock
\showISBNx{9798400702310}
\urldef\tempurl%
\url{https://doi.org/10.1145/3600211.3604722}
\showDOI{\tempurl}


\bibitem[Blythe(2023)]%
        {Blythe:2023}
\bibfield{author}{\bibinfo{person}{Mark Blythe}.} \bibinfo{year}{2023}\natexlab{}.
\newblock \showarticletitle{Artificial Design Fiction: Using AI as a Material for Pastiche Scenarios}. In \bibinfo{booktitle}{\emph{Proceedings of the 26th International Academic Mindtrek Conference}} (Tampere, Finland) \emph{(\bibinfo{series}{Mindtrek '23})}. \bibinfo{publisher}{Association for Computing Machinery}, \bibinfo{address}{New York, NY, USA}, \bibinfo{pages}{195--206}.
\newblock
\showISBNx{9798400708749}
\urldef\tempurl%
\url{https://doi.org/10.1145/3616961.3616987}
\showDOI{\tempurl}


\bibitem[Brand et~al\mbox{.}(2023)]%
        {Brand:CHI2023}
\bibfield{author}{\bibinfo{person}{Nico Brand}, \bibinfo{person}{William Odom}, {and} \bibinfo{person}{Samuel Barnett}.} \bibinfo{year}{2023}\natexlab{}.
\newblock \showarticletitle{Envisioning and Understanding Orientations to Introspective AI: Exploring a Design Space with Meta.Aware}. In \bibinfo{booktitle}{\emph{Proceedings of the 2023 CHI Conference on Human Factors in Computing Systems}} (Hamburg, Germany) \emph{(\bibinfo{series}{CHI '23})}. \bibinfo{publisher}{Association for Computing Machinery}, \bibinfo{address}{New York, NY, USA}, Article \bibinfo{articleno}{175}, \bibinfo{numpages}{18}~pages.
\newblock
\showISBNx{9781450394215}
\urldef\tempurl%
\url{https://doi.org/10.1145/3544548.3581336}
\showDOI{\tempurl}


\bibitem[Brandom(2024)]%
        {Brazil}
\bibfield{author}{\bibinfo{person}{Russell Brandom}.} \bibinfo{year}{2024}\natexlab{}.
\newblock \bibinfo{title}{Brazil's Flood Disaster Set Off a Torrent of AI Misinformation}.
\newblock
\newblock
\urldef\tempurl%
\url{https://restofworld.org/2024/exporter-brazil-floods-ai-misinformation/}
\showURL{%
\tempurl}


\bibitem[Braun and Clarke(2019)]%
        {Braun:2019}
\bibfield{author}{\bibinfo{person}{Virginia Braun} {and} \bibinfo{person}{Victoria Clarke}.} \bibinfo{year}{2019}\natexlab{}.
\newblock \showarticletitle{Reflecting on reflexive thematic analysis}.
\newblock \bibinfo{journal}{\emph{Qualitative Research in Sport, Exercise and Health}} \bibinfo{volume}{11}, \bibinfo{number}{4} (\bibinfo{year}{2019}), \bibinfo{pages}{589--597}.
\newblock
\urldef\tempurl%
\url{https://doi.org/10.1080/2159676X.2019.1628806}
\showDOI{\tempurl}


\bibitem[Brown and Juhlin(2015)]%
        {Brown:2015}
\bibfield{author}{\bibinfo{person}{Barry Brown} {and} \bibinfo{person}{Oskar Juhlin}.} \bibinfo{year}{2015}\natexlab{}.
\newblock \bibinfo{booktitle}{\emph{Enjoying machines}}.
\newblock \bibinfo{publisher}{MIT Press}.
\newblock


\bibitem[Byrne(2022)]%
        {Byrne:2022aa}
\bibfield{author}{\bibinfo{person}{David Byrne}.} \bibinfo{year}{2022}\natexlab{}.
\newblock \showarticletitle{A worked example of Braun and Clarke's approach to reflexive thematic analysis}.
\newblock \bibinfo{journal}{\emph{Quality \& Quantity}} \bibinfo{volume}{56}, \bibinfo{number}{3} (\bibinfo{year}{2022}), \bibinfo{pages}{1391--1412}.
\newblock
\showISBNx{1573-7845}
\urldef\tempurl%
\url{https://doi.org/10.1007/s11135-021-01182-y}
\showDOI{\tempurl}


\bibitem[Capra et~al\mbox{.}(2005)]%
        {Capra:Multimedia2005}
\bibfield{author}{\bibinfo{person}{Mauricio Capra}, \bibinfo{person}{Milena Radenkovic}, \bibinfo{person}{Steve Benford}, \bibinfo{person}{Leif Oppermann}, \bibinfo{person}{Adam Drozd}, {and} \bibinfo{person}{Martin Flintham}.} \bibinfo{year}{2005}\natexlab{}.
\newblock \showarticletitle{The multimedia challenges raised by pervasive games}. In \bibinfo{booktitle}{\emph{Proceedings of the 13th Annual ACM International Conference on Multimedia}} (Hilton, Singapore) \emph{(\bibinfo{series}{MULTIMEDIA '05})}. \bibinfo{publisher}{Association for Computing Machinery}, \bibinfo{address}{New York, NY, USA}, \bibinfo{pages}{89--95}.
\newblock
\showISBNx{1595930442}
\urldef\tempurl%
\url{https://doi.org/10.1145/1101149.1101163}
\showDOI{\tempurl}


\bibitem[Carter and Egliston(2020)]%
        {Carter:2020}
\bibfield{author}{\bibinfo{person}{Marcus Carter} {and} \bibinfo{person}{Ben Egliston}.} \bibinfo{year}{2020}\natexlab{}.
\newblock \bibinfo{booktitle}{\emph{Ethical Implications of Emerging Mixed Reality Technologies}}.
\newblock \bibinfo{type}{{T}echnical {R}eport}. \bibinfo{institution}{the Socio-Tech Futures Lab and Department of Media and Communication, Faculty of Arts and Social Sciences, the University of Sydney NSW}.
\newblock
\urldef\tempurl%
\url{https://hdl.handle.net/2123/22485}
\showURL{%
\tempurl}


\bibitem[Chew et~al\mbox{.}(2021)]%
        {Chew:OzCHI2021}
\bibfield{author}{\bibinfo{person}{Louis Chew}, \bibinfo{person}{Lian Loke}, {and} \bibinfo{person}{Luke Hespanhol}.} \bibinfo{year}{2021}\natexlab{}.
\newblock \showarticletitle{A Preliminary Design Vocabulary for Interactive Urban Play: Analysing and Composing Design Configurations for Playful Digital Placemaking}. In \bibinfo{booktitle}{\emph{Proceedings of the 32nd Australian Conference on Human-Computer Interaction}} (Sydney, NSW, Australia) \emph{(\bibinfo{series}{OzCHI '20})}. \bibinfo{publisher}{Association for Computing Machinery}, \bibinfo{address}{New York, NY, USA}, \bibinfo{pages}{11--24}.
\newblock
\showISBNx{9781450389754}
\urldef\tempurl%
\url{https://doi.org/10.1145/3441000.3441064}
\showDOI{\tempurl}


\bibitem[Chiou et~al\mbox{.}(2023)]%
        {Chiou:2023}
\bibfield{author}{\bibinfo{person}{Li-Yuan Chiou}, \bibinfo{person}{Peng-Kai Hung}, \bibinfo{person}{Rung-Huei Liang}, {and} \bibinfo{person}{Chun-Teng Wang}.} \bibinfo{year}{2023}\natexlab{}.
\newblock \showarticletitle{Designing with AI: An Exploration of Co-Ideation with Image Generators}. In \bibinfo{booktitle}{\emph{Proceedings of the 2023 ACM Designing Interactive Systems Conference}} (Pittsburgh, PA, USA) \emph{(\bibinfo{series}{DIS '23})}. \bibinfo{publisher}{Association for Computing Machinery}, \bibinfo{address}{New York, NY, USA}, \bibinfo{pages}{1941--1954}.
\newblock
\showISBNx{9781450398930}
\urldef\tempurl%
\url{https://doi.org/10.1145/3563657.3596001}
\showDOI{\tempurl}


\bibitem[Chisik et~al\mbox{.}(2022)]%
        {Chisik:2022}
\bibfield{author}{\bibinfo{person}{Yoram Chisik}, \bibinfo{person}{Anton Nijholt}, \bibinfo{person}{Ben Schouten}, {and} \bibinfo{person}{Mattia Thibault}.} \bibinfo{year}{2022}\natexlab{}.
\newblock \showarticletitle{Urban Play and the Playable City: A Critical Perspective}.
\newblock \bibinfo{journal}{\emph{Frontiers in Computer Science}}  \bibinfo{volume}{3} (\bibinfo{year}{2022}).
\newblock
\showISSN{2624-9898}
\urldef\tempurl%
\url{https://doi.org/10.3389/fcomp.2021.806494}
\showDOI{\tempurl}


\bibitem[Chu-Ke and Dong(2024)]%
        {Chu-Ke:2024}
\bibfield{author}{\bibinfo{person}{Chun Chu-Ke} {and} \bibinfo{person}{Yujie Dong}.} \bibinfo{year}{2024}\natexlab{}.
\newblock \showarticletitle{Misinformation and Literacies in the Era of Generative Artificial Intelligence: A Brief Overview and a Call for Future Research}.
\newblock \bibinfo{journal}{\emph{Emerging Media}} \bibinfo{volume}{2}, \bibinfo{number}{1} (\bibinfo{year}{2024}), \bibinfo{pages}{70--85}.
\newblock
\urldef\tempurl%
\url{https://doi.org/10.1177/27523543241240285}
\showDOI{\tempurl}


\bibitem[Dameri and Rosenthal-Sabroux(2014)]%
        {Dameri:2014}
\bibfield{editor}{\bibinfo{person}{Renata~Paola Dameri} {and} \bibinfo{person}{Camille Rosenthal-Sabroux}} (Eds.). \bibinfo{year}{2014}\natexlab{}.
\newblock \bibinfo{booktitle}{\emph{Smart City: How to Create Public and Economic Value with High Technology in Urban Space}}.
\newblock \bibinfo{publisher}{Springer Cham}.
\newblock


\bibitem[Dang et~al\mbox{.}(2022)]%
        {Dang:ArXiv2022}
\bibfield{author}{\bibinfo{person}{Hai Dang}, \bibinfo{person}{Lukas Mecke}, \bibinfo{person}{Florian Lehmann}, \bibinfo{person}{Sven Goller}, {and} \bibinfo{person}{Daniel Buschek}.} \bibinfo{year}{2022}\natexlab{}.
\newblock \showarticletitle{How to Prompt? Opportunities and Challenges of Zero- and Few-Shot Learning for Human-AI Interaction in Creative Applications of Generative Models}.
\newblock \bibinfo{journal}{\emph{ArXiv}}  \bibinfo{volume}{abs/2209.01390} (\bibinfo{year}{2022}).
\newblock


\bibitem[Epstein et~al\mbox{.}(2022)]%
        {Epstein:2022}
\bibfield{author}{\bibinfo{person}{Ziv Epstein}, \bibinfo{person}{Hope Schroeder}, {and} \bibinfo{person}{Dava Newman}.} \bibinfo{year}{2022}\natexlab{}.
\newblock \showarticletitle{When Happy Accidents Spark Creativity: Bringing Collaborative Speculation to Life with Generative AI}. In \bibinfo{booktitle}{\emph{Proceedings of the International Conference on Computational Creativity (ICCC '22)}}. \bibinfo{pages}{334--338}.
\newblock


\bibitem[Fischer et~al\mbox{.}(2010)]%
        {Tobias:2010}
\bibfield{author}{\bibinfo{person}{Patrick~Tobias Fischer}, \bibinfo{person}{Christian Z\"{o}llner}, {and} \bibinfo{person}{Eva Hornecker}.} \bibinfo{year}{2010}\natexlab{}.
\newblock \showarticletitle{VR/Urban: Spread.gun---Design process and challenges in developing a shared encounter for media facades}. In \bibinfo{booktitle}{\emph{Proceedings of British HCI 2010}}.
\newblock


\bibitem[Flintham et~al\mbox{.}(2003)]%
        {Flintham:CHI2003}
\bibfield{author}{\bibinfo{person}{Martin Flintham}, \bibinfo{person}{Steve Benford}, \bibinfo{person}{Rob Anastasi}, \bibinfo{person}{Terry Hemmings}, \bibinfo{person}{Andy Crabtree}, \bibinfo{person}{Chris Greenhalgh}, \bibinfo{person}{Nick Tandavanitj}, \bibinfo{person}{Matt Adams}, {and} \bibinfo{person}{Ju Row-Farr}.} \bibinfo{year}{2003}\natexlab{}.
\newblock \showarticletitle{Where on-line meets on the streets: experiences with mobile mixed reality games}. In \bibinfo{booktitle}{\emph{Proceedings of the SIGCHI Conference on Human Factors in Computing Systems}} (Ft. Lauderdale, Florida, USA) \emph{(\bibinfo{series}{CHI '03})}. \bibinfo{publisher}{Association for Computing Machinery}, \bibinfo{address}{New York, NY, USA}, \bibinfo{pages}{569--576}.
\newblock
\showISBNx{1581136307}
\urldef\tempurl%
\url{https://doi.org/10.1145/642611.642710}
\showDOI{\tempurl}


\bibitem[Fu et~al\mbox{.}(2024)]%
        {Fu:DIS2024}
\bibfield{author}{\bibinfo{person}{Kexue Fu}, \bibinfo{person}{Ruishan Wu}, \bibinfo{person}{Yuying Tang}, \bibinfo{person}{Yixin Chen}, \bibinfo{person}{Bowen Liu}, {and} \bibinfo{person}{RAY LC}.} \bibinfo{year}{2024}\natexlab{}.
\newblock \showarticletitle{"Being Eroded, Piece by Piece": Enhancing Engagement and Storytelling in Cultural Heritage Dissemination by Exhibiting GenAI Co-Creation Artifacts}. In \bibinfo{booktitle}{\emph{Proceedings of the 2024 ACM Designing Interactive Systems Conference}} (Copenhagen, Denmark) \emph{(\bibinfo{series}{DIS '24})}. \bibinfo{publisher}{Association for Computing Machinery}, \bibinfo{address}{New York, NY, USA}, \bibinfo{pages}{2833--2850}.
\newblock
\showISBNx{9798400705830}
\urldef\tempurl%
\url{https://doi.org/10.1145/3643834.3660711}
\showDOI{\tempurl}


\bibitem[Goswami(2023)]%
        {Adobe}
\bibfield{author}{\bibinfo{person}{Rohan Goswami}.} \bibinfo{year}{2023}\natexlab{}.
\newblock \bibinfo{title}{Adobe Launches Firefly Generative A.I., Which Lets Users Type to Edit Images}.
\newblock
\newblock
\urldef\tempurl%
\url{https://www.cnbc.com/2023/03/21/adobe-firefly-generative-ai-lets-you-type-to-edit-images.html}
\showURL{%
\tempurl}


\bibitem[Grajper and Dobiesz(2015)]%
        {urbanimals:2024}
\bibfield{author}{\bibinfo{person}{Anna Grajper} {and} \bibinfo{person}{Sebastian Dobiesz}.} \bibinfo{year}{2015}\natexlab{}.
\newblock \bibinfo{title}{Urbanimals}.
\newblock
\newblock
\urldef\tempurl%
\url{http://lax.com.pl/portfolio_page/urbanimals/}
\showURL{%
\tempurl}


\bibitem[Guridi et~al\mbox{.}(2024)]%
        {Guridi:2024}
\bibfield{author}{\bibinfo{person}{Jose~A. Guridi}, \bibinfo{person}{Cristobal Cheyre}, \bibinfo{person}{Maria Goula}, \bibinfo{person}{Duarte Santo}, \bibinfo{person}{Lee Humphreys}, \bibinfo{person}{Aishwarya Shankar}, {and} \bibinfo{person}{Achilleas Souras}.} \bibinfo{year}{2024}\natexlab{}.
\newblock \showarticletitle{Image Generative AI to Design Public Spaces: a Reflection of how AI Could Improve Co-Design of Public Parks}.
\newblock \bibinfo{journal}{\emph{Digit. Gov.: Res. Pract.}} (\bibinfo{date}{April} \bibinfo{year}{2024}).
\newblock
\urldef\tempurl%
\url{https://doi.org/10.1145/3656588}
\showDOI{\tempurl}
\newblock
\shownote{Just Accepted}.


\bibitem[Hagendorff(2024)]%
        {Hagendorff:CoRR2024}
\bibfield{author}{\bibinfo{person}{Thilo Hagendorff}.} \bibinfo{year}{2024}\natexlab{}.
\newblock \showarticletitle{Mapping the Ethics of Generative AI: A Comprehensive Scoping Review}.
\newblock \bibinfo{journal}{\emph{Minds and Machines}} \bibinfo{volume}{34}, \bibinfo{number}{4} (\bibinfo{year}{2024}), \bibinfo{pages}{39}.
\newblock
\showISBNx{1572-8641}
\urldef\tempurl%
\url{https://doi.org/10.1007/s11023-024-09694-w}
\showDOI{\tempurl}


\bibitem[Harley et~al\mbox{.}(2019)]%
        {Harley:DIS2019}
\bibfield{author}{\bibinfo{person}{Daniel Harley}, \bibinfo{person}{Aneesh~P. Tarun}, \bibinfo{person}{Sara Elsharawy}, \bibinfo{person}{Alexander Verni}, \bibinfo{person}{Tudor Tibu}, \bibinfo{person}{Marko Bilic}, \bibinfo{person}{Alexander Bakogeorge}, {and} \bibinfo{person}{Ali Mazalek}.} \bibinfo{year}{2019}\natexlab{}.
\newblock \showarticletitle{Mobile Realities: Designing for the Medium of Smartphone-VR}. In \bibinfo{booktitle}{\emph{Proceedings of the 2019 on Designing Interactive Systems Conference}} (San Diego, CA, USA) \emph{(\bibinfo{series}{DIS '19})}. \bibinfo{publisher}{Association for Computing Machinery}, \bibinfo{address}{New York, NY, USA}, \bibinfo{pages}{1131--1144}.
\newblock
\showISBNx{9781450358507}
\urldef\tempurl%
\url{https://doi.org/10.1145/3322276.3322341}
\showDOI{\tempurl}


\bibitem[Hung et~al\mbox{.}(2024)]%
        {Hung:DISWiP2024}
\bibfield{author}{\bibinfo{person}{Peng-Kai Hung}, \bibinfo{person}{Janet Yi-Ching Huang}, \bibinfo{person}{Stephan Wensveen}, {and} \bibinfo{person}{Rung-Huei Liang}.} \bibinfo{year}{2024}\natexlab{}.
\newblock \showarticletitle{Re.Dis.Cover Place with Generative AI: Exploring the Experience and Design of City Wandering with image-to-image AI}. In \bibinfo{booktitle}{\emph{Companion Publication of the 2024 ACM Designing Interactive Systems Conference}} (IT University of Copenhagen, Denmark) \emph{(\bibinfo{series}{DIS '24 Companion})}. \bibinfo{publisher}{Association for Computing Machinery}, \bibinfo{address}{New York, NY, USA}, \bibinfo{pages}{219--223}.
\newblock
\showISBNx{9798400706325}
\urldef\tempurl%
\url{https://doi.org/10.1145/3656156.3663691}
\showDOI{\tempurl}


\bibitem[Innocent(2018)]%
        {Innocent:2018}
\bibfield{author}{\bibinfo{person}{Troy Innocent}.} \bibinfo{year}{2018}\natexlab{}.
\newblock \showarticletitle{Play about Place: Placemaking in location-based game design}. In \bibinfo{booktitle}{\emph{Proceedings of the 4th Media Architecture Biennale Conference}} (Beijing, China) \emph{(\bibinfo{series}{MAB '18})}. \bibinfo{publisher}{Association for Computing Machinery}, \bibinfo{address}{New York, NY, USA}, \bibinfo{pages}{137--143}.
\newblock
\showISBNx{9781450364782}
\urldef\tempurl%
\url{https://doi.org/10.1145/3284389.3284493}
\showDOI{\tempurl}


\bibitem[Innocent(2019)]%
        {Innocent:2019}
\bibfield{author}{\bibinfo{person}{Troy Innocent}.} \bibinfo{year}{2019}\natexlab{}.
\newblock \showarticletitle{Citizens of Play: Revisiting the Relationship between Playable and Smart Cities}.
\newblock In \bibinfo{booktitle}{\emph{Making Smart Cities More Playable: Exploring Playable Cities}}. \bibinfo{publisher}{Springer}.
\newblock


\bibitem[Innocent(2021)]%
        {Innocent:2021}
\bibfield{author}{\bibinfo{person}{Troy Innocent}.} \bibinfo{year}{2021}\natexlab{}.
\newblock \showarticletitle{Urban Play in Practice: Seven Lenses Exploring the Sociocultural Value of Playable Cities}.
\newblock In \bibinfo{booktitle}{\emph{Games and Play in the Creative, Smart and Ecological City}}. Number~5. \bibinfo{publisher}{Routledge}, \bibinfo{address}{New York, United States}, \bibinfo{pages}{19}.
\newblock
\showISBNx{9780367441234}


\bibitem[Johnson(2024)]%
        {Googlepixel}
\bibfield{author}{\bibinfo{person}{Allison Johnson}.} \bibinfo{year}{2024}\natexlab{}.
\newblock \bibinfo{title}{Google's AI `Reimagine' tool helped us add wrecks, disasters, and corpses to our photos}.
\newblock
\newblock
\urldef\tempurl%
\url{https://www.theverge.com/2024/8/21/24224084/google-pixel-9-reimagine-ai-photos}
\showURL{%
\tempurl}


\bibitem[Karmann(2023)]%
        {Karmann:2023}
\bibfield{author}{\bibinfo{person}{Bjoern Karmann}.} \bibinfo{year}{2023}\natexlab{}.
\newblock \bibinfo{title}{Paragraphica}.
\newblock
\newblock
\urldef\tempurl%
\url{https://bjoernkarmann.dk/project/paragraphica}
\showURL{%
\tempurl}


\bibitem[Karpashevich et~al\mbox{.}(2016)]%
        {Karpashevich:MUM2016}
\bibfield{author}{\bibinfo{person}{Pavel Karpashevich}, \bibinfo{person}{Eva Hornecker}, \bibinfo{person}{Nana~Kesewaa Dankwa}, \bibinfo{person}{Mohamed Hanafy}, {and} \bibinfo{person}{Julian Fietkau}.} \bibinfo{year}{2016}\natexlab{}.
\newblock \showarticletitle{Blurring boundaries between everyday life and pervasive gaming: an interview study of ingress}. In \bibinfo{booktitle}{\emph{Proceedings of the 15th International Conference on Mobile and Ubiquitous Multimedia}} (Rovaniemi, Finland) \emph{(\bibinfo{series}{MUM '16})}. \bibinfo{publisher}{Association for Computing Machinery}, \bibinfo{address}{New York, NY, USA}, \bibinfo{pages}{217--228}.
\newblock
\showISBNx{9781450348607}
\urldef\tempurl%
\url{https://doi.org/10.1145/3012709.3012716}
\showDOI{\tempurl}


\bibitem[Kasapakis et~al\mbox{.}(2013)]%
        {Kasapakis:PCI2013}
\bibfield{author}{\bibinfo{person}{Vlasios Kasapakis}, \bibinfo{person}{Damianos Gavalas}, {and} \bibinfo{person}{Nikos Bubaris}.} \bibinfo{year}{2013}\natexlab{}.
\newblock \showarticletitle{Pervasive games research: a design aspects-based state of the art report}. In \bibinfo{booktitle}{\emph{Proceedings of the 17th Panhellenic Conference on Informatics}} (Thessaloniki, Greece) \emph{(\bibinfo{series}{PCI '13})}. \bibinfo{publisher}{Association for Computing Machinery}, \bibinfo{address}{New York, NY, USA}, \bibinfo{pages}{152--157}.
\newblock
\showISBNx{9781450319690}
\urldef\tempurl%
\url{https://doi.org/10.1145/2491845.2491874}
\showDOI{\tempurl}


\bibitem[Kenthapadi et~al\mbox{.}(2023)]%
        {Kenthapadi:KDD2023}
\bibfield{author}{\bibinfo{person}{Krishnaram Kenthapadi}, \bibinfo{person}{Himabindu Lakkaraju}, {and} \bibinfo{person}{Nazneen Rajani}.} \bibinfo{year}{2023}\natexlab{}.
\newblock \showarticletitle{Generative AI meets Responsible AI: Practical Challenges and Opportunities}. In \bibinfo{booktitle}{\emph{Proceedings of the 29th ACM SIGKDD Conference on Knowledge Discovery and Data Mining}} (Long Beach, CA, USA) \emph{(\bibinfo{series}{KDD '23})}. \bibinfo{publisher}{Association for Computing Machinery}, \bibinfo{address}{New York, NY, USA}, \bibinfo{pages}{5805--5806}.
\newblock
\showISBNx{9798400701030}
\urldef\tempurl%
\url{https://doi.org/10.1145/3580305.3599557}
\showDOI{\tempurl}


\bibitem[Korte and Ferri(2018)]%
        {Korte:2018}
\bibfield{author}{\bibinfo{person}{Gen{\`e}vi{\'e}ve Korte} {and} \bibinfo{person}{Gabriele Ferri}.} \bibinfo{year}{2018}\natexlab{}.
\newblock \showarticletitle{Research Through Game Design. Interactive Stories from a Submerged Amsterdam}.
\newblock \bibinfo{journal}{\emph{Ocula}} \bibinfo{volume}{19}, \bibinfo{number}{19} (\bibinfo{year}{2018}), \bibinfo{pages}{110--128}.
\newblock
\showISSN{1724-7810}
\urldef\tempurl%
\url{https://doi.org/10.12977/ocula2018-14}
\showDOI{\tempurl}


\bibitem[Lancel et~al\mbox{.}(2019)]%
        {Lancel:2019}
\bibfield{author}{\bibinfo{person}{Karen Lancel}, \bibinfo{person}{Frances Brazier}, {and} \bibinfo{person}{Hermen Maat}.} \bibinfo{year}{2019}\natexlab{}.
\newblock \bibinfo{booktitle}{\emph{Saving Face: Shared experience and dialogue on social touch, in playful smart public space}}.
\newblock \bibinfo{publisher}{Springer}, \bibinfo{pages}{179--203}.
\newblock
\showISBNx{978-981-13-9764-6}
\urldef\tempurl%
\url{https://doi.org/10.1007/978-981-13-9765-3_9}
\showDOI{\tempurl}


\bibitem[Lange(2015)]%
        {DeLange:2015}
\bibfield{author}{\bibinfo{person}{Michiel~De Lange}.} \bibinfo{year}{2015}\natexlab{}.
\newblock \showarticletitle{The Playful City: Using Play and Games to Foster Citizen Participation}.
\newblock In \bibinfo{booktitle}{\emph{Social Technologies and Collective Intelligence}}. \bibinfo{publisher}{Vilnius: Mykolas Romeris Universit}, \bibinfo{pages}{426--434}.
\newblock


\bibitem[Lau et~al\mbox{.}(2024)]%
        {Lau:DIS2024}
\bibfield{author}{\bibinfo{person}{Tatiana Lau}, \bibinfo{person}{Scott Carter}, \bibinfo{person}{Francine Chen}, \bibinfo{person}{Brandon Huynh}, \bibinfo{person}{Everlyne Kimani}, \bibinfo{person}{Matthew~L Lee}, {and} \bibinfo{person}{Kate~A Sieck}.} \bibinfo{year}{2024}\natexlab{}.
\newblock \showarticletitle{Democratizing Design through Generative AI}. In \bibinfo{booktitle}{\emph{Companion Publication of the 2024 ACM Designing Interactive Systems Conference}} (IT University of Copenhagen, Denmark) \emph{(\bibinfo{series}{DIS '24 Companion})}. \bibinfo{publisher}{Association for Computing Machinery}, \bibinfo{address}{New York, NY, USA}, \bibinfo{pages}{239--244}.
\newblock
\showISBNx{9798400706325}
\urldef\tempurl%
\url{https://doi.org/10.1145/3656156.3663703}
\showDOI{\tempurl}


\bibitem[Li et~al\mbox{.}(2020)]%
        {Li:CHIEA2020}
\bibfield{author}{\bibinfo{person}{Zhuying Li}, \bibinfo{person}{Yan Wang}, \bibinfo{person}{Wei Wang}, \bibinfo{person}{Stefan Greuter}, {and} \bibinfo{person}{Florian~'Floyd' Mueller}.} \bibinfo{year}{2020}\natexlab{}.
\newblock \showarticletitle{Empowering a Creative City: Engage Citizens in Creating Street Art through Human-AI Collaboration}. In \bibinfo{booktitle}{\emph{Extended Abstracts of the 2020 CHI Conference on Human Factors in Computing Systems}} (Honolulu, HI, USA) \emph{(\bibinfo{series}{CHI EA '20})}. \bibinfo{publisher}{Association for Computing Machinery}, \bibinfo{address}{New York, NY, USA}, \bibinfo{pages}{1--8}.
\newblock
\showISBNx{9781450368193}
\urldef\tempurl%
\url{https://doi.org/10.1145/3334480.3382976}
\showDOI{\tempurl}


\bibitem[Lin and Long(2023)]%
        {Lin:CC2023}
\bibfield{author}{\bibinfo{person}{Lauren Lin} {and} \bibinfo{person}{Duri Long}.} \bibinfo{year}{2023}\natexlab{}.
\newblock \showarticletitle{Generative AI Futures: A Speculative Design Exploration}. In \bibinfo{booktitle}{\emph{Proceedings of the 15th Conference on Creativity and Cognition}} (Virtual Event, USA) \emph{(\bibinfo{series}{C\&C '23})}. \bibinfo{publisher}{Association for Computing Machinery}, \bibinfo{address}{New York, NY, USA}, \bibinfo{pages}{380--383}.
\newblock
\showISBNx{9798400701801}
\urldef\tempurl%
\url{https://doi.org/10.1145/3591196.3596616}
\showDOI{\tempurl}


\bibitem[Long et~al\mbox{.}(2019)]%
        {Long:2019}
\bibfield{author}{\bibinfo{person}{Duri Long}, \bibinfo{person}{Mikhail Jacob}, {and} \bibinfo{person}{Brian Magerko}.} \bibinfo{year}{2019}\natexlab{}.
\newblock \showarticletitle{Designing Co-Creative AI for Public Spaces}. In \bibinfo{booktitle}{\emph{Proceedings of the 2019 Conference on Creativity and Cognition}} (San Diego, CA, USA) \emph{(\bibinfo{series}{C\&C '19})}. \bibinfo{publisher}{Association for Computing Machinery}, \bibinfo{address}{New York, NY, USA}, \bibinfo{pages}{271--284}.
\newblock
\showISBNx{9781450359177}
\urldef\tempurl%
\url{https://doi.org/10.1145/3325480.3325504}
\showDOI{\tempurl}


\bibitem[Luusua and Ylipulli(2020)]%
        {Luusua:2020}
\bibfield{author}{\bibinfo{person}{Aale Luusua} {and} \bibinfo{person}{Johanna Ylipulli}.} \bibinfo{year}{2020}\natexlab{}.
\newblock \showarticletitle{Urban AI: Formulating an Agenda for the Interdisciplinary Research of Artificial Intelligence in Cities}. In \bibinfo{booktitle}{\emph{Companion Publication of the 2020 ACM Designing Interactive Systems Conference}} (Eindhoven, Netherlands) \emph{(\bibinfo{series}{DIS' 20 Companion})}. \bibinfo{publisher}{Association for Computing Machinery}, \bibinfo{address}{New York, NY, USA}, \bibinfo{pages}{373--376}.
\newblock
\showISBNx{9781450379878}
\urldef\tempurl%
\url{https://doi.org/10.1145/3393914.3395905}
\showDOI{\tempurl}


\bibitem[Luusua et~al\mbox{.}(2023)]%
        {Luusua:2023aa}
\bibfield{author}{\bibinfo{person}{Aale Luusua}, \bibinfo{person}{Johanna Ylipulli}, \bibinfo{person}{Marcus Foth}, {and} \bibinfo{person}{Alessandro Aurigi}.} \bibinfo{year}{2023}\natexlab{}.
\newblock \showarticletitle{Urban AI: understanding the emerging role of artificial intelligence in smart cities}.
\newblock \bibinfo{journal}{\emph{AI \& SOCIETY}} \bibinfo{volume}{38}, \bibinfo{number}{3} (\bibinfo{year}{2023}), \bibinfo{pages}{1039--1044}.
\newblock
\showISBNx{1435-5655}
\urldef\tempurl%
\url{https://doi.org/10.1007/s00146-022-01537-5}
\showDOI{\tempurl}


\bibitem[Malsattar et~al\mbox{.}(2019)]%
        {Malsattar:2019}
\bibfield{author}{\bibinfo{person}{Nirav Malsattar}, \bibinfo{person}{Tomo Kihara}, {and} \bibinfo{person}{Elisa Giaccardi}.} \bibinfo{year}{2019}\natexlab{}.
\newblock \showarticletitle{Designing and Prototyping from the Perspective of AI in the Wild}. In \bibinfo{booktitle}{\emph{Proceedings of the 2019 on Designing Interactive Systems Conference}} (San Diego, CA, USA) \emph{(\bibinfo{series}{DIS '19})}. \bibinfo{publisher}{Association for Computing Machinery}, \bibinfo{address}{New York, NY, USA}, \bibinfo{pages}{1083--1088}.
\newblock
\showISBNx{9781450358507}
\urldef\tempurl%
\url{https://doi.org/10.1145/3322276.3322351}
\showDOI{\tempurl}


\bibitem[Mlynar et~al\mbox{.}(2022)]%
        {Mlynar:CHI2022}
\bibfield{author}{\bibinfo{person}{Jakub Mlynar}, \bibinfo{person}{Farzaneh Bahrami}, \bibinfo{person}{Andr\'{e} Ourednik}, \bibinfo{person}{Nico Mutzner}, \bibinfo{person}{Himanshu Verma}, {and} \bibinfo{person}{Hamed Alavi}.} \bibinfo{year}{2022}\natexlab{}.
\newblock \showarticletitle{AI beyond Deus ex Machina -- Reimagining Intelligence in Future Cities with Urban Experts}. In \bibinfo{booktitle}{\emph{Proceedings of the 2022 CHI Conference on Human Factors in Computing Systems}} (New Orleans, LA, USA) \emph{(\bibinfo{series}{CHI '22})}. \bibinfo{publisher}{Association for Computing Machinery}, \bibinfo{address}{New York, NY, USA}, Article \bibinfo{articleno}{370}, \bibinfo{numpages}{13}~pages.
\newblock
\showISBNx{9781450391573}
\urldef\tempurl%
\url{https://doi.org/10.1145/3491102.3517502}
\showDOI{\tempurl}


\bibitem[Mora et~al\mbox{.}(2017)]%
        {Mora:2017}
\bibfield{author}{\bibinfo{person}{Luca Mora}, \bibinfo{person}{Roberto Bolici}, {and} \bibinfo{person}{Mark Deakin}.} \bibinfo{year}{2017}\natexlab{}.
\newblock \showarticletitle{The First Two Decades of Smart-City Research: A Bibliometric Analysis}.
\newblock \bibinfo{journal}{\emph{Journal of Urban Technology}} \bibinfo{volume}{24}, \bibinfo{number}{1} (\bibinfo{year}{2017}), \bibinfo{pages}{3--27}.
\newblock
\urldef\tempurl%
\url{https://doi.org/10.1080/10630732.2017.1285123}
\showDOI{\tempurl}


\bibitem[Moreno-Ibarra et~al\mbox{.}(2024)]%
        {Moreno-Ibarra2024}
\bibfield{author}{\bibinfo{person}{Marco Moreno-Ibarra}, \bibinfo{person}{Magdalena Salda{\~n}a-Perez}, \bibinfo{person}{Samuel~P{\'e}rez Rodr{\'\i}guez}, {and} \bibinfo{person}{Emmanuel~Ju{\'a}rez Carbajal}.} \bibinfo{year}{2024}\natexlab{}.
\newblock \bibinfo{booktitle}{\emph{Generative AI (GenAI) and smart cities: Efficiency, cohesion, and sustainability}}.
\newblock \bibinfo{publisher}{Routledge}, Chapter~8, \bibinfo{pages}{118--129}.
\newblock
\urldef\tempurl%
\url{https://doi.org/10.1201/9781003415930-11}
\showDOI{\tempurl}


\bibitem[Muller et~al\mbox{.}(2022)]%
        {Muller:GenAICHI2022}
\bibfield{author}{\bibinfo{person}{Michael Muller}, \bibinfo{person}{Lydia~B Chilton}, \bibinfo{person}{Anna Kantosalo}, \bibinfo{person}{Charles~Patrick Martin}, {and} \bibinfo{person}{Greg Walsh}.} \bibinfo{year}{2022}\natexlab{}.
\newblock \showarticletitle{GenAICHI: Generative AI and HCI}. In \bibinfo{booktitle}{\emph{Extended Abstracts of the 2022 CHI Conference on Human Factors in Computing Systems}} (New Orleans, LA, USA) \emph{(\bibinfo{series}{CHI EA '22})}. \bibinfo{publisher}{Association for Computing Machinery}, \bibinfo{address}{New York, NY, USA}, Article \bibinfo{articleno}{110}, \bibinfo{numpages}{7}~pages.
\newblock
\showISBNx{9781450391566}
\urldef\tempurl%
\url{https://doi.org/10.1145/3491101.3503719}
\showDOI{\tempurl}


\bibitem[Nijholt(2017)]%
        {Nijholt:Springer2017}
\bibfield{editor}{\bibinfo{person}{Anton Nijholt}} (Ed.). \bibinfo{year}{2017}\natexlab{}.
\newblock \bibinfo{booktitle}{\emph{Playable Cities: The City as a Digital Playground}}.
\newblock \bibinfo{publisher}{Springer}, \bibinfo{address}{Germany}.
\newblock
\showISBNx{978-981-10-1961-6}
\urldef\tempurl%
\url{https://doi.org/10.1007/978-981-10-1962-3}
\showDOI{\tempurl}


\bibitem[Nijholt(2020)]%
        {Nijholt2020}
\bibfield{author}{\bibinfo{person}{Anton Nijholt}.} \bibinfo{year}{2020}\natexlab{}.
\newblock \bibinfo{booktitle}{\emph{City Residents as Videogame Characters in Smart Urban Environments}}.
\newblock \bibinfo{publisher}{Springer Singapore}, \bibinfo{address}{Singapore}, \bibinfo{pages}{355--377}.
\newblock
\showISBNx{978-981-13-9765-3}
\urldef\tempurl%
\url{https://doi.org/10.1007/978-981-13-9765-3_16}
\showDOI{\tempurl}


\bibitem[of~Tourism \&~Conventions(2024)]%
        {DutchCyclingLifestyle2024}
\bibfield{author}{\bibinfo{person}{Netherlands~Board of Tourism \&~Conventions}.} \bibinfo{year}{2024}\natexlab{}.
\newblock \bibinfo{title}{Dutch Cycling Lifestyle}.
\newblock
\newblock
\urldef\tempurl%
\url{https://dutchcyclinglifestyle.com/}
\showURL{%
\tempurl}


\bibitem[O'Hara(2008)]%
        {OHara:CHI2008}
\bibfield{author}{\bibinfo{person}{Kenton O'Hara}.} \bibinfo{year}{2008}\natexlab{}.
\newblock \showarticletitle{Understanding geocaching practices and motivations}. In \bibinfo{booktitle}{\emph{Proceedings of the SIGCHI Conference on Human Factors in Computing Systems}} (Florence, Italy) \emph{(\bibinfo{series}{CHI '08})}. \bibinfo{publisher}{Association for Computing Machinery}, \bibinfo{address}{New York, NY, USA}, \bibinfo{pages}{1177--1186}.
\newblock
\showISBNx{9781605580111}
\urldef\tempurl%
\url{https://doi.org/10.1145/1357054.1357239}
\showDOI{\tempurl}


\bibitem[Pang et~al\mbox{.}(2020)]%
        {Pang:2020}
\bibfield{author}{\bibinfo{person}{Carolyn Pang}, \bibinfo{person}{Carman Neustaedter}, \bibinfo{person}{Karyn Moffatt}, \bibinfo{person}{Kate Hennessy}, {and} \bibinfo{person}{Rui Pan}.} \bibinfo{year}{2020}\natexlab{}.
\newblock \showarticletitle{The role of a location-based city exploration game in digital placemaking}.
\newblock \bibinfo{journal}{\emph{Behaviour \& Information Technology}} \bibinfo{volume}{39}, \bibinfo{number}{6} (\bibinfo{year}{2020}), \bibinfo{pages}{624--647}.
\newblock
\urldef\tempurl%
\url{https://doi.org/10.1080/0144929X.2019.1697899}
\showDOI{\tempurl}


\bibitem[Paulos and Jenkins(2005)]%
        {Paulos:2005}
\bibfield{author}{\bibinfo{person}{Eric Paulos} {and} \bibinfo{person}{Tom Jenkins}.} \bibinfo{year}{2005}\natexlab{}.
\newblock \showarticletitle{Urban probes: encountering our emerging urban atmospheres}. In \bibinfo{booktitle}{\emph{Proceedings of the SIGCHI Conference on Human Factors in Computing Systems}} (Portland, Oregon, USA) \emph{(\bibinfo{series}{CHI '05})}. \bibinfo{publisher}{Association for Computing Machinery}, \bibinfo{address}{New York, NY, USA}, \bibinfo{pages}{341--350}.
\newblock
\showISBNx{1581139985}
\urldef\tempurl%
\url{https://doi.org/10.1145/1054972.1055020}
\showDOI{\tempurl}


\bibitem[Peng-Kai~Hung and Kong(2024)]%
        {Hung:2024}
\bibfield{author}{\bibinfo{person}{Shih-Yu~Ma Peng-Kai~Hung, Rung-Huei~Liang} {and} \bibinfo{person}{Bo-Wen Kong}.} \bibinfo{year}{2024}\natexlab{}.
\newblock \showarticletitle{Exploring the Experience of Traveling to Familiar Places in VR: An Empirical Study Using Google Earth VR}.
\newblock \bibinfo{journal}{\emph{International Journal of Human--Computer Interaction}} \bibinfo{volume}{40}, \bibinfo{number}{2} (\bibinfo{year}{2024}), \bibinfo{pages}{255--277}.
\newblock
\urldef\tempurl%
\url{https://doi.org/10.1080/10447318.2022.2114141}
\showDOI{\tempurl}


\bibitem[Petridis et~al\mbox{.}(2024)]%
        {Petridis:2024}
\bibfield{author}{\bibinfo{person}{Savvas Petridis}, \bibinfo{person}{Michael~Xieyang Liu}, \bibinfo{person}{Alexander~J. Fiannaca}, \bibinfo{person}{Vivian Tsai}, \bibinfo{person}{Michael Terry}, {and} \bibinfo{person}{Carrie~J. Cai}.} \bibinfo{year}{2024}\natexlab{}.
\newblock \showarticletitle{In Situ AI Prototyping: Infusing Multimodal Prompts into Mobile Settings with MobileMaker}. In \bibinfo{booktitle}{\emph{2024 IEEE Symposium on Visual Languages and Human-Centric Computing (VL/HCC)}}. \bibinfo{pages}{121--133}.
\newblock
\urldef\tempurl%
\url{https://doi.org/10.1109/VL/HCC60511.2024.00023}
\showDOI{\tempurl}


\bibitem[Poplin(2011)]%
        {Poplin:ICCSA2011}
\bibfield{author}{\bibinfo{person}{Alenka Poplin}.} \bibinfo{year}{2011}\natexlab{}.
\newblock \showarticletitle{Games and Serious Games in Urban Planning: Study Cases}. In \bibinfo{booktitle}{\emph{International Conference on Computational Science and Its Applications (ICCSA2011)}}. \bibinfo{publisher}{Springer, Berlin, Heidelberg}, \bibinfo{pages}{1--14}.
\newblock


\bibitem[Procyk and Neustaedter(2014)]%
        {Procyk:2014}
\bibfield{author}{\bibinfo{person}{Jason Procyk} {and} \bibinfo{person}{Carman Neustaedter}.} \bibinfo{year}{2014}\natexlab{}.
\newblock \showarticletitle{GEMS: the design and evaluation of a location-based storytelling game}. In \bibinfo{booktitle}{\emph{Proceedings of the 17th ACM Conference on Computer Supported Cooperative Work \& Social Computing}} (Baltimore, Maryland, USA) \emph{(\bibinfo{series}{CSCW '14})}. \bibinfo{publisher}{Association for Computing Machinery}, \bibinfo{address}{New York, NY, USA}, \bibinfo{pages}{1156--1166}.
\newblock
\showISBNx{9781450325400}
\urldef\tempurl%
\url{https://doi.org/10.1145/2531602.2531701}
\showDOI{\tempurl}


\bibitem[Rafner et~al\mbox{.}(2021)]%
        {Rafner:CC2021}
\bibfield{author}{\bibinfo{person}{Janet Rafner}, \bibinfo{person}{Steven Langsford}, \bibinfo{person}{Arthur Hjorth}, \bibinfo{person}{Miroslav Gajdacz}, \bibinfo{person}{Lotte Philipsen}, \bibinfo{person}{Sebastian Risi}, \bibinfo{person}{Joel Simon}, {and} \bibinfo{person}{Jacob Sherson}.} \bibinfo{year}{2021}\natexlab{}.
\newblock \showarticletitle{Utopian or Dystopian?: using a ML-assisted image generation game to empower the general public to envision the future}. In \bibinfo{booktitle}{\emph{Proceedings of the 13th Conference on Creativity and Cognition}} (Virtual Event, Italy) \emph{(\bibinfo{series}{C\&C '21})}. \bibinfo{publisher}{Association for Computing Machinery}, \bibinfo{address}{New York, NY, USA}, Article \bibinfo{articleno}{53}, \bibinfo{numpages}{5}~pages.
\newblock
\showISBNx{9781450383769}
\urldef\tempurl%
\url{https://doi.org/10.1145/3450741.3466815}
\showDOI{\tempurl}


\bibitem[Ruiz et~al\mbox{.}(2011)]%
        {Ruiz:2011}
\bibfield{author}{\bibinfo{person}{Susana B.~Diaz Ruiz}, \bibinfo{person}{Benjamin Stokes}, {and} \bibinfo{person}{Jeff Watson}.} \bibinfo{year}{2011}\natexlab{}.
\newblock \showarticletitle{Mobile and Locative Games in the ``Civic Tripod'': Activism, Art and Learning}.
\newblock \bibinfo{journal}{\emph{International Journal of Learning and Media}}  \bibinfo{volume}{3} (\bibinfo{year}{2011}).
\newblock


\bibitem[Shi et~al\mbox{.}(2024)]%
        {shi2024}
\bibfield{author}{\bibinfo{person}{Yichun Shi}, \bibinfo{person}{Peng Wang}, {and} \bibinfo{person}{Weilin Huang}.} \bibinfo{year}{2024}\natexlab{}.
\newblock \bibinfo{title}{SeedEdit: Align Image Re-Generation to Image Editing}.
\newblock
\newblock
\showeprint[arxiv]{2411.06686}~[cs.CV]
\urldef\tempurl%
\url{https://arxiv.org/abs/2411.06686}
\showURL{%
\tempurl}


\bibitem[Sivertsen and L\o{}vlie(2024)]%
        {Sivertsen:DIS2024}
\bibfield{author}{\bibinfo{person}{Christian Sivertsen} {and} \bibinfo{person}{Anders~Sundnes L\o{}vlie}.} \bibinfo{year}{2024}\natexlab{}.
\newblock \showarticletitle{Exploring Aesthetic Qualities of Deep Generative Models through Technological (Art) Mediation}. In \bibinfo{booktitle}{\emph{Proceedings of the 2024 ACM Designing Interactive Systems Conference}} (Copenhagen, Denmark) \emph{(\bibinfo{series}{DIS '24})}. \bibinfo{publisher}{Association for Computing Machinery}, \bibinfo{address}{New York, NY, USA}, \bibinfo{pages}{2738--2752}.
\newblock
\showISBNx{9798400705830}
\urldef\tempurl%
\url{https://doi.org/10.1145/3643834.3661498}
\showDOI{\tempurl}


\bibitem[Stenros et~al\mbox{.}(2007)]%
        {Stenros:2007}
\bibfield{author}{\bibinfo{person}{Jaakko Stenros}, \bibinfo{person}{Markus Montola}, {and} \bibinfo{person}{Frans M\"{a}yr\"{a}}.} \bibinfo{year}{2007}\natexlab{}.
\newblock \showarticletitle{Pervasive games in ludic society}. In \bibinfo{booktitle}{\emph{Proceedings of the 2007 Conference on Future Play}} (Toronto, Canada) \emph{(\bibinfo{series}{Future Play '07})}. \bibinfo{publisher}{Association for Computing Machinery}, \bibinfo{address}{New York, NY, USA}, \bibinfo{pages}{30--37}.
\newblock
\showISBNx{9781595939432}
\urldef\tempurl%
\url{https://doi.org/10.1145/1328202.1328209}
\showDOI{\tempurl}


\bibitem[Stokes et~al\mbox{.}(2020)]%
        {Stokes:CHIPLAY2020}
\bibfield{author}{\bibinfo{person}{Benjamin Stokes}, \bibinfo{person}{Hazel Arroyo}, \bibinfo{person}{Mitchell Loewen}, \bibinfo{person}{Tambra Stevenson}, {and} \bibinfo{person}{Chris~J. Karr}.} \bibinfo{year}{2020}\natexlab{}.
\newblock \showarticletitle{A Playful City in the Cards: Sharing Power in Game Design by Extending the Card Metaphor}. In \bibinfo{booktitle}{\emph{Extended Abstracts of the 2020 Annual Symposium on Computer-Human Interaction in Play}} (Virtual Event, Canada) \emph{(\bibinfo{series}{CHI PLAY '20})}. \bibinfo{publisher}{Association for Computing Machinery}, \bibinfo{address}{New York, NY, USA}, \bibinfo{pages}{375--378}.
\newblock
\showISBNx{9781450375870}
\urldef\tempurl%
\url{https://doi.org/10.1145/3383668.3419873}
\showDOI{\tempurl}


\bibitem[Studio et~al\mbox{.}(2024)]%
        {helloLampPost}
\bibfield{author}{\bibinfo{person}{PAN Studio}, \bibinfo{person}{Tom Armitage}, {and} \bibinfo{person}{Gyorgyi Galik}.} \bibinfo{year}{2024}\natexlab{}.
\newblock \bibinfo{title}{About Hello Lamp Post}.
\newblock
\newblock
\urldef\tempurl%
\url{http://www.hellolamppost.co.uk/about}
\showURL{%
\tempurl}


\bibitem[Suh et~al\mbox{.}(2021)]%
        {Suh:CHI2021}
\bibfield{author}{\bibinfo{person}{Minhyang~(Mia) Suh}, \bibinfo{person}{Emily Youngblom}, \bibinfo{person}{Michael Terry}, {and} \bibinfo{person}{Carrie~J Cai}.} \bibinfo{year}{2021}\natexlab{}.
\newblock \showarticletitle{AI as Social Glue: Uncovering the Roles of Deep Generative AI during Social Music Composition}. In \bibinfo{booktitle}{\emph{Proceedings of the 2021 CHI Conference on Human Factors in Computing Systems}} (Yokohama, Japan) \emph{(\bibinfo{series}{CHI '21})}. \bibinfo{publisher}{Association for Computing Machinery}, \bibinfo{address}{New York, NY, USA}, Article \bibinfo{articleno}{582}, \bibinfo{numpages}{11}~pages.
\newblock
\showISBNx{9781450380966}
\urldef\tempurl%
\url{https://doi.org/10.1145/3411764.3445219}
\showDOI{\tempurl}


\bibitem[Sun et~al\mbox{.}(2024)]%
        {Sun:CHI2024}
\bibfield{author}{\bibinfo{person}{Yuan Sun}, \bibinfo{person}{Eunchae Jang}, \bibinfo{person}{Fenglong Ma}, {and} \bibinfo{person}{Ting Wang}.} \bibinfo{year}{2024}\natexlab{}.
\newblock \showarticletitle{Generative AI in the Wild: Prospects, Challenges, and Strategies}. In \bibinfo{booktitle}{\emph{Proceedings of the CHI Conference on Human Factors in Computing Systems}} (Honolulu, HI, USA) \emph{(\bibinfo{series}{CHI '24})}. \bibinfo{publisher}{Association for Computing Machinery}, \bibinfo{address}{New York, NY, USA}, Article \bibinfo{articleno}{747}, \bibinfo{numpages}{16}~pages.
\newblock
\showISBNx{9798400703300}
\urldef\tempurl%
\url{https://doi.org/10.1145/3613904.3642160}
\showDOI{\tempurl}


\bibitem[Sun et~al\mbox{.}(2023)]%
        {Sun:CHIEA2023}
\bibfield{author}{\bibinfo{person}{Yuqian Sun}, \bibinfo{person}{Ying Xu}, \bibinfo{person}{Chenhang Cheng}, \bibinfo{person}{Yihua Li}, \bibinfo{person}{Chang~Hee Lee}, {and} \bibinfo{person}{Ali Asadipour}.} \bibinfo{year}{2023}\natexlab{}.
\newblock \showarticletitle{Explore the Future Earth with Wander 2.0: AI Chatbot Driven By Knowledge-base Story Generation and Text-to-image Model}. In \bibinfo{booktitle}{\emph{Extended Abstracts of the 2023 CHI Conference on Human Factors in Computing Systems}} (Hamburg, Germany) \emph{(\bibinfo{series}{CHI EA '23})}. \bibinfo{publisher}{Association for Computing Machinery}, \bibinfo{address}{New York, NY, USA}, Article \bibinfo{articleno}{450}, \bibinfo{numpages}{5}~pages.
\newblock
\showISBNx{9781450394222}
\urldef\tempurl%
\url{https://doi.org/10.1145/3544549.3583931}
\showDOI{\tempurl}


\bibitem[Thibault(2019)]%
        {Thibault:2019}
\bibfield{author}{\bibinfo{person}{Mattia Thibault}.} \bibinfo{year}{2019}\natexlab{}.
\newblock \showarticletitle{Towards a Typology of Urban Gamification}. In \bibinfo{booktitle}{\emph{Proceedings of the 52nd Hawaii International Conference on System Sciences}}. \bibinfo{publisher}{HICSS}, \bibinfo{pages}{1216--1225}.
\newblock


\bibitem[UNESCO(2022)]%
        {UNESCO:ethics}
\bibfield{author}{\bibinfo{person}{UNESCO}.} \bibinfo{year}{2022}\natexlab{}.
\newblock \bibinfo{title}{Recommendation on the Ethics of Artificial Intelligence}.
\newblock
\newblock
\urldef\tempurl%
\url{https://unesdoc.unesco.org/ark:/48223/pf0000381137}
\showURL{%
\tempurl}


\bibitem[Wagner-Greene et~al\mbox{.}(2017)]%
        {Wagner-Greene:2017aa}
\bibfield{author}{\bibinfo{person}{Victoria~R Wagner-Greene}, \bibinfo{person}{Amy~J Wotring}, \bibinfo{person}{Thomas Castor}, \bibinfo{person}{Jessica Kruger}, \bibinfo{person}{Sarah Mortemore}, {and} \bibinfo{person}{Joseph~A Dake}.} \bibinfo{year}{2017}\natexlab{}.
\newblock \showarticletitle{Pok{\'e}mon GO: Healthy or Harmful?}
\newblock \bibinfo{journal}{\emph{Am J Public Health}} \bibinfo{volume}{107}, \bibinfo{number}{1} (\bibinfo{date}{Jan} \bibinfo{year}{2017}), \bibinfo{pages}{35--36}.
\newblock
\showISSN{1541-0048 (Electronic); 0090-0036 (Print); 0090-0036 (Linking)}
\urldef\tempurl%
\url{https://doi.org/10.2105/AJPH.2016.303548}
\showDOI{\tempurl}


\bibitem[Walsh and Wronsky(2019)]%
        {Walsh:CSCWWiP2019}
\bibfield{author}{\bibinfo{person}{Greg Walsh} {and} \bibinfo{person}{Eric Wronsky}.} \bibinfo{year}{2019}\natexlab{}.
\newblock \showarticletitle{AI + Co-Design: Developing a Novel Computer-supported Approach to Inclusive Design}. In \bibinfo{booktitle}{\emph{Companion Publication of the 2019 Conference on Computer Supported Cooperative Work and Social Computing}} (Austin, TX, USA) \emph{(\bibinfo{series}{CSCW '19 Companion})}. \bibinfo{publisher}{Association for Computing Machinery}, \bibinfo{address}{New York, NY, USA}, \bibinfo{pages}{408--412}.
\newblock
\showISBNx{9781450366922}
\urldef\tempurl%
\url{https://doi.org/10.1145/3311957.3359456}
\showDOI{\tempurl}


\bibitem[Wang et~al\mbox{.}(2019)]%
        {Wang:RTD2019}
\bibfield{author}{\bibinfo{person}{Po-Hao Wang}, \bibinfo{person}{Yu-Ting Cheng}, \bibinfo{person}{Wenn-Chieh Tsai}, {and} \bibinfo{person}{Rung-Huei Liang}.} \bibinfo{year}{2019}\natexlab{}.
\newblock \showarticletitle{Fl\^{a}neur's Phonograph: A Fl\^{a}neur Shift in Urban Exploration}. In \bibinfo{booktitle}{\emph{RTD2019 Research Through Design}}.
\newblock


\bibitem[Weidinger et~al\mbox{.}(2022)]%
        {Laura:FACCT2022}
\bibfield{author}{\bibinfo{person}{Laura Weidinger}, \bibinfo{person}{Jonathan Uesato}, \bibinfo{person}{Maribeth Rauh}, \bibinfo{person}{Conor Griffin}, \bibinfo{person}{Po-Sen Huang}, \bibinfo{person}{John Mellor}, \bibinfo{person}{Amelia Glaese}, \bibinfo{person}{Myra Cheng}, \bibinfo{person}{Borja Balle}, \bibinfo{person}{Atoosa Kasirzadeh}, \bibinfo{person}{Courtney Biles}, \bibinfo{person}{Sasha Brown}, \bibinfo{person}{Zac Kenton}, \bibinfo{person}{Will Hawkins}, \bibinfo{person}{Tom Stepleton}, \bibinfo{person}{Abeba Birhane}, \bibinfo{person}{Lisa~Anne Hendricks}, \bibinfo{person}{Laura Rimell}, \bibinfo{person}{William Isaac}, \bibinfo{person}{Julia Haas}, \bibinfo{person}{Sean Legassick}, \bibinfo{person}{Geoffrey Irving}, {and} \bibinfo{person}{Iason Gabriel}.} \bibinfo{year}{2022}\natexlab{}.
\newblock \showarticletitle{Taxonomy of Risks posed by Language Models}. In \bibinfo{booktitle}{\emph{Proceedings of the 2022 ACM Conference on Fairness, Accountability, and Transparency}} (Seoul, Republic of Korea) \emph{(\bibinfo{series}{FAccT '22})}. \bibinfo{publisher}{Association for Computing Machinery}, \bibinfo{address}{New York, NY, USA}, \bibinfo{pages}{214--229}.
\newblock
\showISBNx{9781450393522}
\urldef\tempurl%
\url{https://doi.org/10.1145/3531146.3533088}
\showDOI{\tempurl}


\bibitem[Weisz et~al\mbox{.}(2024)]%
        {Weisz:CHI2024}
\bibfield{author}{\bibinfo{person}{Justin~D. Weisz}, \bibinfo{person}{Jessica He}, \bibinfo{person}{Michael Muller}, \bibinfo{person}{Gabriela Hoefer}, \bibinfo{person}{Rachel Miles}, {and} \bibinfo{person}{Werner Geyer}.} \bibinfo{year}{2024}\natexlab{}.
\newblock \showarticletitle{Design Principles for Generative AI Applications}. In \bibinfo{booktitle}{\emph{Proceedings of the CHI Conference on Human Factors in Computing Systems}} (Honolulu, HI, USA) \emph{(\bibinfo{series}{CHI '24})}. \bibinfo{publisher}{Association for Computing Machinery}, \bibinfo{address}{New York, NY, USA}, Article \bibinfo{articleno}{378}, \bibinfo{numpages}{22}~pages.
\newblock
\showISBNx{9798400703300}
\urldef\tempurl%
\url{https://doi.org/10.1145/3613904.3642466}
\showDOI{\tempurl}


\bibitem[Weisz et~al\mbox{.}(2023)]%
        {Weisz:CoRR2023}
\bibfield{author}{\bibinfo{person}{Justin~D. Weisz}, \bibinfo{person}{Michael~J. Muller}, \bibinfo{person}{Jessica He}, {and} \bibinfo{person}{Stephanie Houde}.} \bibinfo{year}{2023}\natexlab{}.
\newblock \showarticletitle{Toward General Design Principles for Generative {AI} Applications}.
\newblock \bibinfo{journal}{\emph{CoRR}}  \bibinfo{volume}{abs/2301.05578} (\bibinfo{year}{2023}).
\newblock
\showeprint[arXiv]{2301.05578}


\bibitem[Wetzel et~al\mbox{.}(2008)]%
        {Wetzel:2008}
\bibfield{author}{\bibinfo{person}{Richard Wetzel}, \bibinfo{person}{Rod McCall}, \bibinfo{person}{Anne-Kathrin Braun}, {and} \bibinfo{person}{Wolfgang Broll}.} \bibinfo{year}{2008}\natexlab{}.
\newblock \showarticletitle{Guidelines for designing augmented reality games}. In \bibinfo{booktitle}{\emph{Proceedings of the 2008 Conference on Future Play: Research, Play, Share}} (Toronto, Ontario, Canada) \emph{(\bibinfo{series}{Future Play '08})}. \bibinfo{publisher}{Association for Computing Machinery}, \bibinfo{address}{New York, NY, USA}, \bibinfo{pages}{173--180}.
\newblock
\showISBNx{9781605582184}
\urldef\tempurl%
\url{https://doi.org/10.1145/1496984.1497013}
\showDOI{\tempurl}


\bibitem[Wetzel et~al\mbox{.}(2016)]%
        {Wetzel:2016}
\bibfield{author}{\bibinfo{person}{Richard Wetzel}, \bibinfo{person}{Tom Rodden}, {and} \bibinfo{person}{Steve Benford}.} \bibinfo{year}{2016}\natexlab{}.
\newblock \showarticletitle{Developing Ideation Cards for Mixed Reality Game Design}. In \bibinfo{booktitle}{\emph{Proceedings of the First Joint International Conference of Digital Games Research Association and Foundation of Digital Games (DiGRA/FDG 2016)}}.
\newblock


\bibitem[Xu et~al\mbox{.}(2023)]%
        {Xu:2023}
\bibfield{author}{\bibinfo{person}{Fengli Xu}, \bibinfo{person}{Jun Zhang}, \bibinfo{person}{Chen Gao}, \bibinfo{person}{Jie Feng}, {and} \bibinfo{person}{Yong Li}.} \bibinfo{year}{2023}\natexlab{}.
\newblock \bibinfo{title}{Urban Generative Intelligence (UGI): A Foundational Platform for Agents in Embodied City Environment}.
\newblock
\newblock
\urldef\tempurl%
\url{https://arxiv.org/abs/2312.11813}
\showURL{%
\tempurl}


\bibitem[Xu et~al\mbox{.}(2024)]%
        {Xu:2024}
\bibfield{author}{\bibinfo{person}{Haowen Xu}, \bibinfo{person}{Femi Omitaomu}, \bibinfo{person}{Soheil Sabri}, \bibinfo{person}{Sisi Zlatanova}, \bibinfo{person}{Xiao Li}, {and} \bibinfo{person}{Yongze Song}.} \bibinfo{year}{2024}\natexlab{}.
\newblock \showarticletitle{Leveraging generative AI for urban digital twins: a scoping review on the autonomous generation of urban data, scenarios, designs, and 3D city models for smart city advancement}.
\newblock \bibinfo{journal}{\emph{Urban Informatics}} \bibinfo{volume}{3}, \bibinfo{number}{1} (\bibinfo{year}{2024}), \bibinfo{pages}{29}.
\newblock
\showISBNx{2731-6963}
\urldef\tempurl%
\url{https://doi.org/10.1007/s44212-024-00060-w}
\showDOI{\tempurl}


\bibitem[Zhu(2023)]%
        {zhu:2023}
\bibfield{author}{\bibinfo{person}{Dongwen Zhu}.} \bibinfo{year}{2023}\natexlab{}.
\newblock \showarticletitle{City Ai: A strategic framework for artificial intelligence integration in city development}.
\newblock \bibinfo{journal}{\emph{Available at SSRN 4559021}} (\bibinfo{year}{2023}).
\newblock
\urldef\tempurl%
\url{http://dx.doi.org/10.2139/ssrn.4559021}
\showURL{%
\tempurl}


\end{thebibliography}

\appendix

\end{document}